\documentclass[12pt,letterpaper]{article}

\usepackage[utf8]{inputenc}
\usepackage[T1]{fontenc}
\usepackage{lmodern}

\usepackage{geometry}
\usepackage{setspace}
\usepackage{titlesec}
\titleformat{\section}{\normalfont\large\bfseries}{\thesection}{1em}{}[\vspace{0.5ex}]
\titleformat{\subsection}{\normalfont\normalsize\bfseries}{\thesubsection}{1em}{}[\vspace{0.3ex}]
\titlespacing*{\section}{0pt}{1.5ex plus 0.5ex minus 0.2ex}{1ex plus 0.2ex}
\titlespacing*{\subsection}{0pt}{1.2ex plus 0.4ex minus 0.2ex}{0.8ex plus 0.2ex}

\usepackage{fancyhdr}
\usepackage{abstract}
\renewenvironment{abstract}
  {\small
   \begin{center}
   \textbf{Abstract}
   \end{center}
   \quotation\noindent\ignorespaces}
  {\endquotation\clearpage}

\makeatletter
\newcommand{\@institutions}{}
\newcommand{\@corresponding}{}

\newcommand{\institutions}[1]{\renewcommand{\@institutions}{#1}}
\newcommand{\corresponding}[1]{\renewcommand{\@corresponding}{#1}}

\renewcommand\maketitle{
  \begin{center}
    {\Large\bfseries\@title \par}
    \vskip 1.5em
    {\normalsize\@author \par}
    \vskip 1em
    {\small\@institutions \par}
    \vskip 0.8em
    {\small\@corresponding \par}
    \vskip 1em
    {\small\@date \par}
  \end{center}
  \vskip 2em
}
\makeatother

\usepackage{graphicx}
\usepackage{amsmath}
\usepackage{amssymb}
\usepackage{appendix}
\usepackage{float}
\usepackage{subcaption}
\usepackage[hidelinks]{hyperref}

\title{Statistical methods for reference-free single-molecule localisation microscopy analysis}
\author{Jack Peyton$^{1^*}$, Benjamin Davis$^{2^*}$, Emily Gribbin$^{1}$, Daniel Rolfe$^{2}$, Hannah Mitchell$^{1}$}
\institutions{
$^{1}$Mathematical Sciences Research Centre, School of Mathematics and Physics, Queen's University Belfast, University Road, Belfast, BT7 1NN, United Kingdom. \\
$^{2}$OCTOPUS Group, Central Laser Facility, Research Complex at Harwell, Science and Technologies Facilities Council, Appleton Laboratory, Oxfordshire, OX11 0FA, United Kingdom.
}
\corresponding{$^{*}$Corresponding authors: jpeyton01@qub.ac.uk, benjamin.davis@stfc.ac.uk}

\usepackage{booktabs}
\usepackage{placeins}
\usepackage{xcolor}

\begin{document}

\maketitle

\begin{abstract}
MINFLUX (Minimal Photon Flux) is a single-molecule imaging technique capable of resolving fluorophores at a precision of $<5 $ nm. Interpretation of the point patterns generated by this technique presents challenges due to variable emitter density, incomplete bio-labelling of target molecules and their detection, error prone measurement processes, and the presence of spurious (non-structure associated) fluorescent detections. Together, these challenges ensure structural inferences from single-molecule imaging datasets are non-trivial in the absence of strong \textit{a priori} information, for all but the smallest of point patterns. In addition, current methods often require subjective parameter tuning and presuppose known structural templates, limiting reference-free discovery. We present a statistically grounded, end-to-end analysis framework. Focusing on MINFLUX derived datasets and leveraging Bayesian and spatial statistical methods, a pipeline is presented that demonstrates 1) uncertainty aware clustering of measurements into emitter groups that performs better than current gold standards, 2) rapid identification of molecular structure supergroups, and 3) reconstruction of repeating structures within the dataset without substantial prior knowledge. This pipeline is demonstrated using simulated and real MINFLUX datasets, where emitter clustering and centre detection maintain high performance (emitter subset assignment accuracy $ > 0.75$) across all conditions evaluated, while structural inference achieves reliable discrimination ($F1 \approx 0.9$) at high labelling efficiency. Template-free reconstruction of Nup96 and DNA-Origami \(3\times3\) grids are achieved.
\end{abstract}

\section{Introduction}
MINFLUX is a single-molecule localisation microscopy (SMLM) technique first demonstrated by Balzarotti \textit{et al.} in 2017 with a reported nanometre precision \cite{balzarotti_nanometer_2017, gwosch_minflux_2020}. The derivation of biological insights from this valuable and innovative technique are, however, hampered by frequent methodological issues in post-hoc analysis of the resulting point patterns. Criticisms levied against the technique by Prakash \cite{prakash_at_2022} highlight problems typical of analysis of SMLM data: tunable parameters at multiple stages of analysis, opaque filtering and clustering steps, and a final phase of matching known template structures resulting in a ``what you already know is what you get'' paradigm. A recent dialogue between Gwosch \textit{et al.} and Prakash \textit{et al.} \cite{gwosch_minflux_2020, prakash_assessment_2021, gwosch_assessment_2022, prakash_at_2022} recognise a problem with validation of the interpretation of SMLM data, in that the circularity of presupposed parameters and template alignment risks a systematic confirmation bias across SMLM techniques.

The discussion regarding SMLM analysis highlights the requirement for an end-to-end, statistically grounded framework, that can be applied cross-technique to more objectively interpret this valuable nanometre scale data. Such a framework involves clustering events - or measurements - into emitters, grouping emitters into structurally informative sets, and inferring structural detail from these emitter sets. Each of these steps is impacted by the underlying limitations of the microscopy technique itself: bio-labelling that masks structure, and spurious fluorescent detections difficult to parse from meaningful fluorophores. 

The first stage, clustering measurements into emitters, is often performed with density-based clustering algorithms, such as DBSCAN \cite{10.5555/3001460.3001507, hammer_density-based_2024} or HDBSCAN \cite{campello2013density, mcinnes2017accelerated, malzer2020hybrid}. While these algorithms are rapid and scalable, they introduce user-tunable parameters that can substantially impact clustering \cite{khater_review_2020}, affect downstream analysis, and are not measurement uncertainty-aware. Bayesian alternatives to emitter estimations, such as BaGoL, \cite{fazel_bayesian_2019, fazel_high-precision_2022} are powerful yet computationally expensive and prior-sensitive.

At the structural inference stage, the analysis framework for SMLM is further exposed to the limitations of the field: incomplete bio-labelling and bio-label detection masking structure, dense packing of emitters into regions of interest (ROIs), and structural heterogeneity \cite{lelek_single-molecule_2021} (e.g. multiple distinct or nested structures within the same ROI). Template-based approaches attempt to solve this by aligning sampled data subsets to models known \textit{a priori} \cite{gwosch_minflux_2020} however missingness in true structure, owing to under-labelling and spurious detections \cite{prakash_assessment_2021}, can result in hallucinations of structure that do not exist in the data, similar to the well documented ``Einstein from noise'' phenomenon \cite{balanov2024einstein}. Alternative approaches to structural inference include topology \cite{wu_facam_2023, edelsbrunner_three-dimensional_1994}, graph- and machine learning-based approaches inspired by network analysis in systems biology \cite{milenkoviae_uncovering_2008, pineda_spatial_2024, khater_super_2018}. These techniques similarly depend on the choice of scale, opaque data pre-processing, tunable heuristics and frequently neglect the estimation of uncertainty in the discovered structure.

Together, these issues point to the need for an end-to-end, statistically grounded analysis pipeline that (i) treats localisation uncertainty as first-order information, (ii) reduces subjective parameter tuning, (iii) scales beyond single structures, and (iv) avoids the hard-wiring of specific templates into the analysis. Such a framework should provide a transparent chain from raw measurements to emitters, from emitters to structural centres, and from centres to interpretable molecular architectures, while remaining applicable across localisation-based super-resolution techniques, and establish performance boundaries - documenting not only where methods succeed but also where they approach fundamental information-theoretic limits - to enable informed experimental design and avoid over-interpretation of sparse or noisy data.

\begin{figure}[ht]
    \centering
    \subfloat[]{\includegraphics[width=0.33\linewidth]{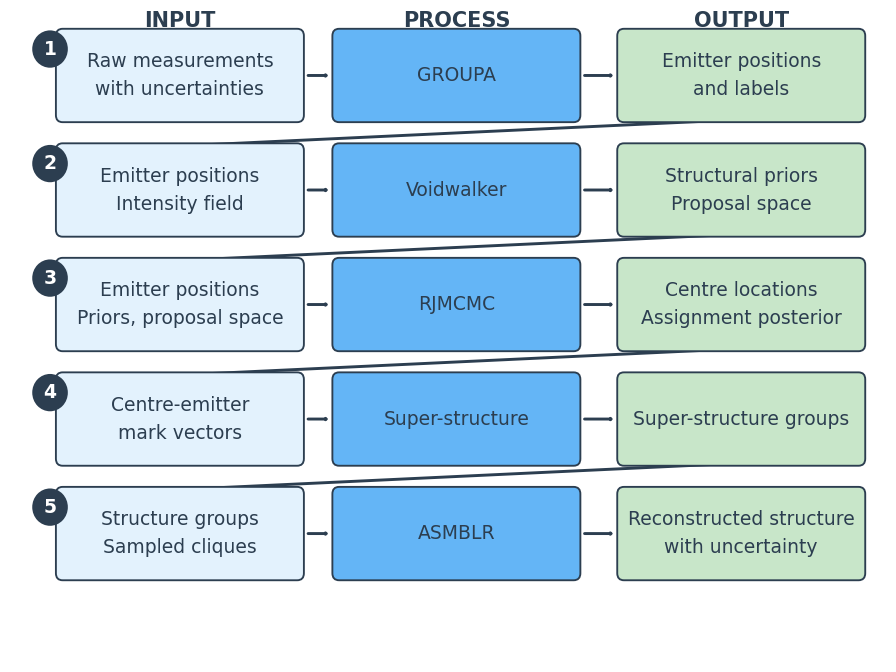}}
    \subfloat[]{\includegraphics[width=0.33\linewidth]{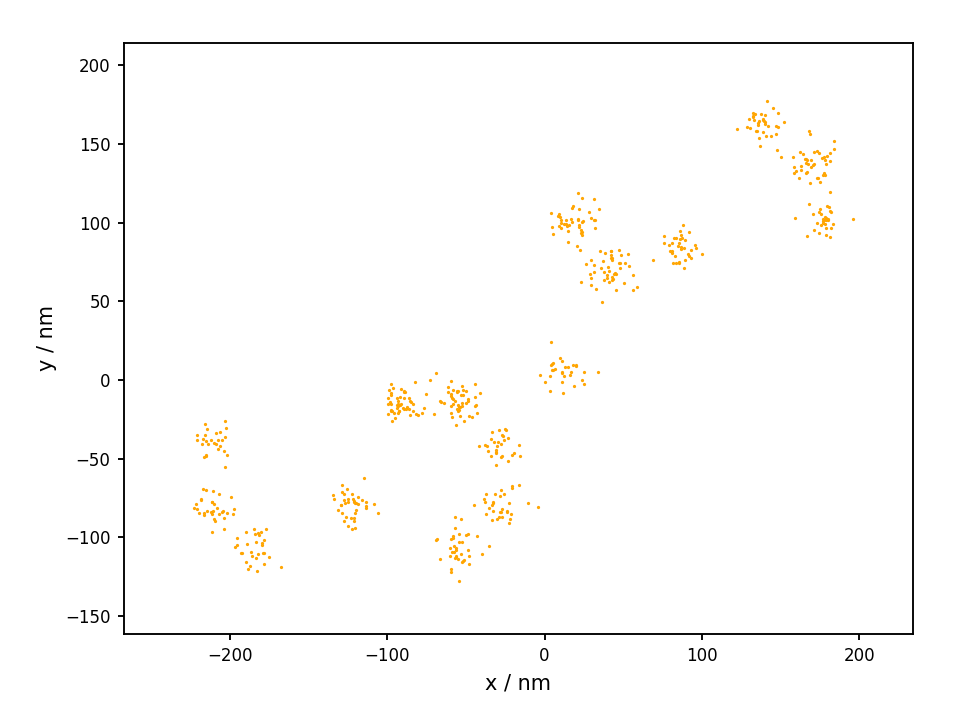}}
    \subfloat[]{\includegraphics[width=0.33\linewidth]{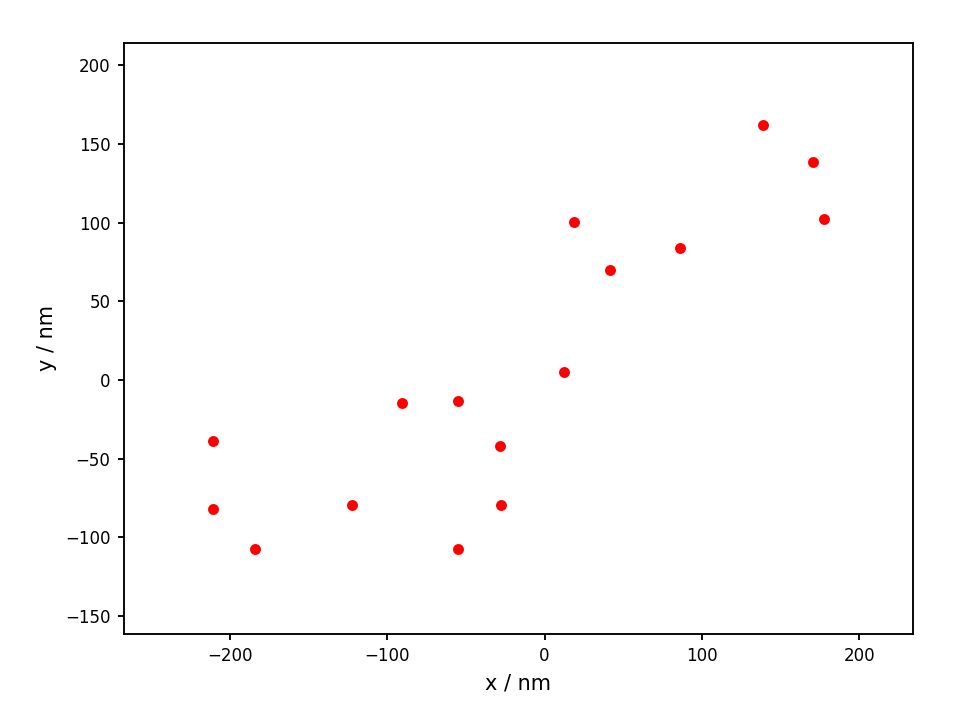}} \\
    \subfloat[]{\includegraphics[width=0.33\linewidth]{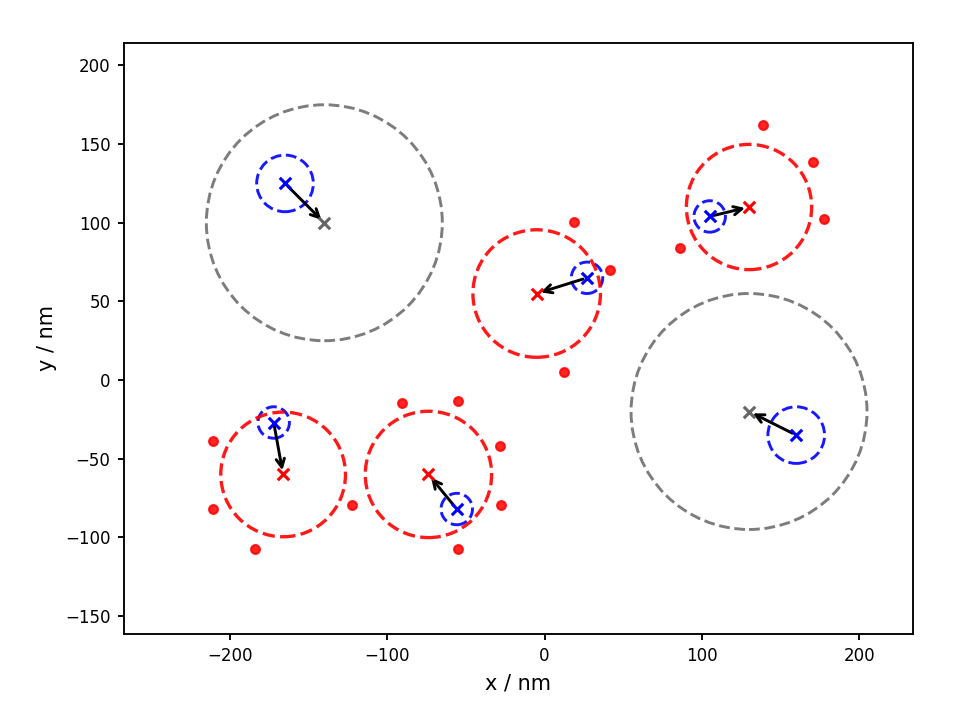}} 
    \subfloat[]{\includegraphics[width=0.33\linewidth]{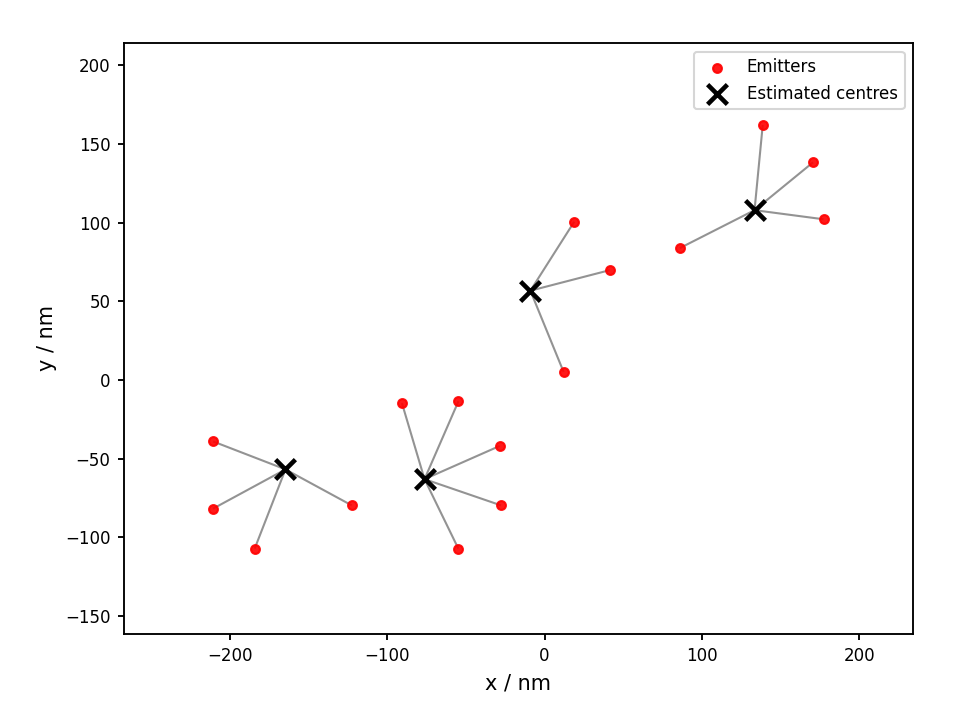}}
    \subfloat[]{\includegraphics[width=0.33\linewidth]{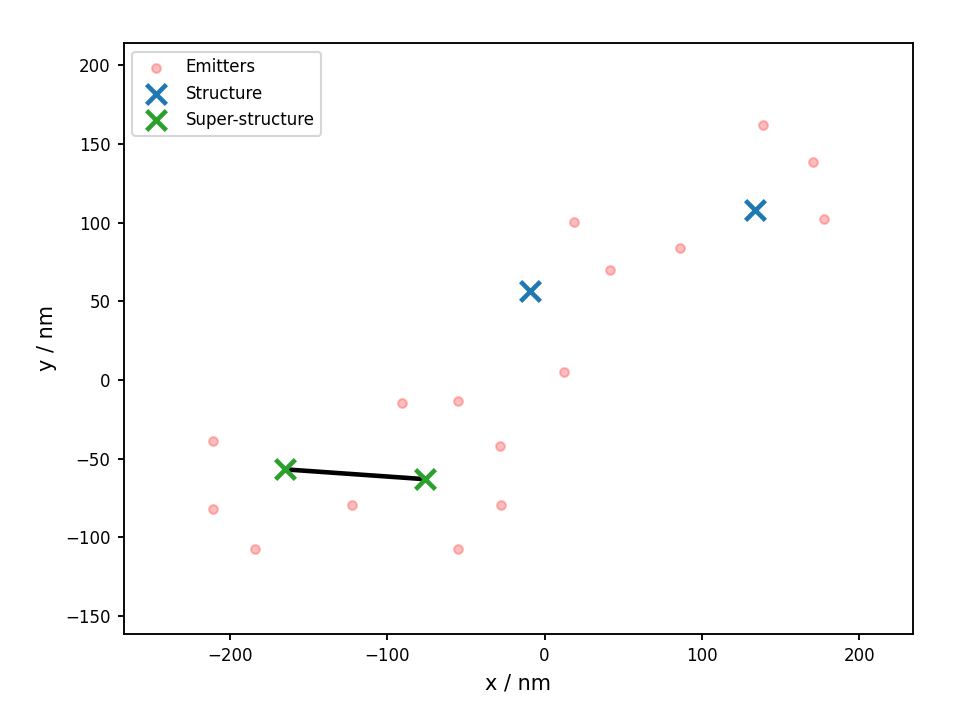}} \\
    \subfloat[]{\includegraphics[width=0.33\linewidth]{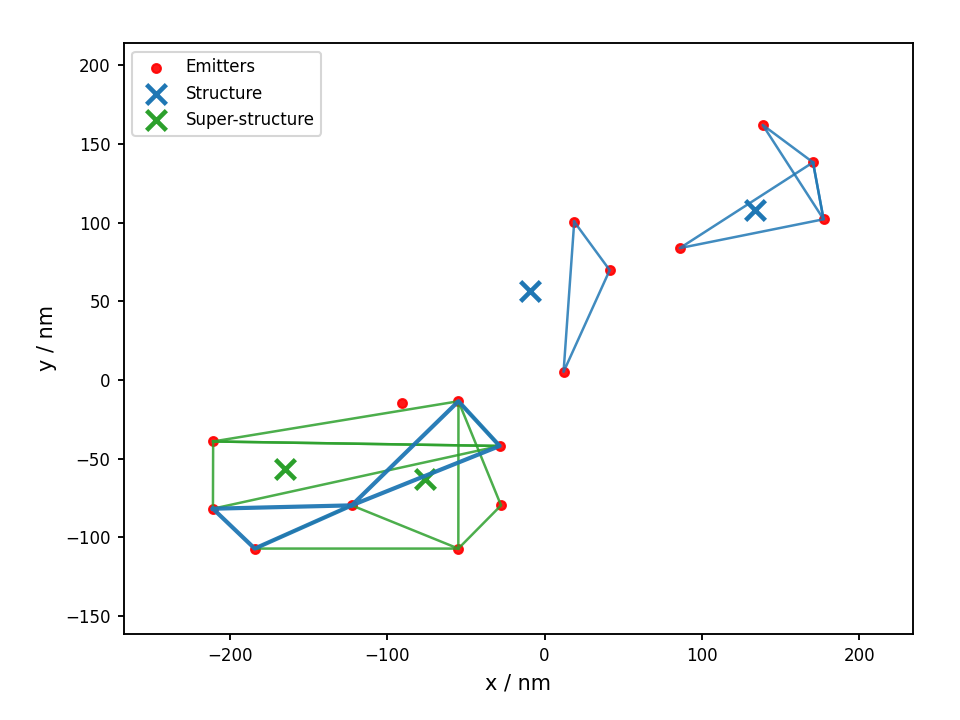}} 
    \subfloat[]{\includegraphics[width=0.33\linewidth]{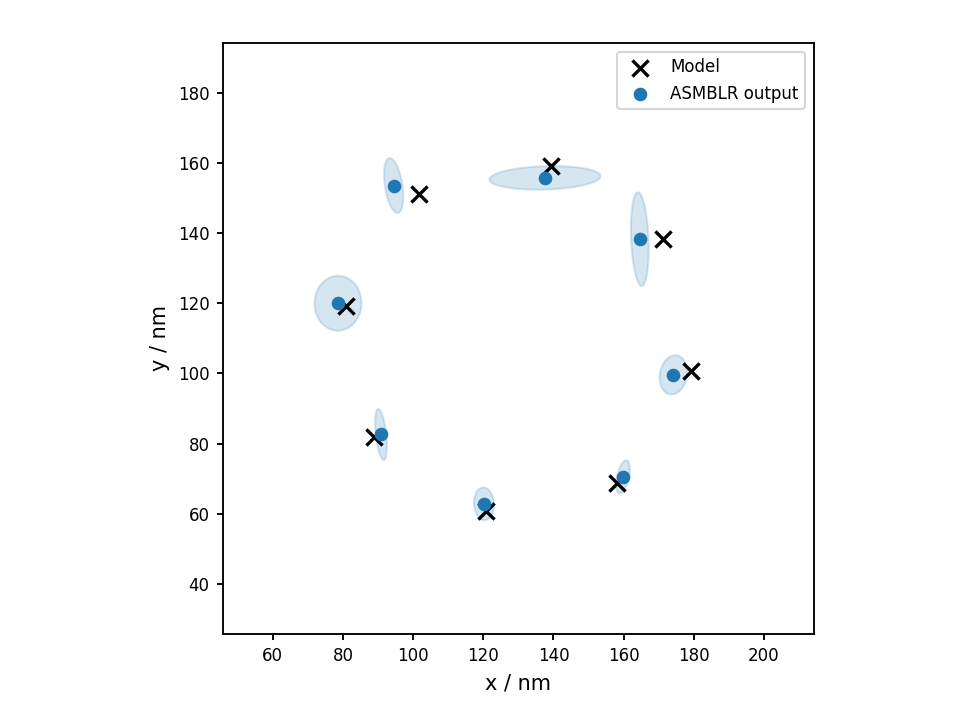}}
    \subfloat[]{\includegraphics[width=0.33\linewidth]{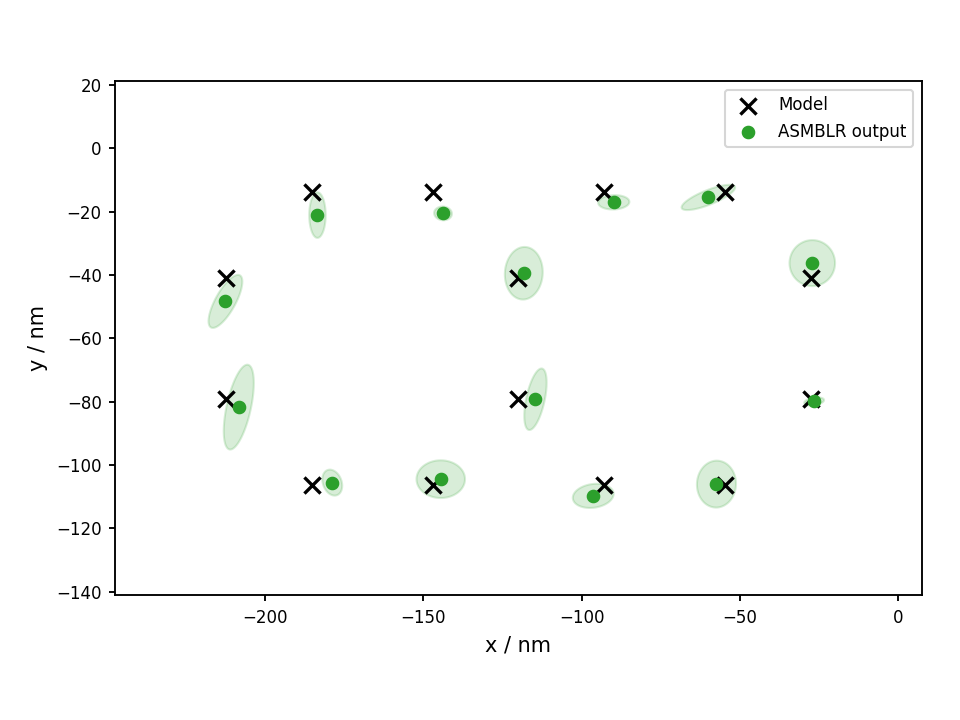}}
    \caption{(a) End-to-end input/output overview for each stage of the framework. (b) Raw localisations are clustered into (c) emitters using GROUPA. (d) Voidwalker distinguishes statistically significant empty space (dashed red circles) that inform priors and proposal space. (e) RJMCMC sampler assigns emitters to structural centres, yielding per-emitter probability distributions over centre assignments. (f) Assignment distributions define marks in a marked point process to identify structure (blue) and super-structure (green). (g) Cliques are sampled from co-assigned emitter populations of both structure (blue) and super-structure (green). ASMBLR reconstructs molecular (h) structure and (i) super-structure from the inner space of sampled cliques.}
    \label{fig:pipeline}
\end{figure}

Here, we outline one such framework for MINFLUX and localisation-based SMLM (Fig. \ref{fig:pipeline}(a)). First, in Grouping Observations Under Pairwise Associations (GROUPA) (Fig. \ref{fig:pipeline}(c)), we replace heuristic density clustering with a Bayesian measurement-to-emitter model that couples pairwise Bayes factors and the Infomap community detection algorithm \cite{Smiljanic2023MapEquation} to infer emitter assignments without user-tuned distance thresholds, making use of known measurement uncertainty (SI Sec. 1.1). Second, Voidwalker (Fig. \ref{fig:pipeline}(d)), an empty-space–seeking algorithm grounded in spatial point process theory, identifies statistically significant voids in the emitter pattern and uses them to define a data-driven proposal space and priors (SI Sec. 1.2) for downstream Reversible Jump Markov Chain Monte Carlo (RJMCMC) \cite{green_reversible_1995}. Third, we model structural centres as a Gibbs point process and use an RJMCMC sampler with a BaGoL-inspired move set \cite{fazel_high-precision_2022} to infer centre locations and provide, for each emitter, an uncertainty-aware distribution over centre assignments (Fig. \ref{fig:pipeline}(e)). Fourth, we interpret these emitter-centre assignments as a marked point process to identify super-structure arrangements of the inferred structures, partitioning the dataset into individual molecular units (Fig. \ref{fig:pipeline}(f)). This RJMCMC and super-structure discovery allows guided sampling of fully connected subgraphs, or cliques (Fig. \ref{fig:pipeline}(g)), that improves the probability of sampling structurally representative cliques (SI Sec. 2.2). Lastly, for each unit, we apply a molecular reconstruction algorithm, Assembling Structured Molecular Building blocks from Localisation Reconstructions (ASMBLR), which uses a population of sampled cliques from co-assigned emitter populations and uses their internal geometry to rapidly reconstruct repeating structural motifs without strong \textit{a priori} information (Fig. \ref{fig:pipeline}(h, i)).

We simulated data representing the 2D projection of Nup96, a nucleoporin commonly used as an SMLM benchmark \cite{thevathasan_nuclear_2019}, motivating its use here. Each stage of the pipeline was validated on this simulated Nup96 nuclear pore complex data, spanning labelling efficiencies (probabilities of successful fluorophore labelling) of \(l=0.3, 0.6, 0.9\), and 1.0, with clutter (spurious, non-structure associated emitters) proportions of 0, 0.1, 0.2, and 0.3 across 100 replicate datasets of each permutation of these conditions. GROUPA is benchmarked against DBSCAN and HDBSCAN for emitter clustering under a range of measurement uncertainties. Voidwalker-Gibbs is assessed for centre-detection accuracy and emitter-to-centre assignment under incomplete labelling and spurious detections. Super-structure detection via marked point processes is evaluated for its ability to distinguish true super-structure and higher-order assemblies from chance spatial proximity. We demonstrate the framework on experimental MINFLUX datasets of Nup96, showing template-free identification of known nuclear pore stoichiometry. The framework is additionally applied to a synthetic DNA-Origami \(3\times3\) grid image, demonstrating its applicability for multi-component structures beyond Nup96. It is shown that template-free, full reconstruction of molecular structure is possible under minimal \textit{a priori} assumptions regarding the structure itself, improving the reproducibility of analysis of experimental results.

\FloatBarrier
\section{Results}
The analysis pipeline is benchmarked against 1600 synthetic Nup96 MINFLUX datasets. These datasets are carried through all stages of the pipeline (example figures demonstrating this are found in SI Sec. 3). For GROUPA benchmarking, additional datasets under expanding measurement uncertainties are simulated. 

\subsection{GROUPA}
In the synthetic Nup96 data, each emitter is tagged with a ground truth ID. Measurements are linked to their parent ID through the same tag. This allows computation of Adjusted Rand Index (ARI)\cite{hubert1985comparing}, Normalised Mutual Information (NMI)\cite{vinh2010information} and Fowlkes–Mallows Index (FMI)\cite{fowlkes1983method} directly between the inferred emitter labels and the ground truth. For this emitter-level benchmark we fix labelling and clutter to isolate the effects of measurement uncertainty. We compare GROUPA against DBSCAN and HDBSCAN across measurement uncertainty values \(\sigma\in\{0.5, 1, 2.5,5, 7,9,12,15,20\}\text{nm}\).

\begin{figure}[ht]
    \centering
    \subfloat[]{\includegraphics[width=0.33\linewidth]{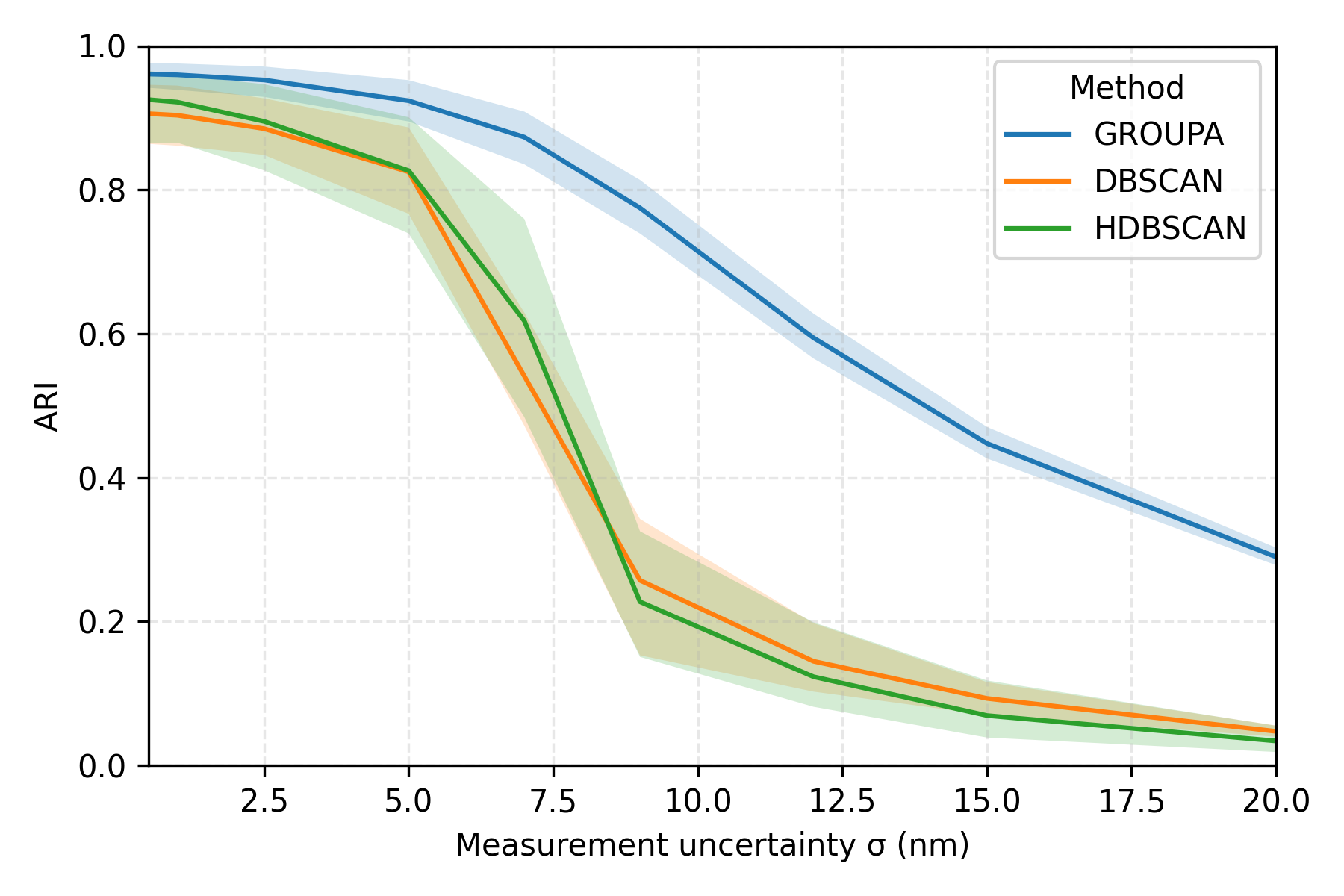}}
    \subfloat[]{\includegraphics[width=0.33\linewidth]{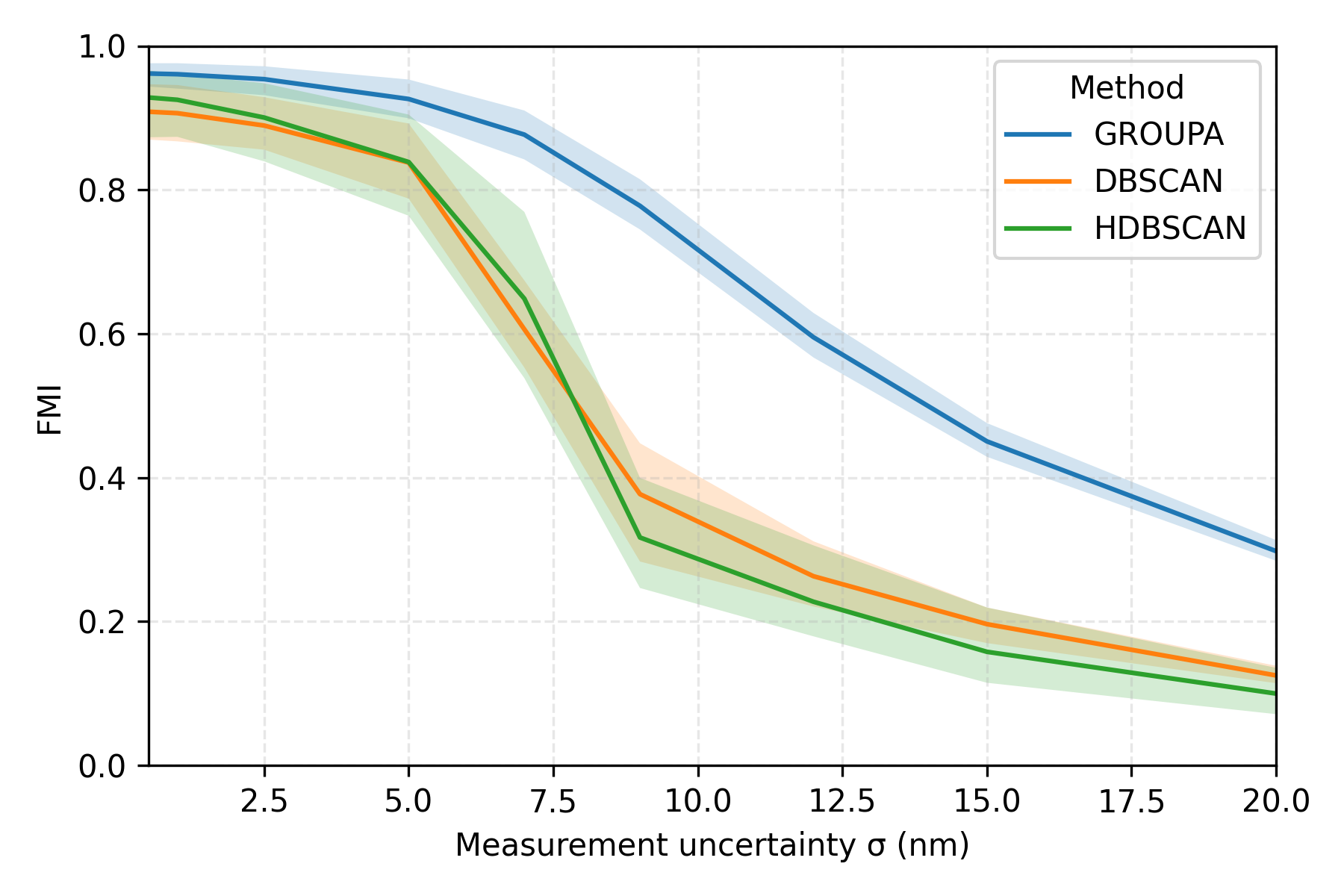}}
    \subfloat[]{\includegraphics[width=0.33\linewidth]{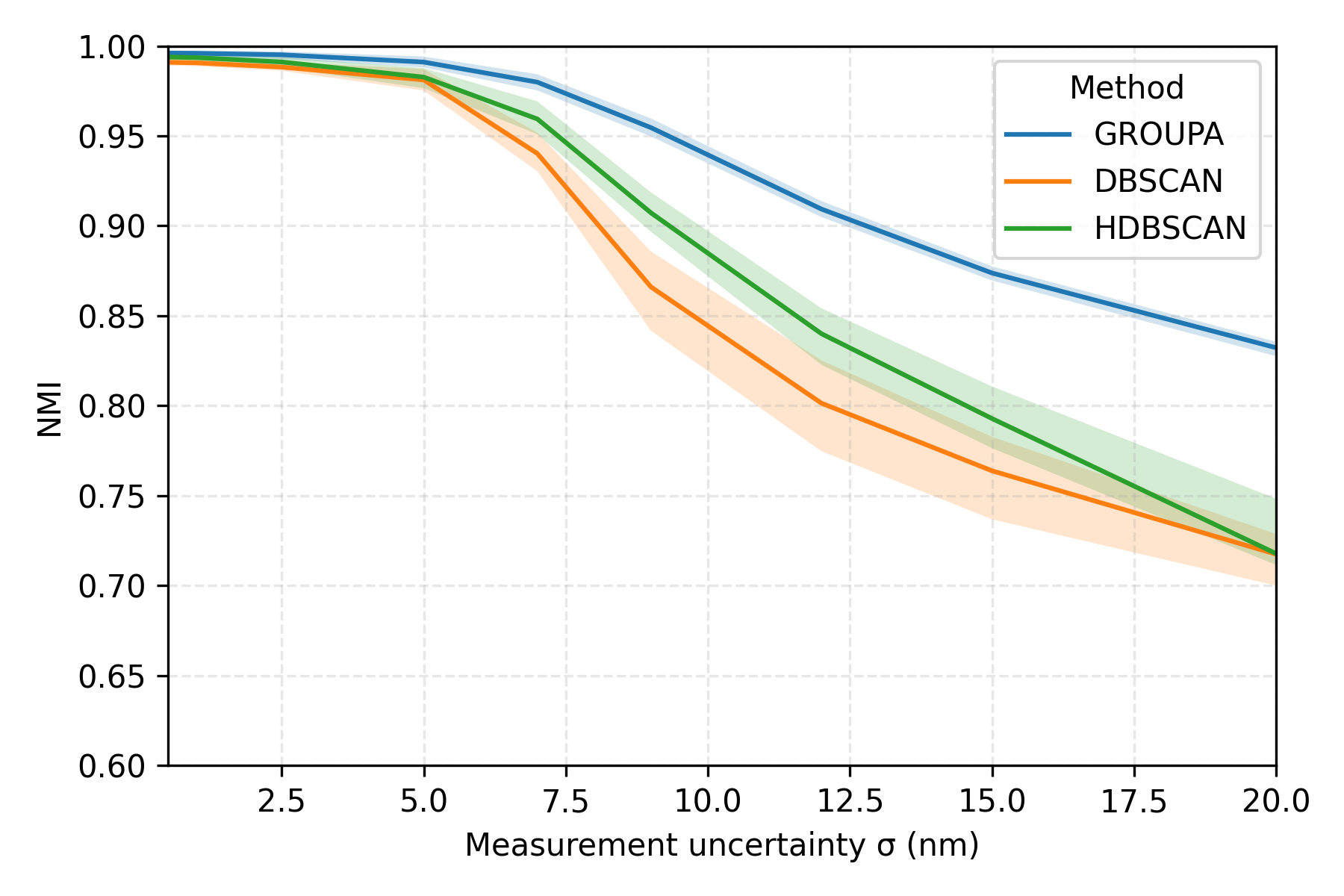}}
    \caption{Clustering performance versus localisation uncertainty $\sigma$ for GROUPA, DBSCAN, and HDBSCAN, evaluated on synthetic Nup96 data with 50 replicates per \(\sigma\). Shaded bands: 2.5-97.5th percentiles across replicates. (a) ARI, (b) FMI, (c) NMI. DBSCAN/HDBSCAN were tuned per \(\sigma\); GROUPA required no parameter tuning.}
    \label{fig:groupa-res}
\end{figure}

Fig. \ref{fig:groupa-res} compares GROUPA with DBSCAN and HDBSCAN baselines across a range of uncertainty values. For each uncertainty level, DBSCAN and HDBSCAN hyperparameters were tuned to maximise performance on that condition; these curves therefore represent condition-optimised baselines rather than a single fixed parametrisation. ARI and FMI show steep decline across all algorithms as the measurement uncertainty increases. NMI decays more slowly, consistent with partial retention of coarse partition structure even as individual emitters become unresolved. GROUPA maintains ARI\(\geq0.75\) up to \(\sigma=10\) nm, where DBSCAN/HDBSCAN hold ARI\(\approx0.2\). All methods degrade substantially beyond this point, highlighting a fundamental information limit independent of algorithm choice. The improved robustness comes at the cost of higher computational overhead, yet GROUPA remains applicable to arbitrarily large point patterns in a way that per-ROI RJMCMC methods such as BaGoL \cite{fazel_high-precision_2022} are not. We have not compared GROUPA directly to BaGoL, as the methods address different analytical scales (SI Sec. 2.1).

\subsection{Voidwalker-Gibbs}\label{sec:vw-g}
Each emitter in the synthetic data is linked to its parent structural centre via a group ID. Voidwalker-inferred centres are aligned to the ground truth via Hungarian algorithm \cite{kuhn1955hungarian}. An emitter is correctly assigned if the sampled centre label from the Voidwalker process matches the true group ID in the ground truth.

Bio-labelling and detection efficiency were represented through application of combined binomial probability to each emitter. This resulted in the partial observation of some structures in the localisation set, and may lead to the complete absence of a structure if the observation probability is sufficiently low. A centre that is present in the ground truth but has no labelled emitters is unrecoverable from the data. Moreover, spurious emitters will be estimated via additional centres, further deflating the \(F1\). Thus, one should note that \(F1\) is bounded by unobserved features and high clutter levels, and while this remains a useful metric, the labelling and clutter agnostic assignment accuracy is a more valuable measure to judge the results. We evaluate the inferred global radius parameter against its ground truth value of 50nm. The ground-truth emitter positions themselves are perturbed by a radial spread of $\pm 1.5\,\mathrm{nm}$, so relative radius errors below $\sim3\%$ are within the generative noise of the simulator.

\begin{figure}[ht]
    \centering
    \subfloat[]{\includegraphics[width=0.33\linewidth]{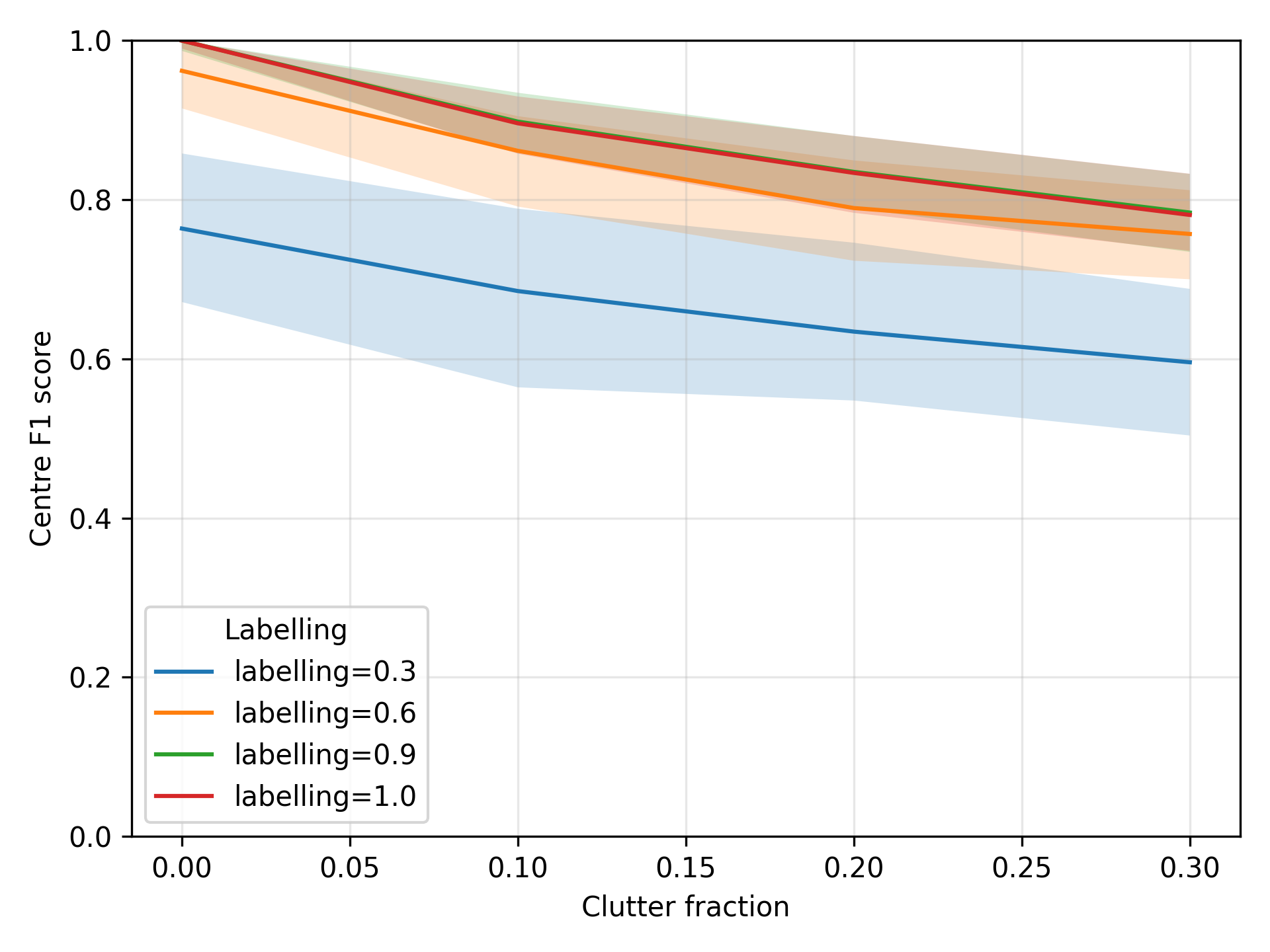}}
    \subfloat[]{\includegraphics[width=0.33\linewidth]{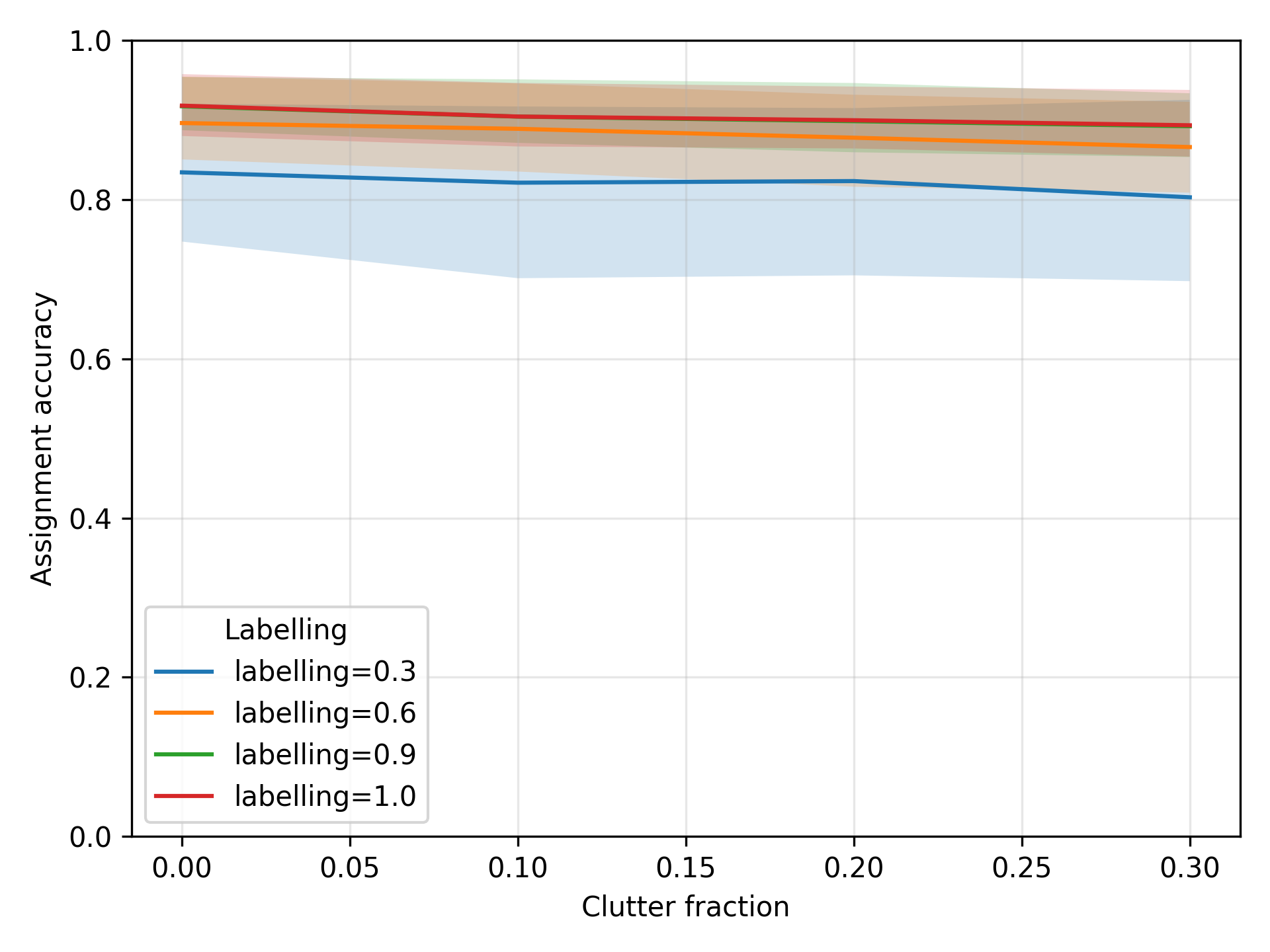}} 
    \subfloat[]{\includegraphics[width=0.33\linewidth]{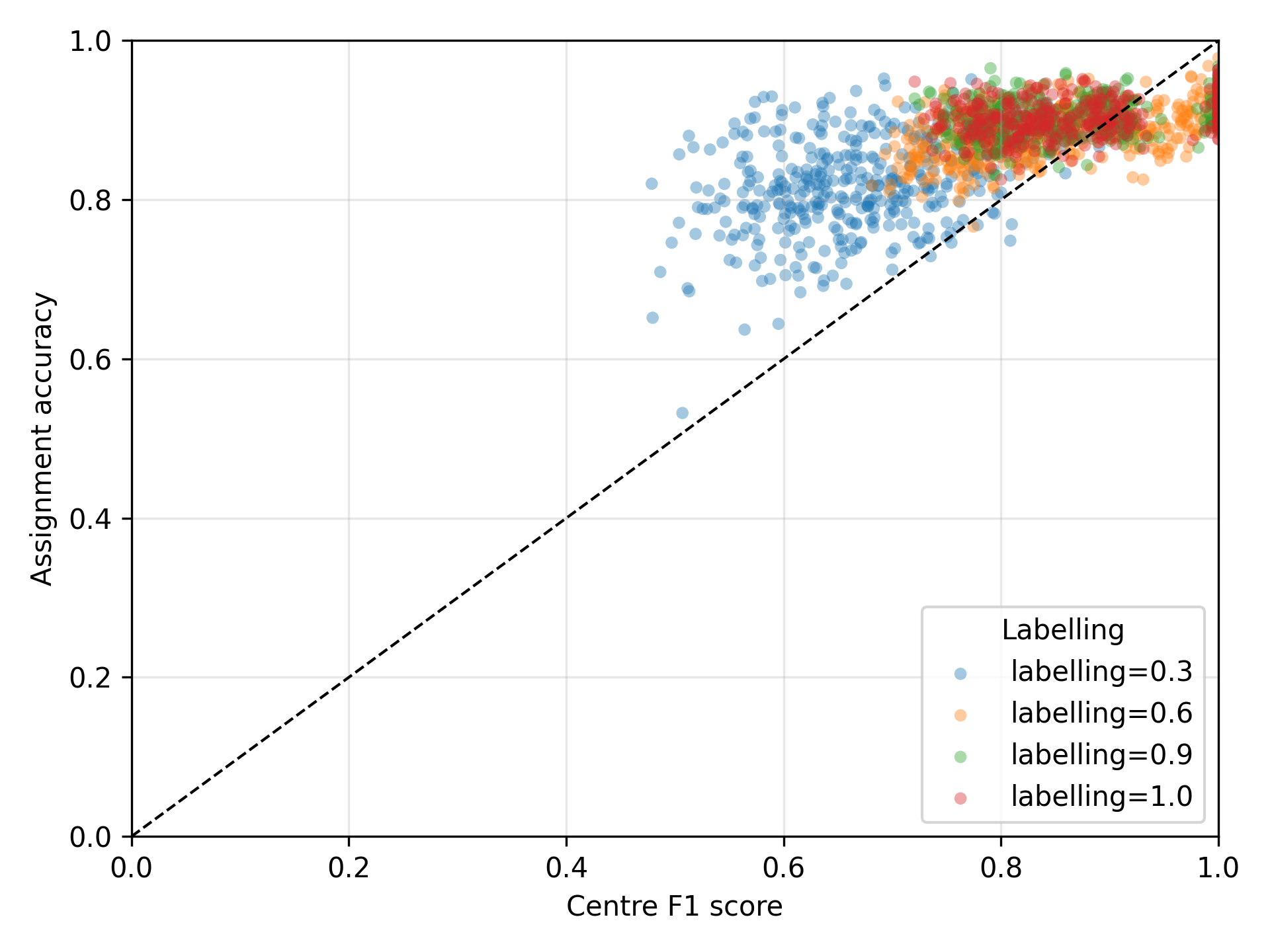}} \\
    \subfloat[]{\includegraphics[width=0.33\linewidth]{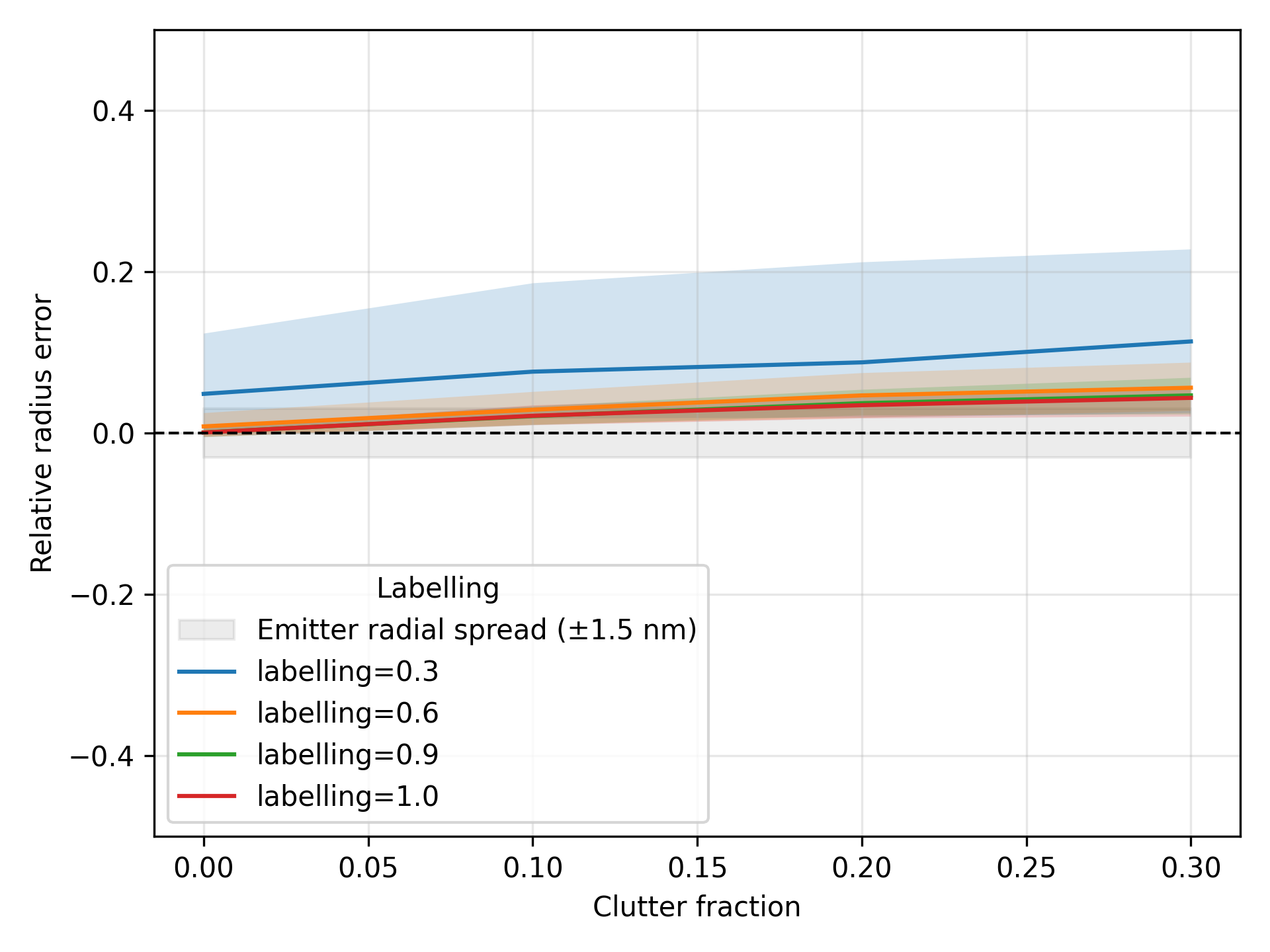}}
    \subfloat[]{\includegraphics[width=0.33\linewidth]{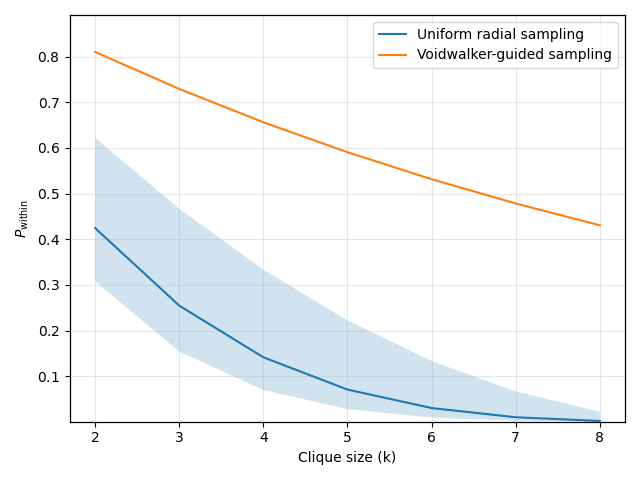}}
    \subfloat[]{\includegraphics[width=0.33\linewidth]{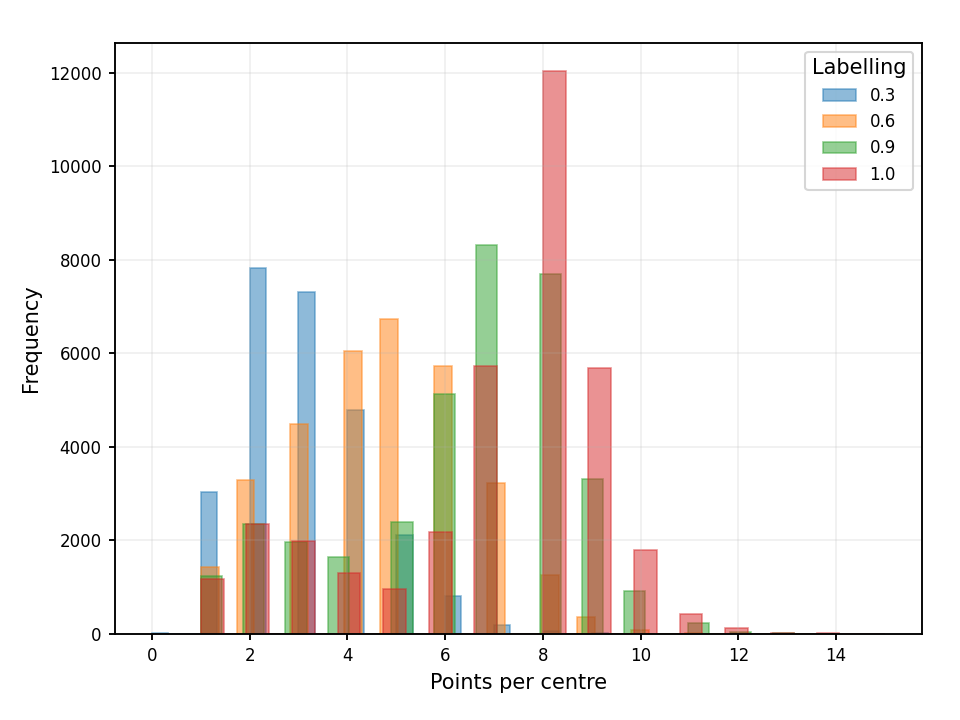}}
    \caption{Voidwalker-Gibbs achieves centre detection and emitter assignment across data quality regimes. Performance on synthetic Nup96 structures across labelling efficiencies of 0.3, 0.6, 0.9, 1.0 and clutter levels of $0-30\%$. Curves show median performance over 100 replicates with 2.5-97.5th percentile bands. (a) Centre-level F1 score versus clutter. (b) Emitter-centre assignment accuracy versus clutter. (c) Joint distribution of F1 and assignment accuracy across all datasets. (d) Relative radius bias versus clutter, with ground truth emitter uncertainty. (e) Probability of sampling a structurally representative clique versus clique size, comparing radial uniform and Voidwalker-guided sampling. Uniform sampling assumes 8 true emitters with 2-6 spurious neighbours; Voidwalker-guided sampling assumes 90\% per-emitter assignment accuracy. (f) No. emitters assigned per centre across labelling efficiencies.}
    \label{fig:vw-g-res}
\end{figure}

Overall, Fig.~\ref{fig:vw-g-res} shows that structurally significant subsets of emitters are well-recovered under Voidwalker-Gibbs, across a range of labelling and clutter conditions. Fig.~\ref{fig:vw-g-res}(a) confirms centre detection degrades with increasing clutter, most severely at 0.3 labelling where median F1 drops from 0.75 to 0.6. Above 0.6 labelling, F1 remains \(\geq0.75\) with narrow uncertainty bands. Emitter-centre assignment accuracy (Fig. \ref{fig:vw-g-res}(b)) proves more robust than centre-level F1, remaining above 0.8 even at 0.3 labelling. Fig. \ref{fig:vw-g-res}(c) confirms assignment accuracy systematically exceeds F1, indicating strict F1 scoring penalises unobservable centres rather than widespread misassignment. Inferred radius (Fig. \ref{fig:vw-g-res}(d)) shows small positive bias within the \(\pm1.5\) nm ground-truth uncertainty for moderate-to-high labelling. At 0.3 labelling, radius is overestimated under high clutter but remains modest. Convergence diagnostics (Gelman-Rubin \cite{brooks1998} \(\hat{R}<1.1\)) confirmed stable mixing.

For our purposes, structural inference concerns local-scale information, such as protein oligomers < 200 nm in size. We previously developed FLImP \cite{needham2016egfr, kingwell2018antibacterial, iyer2024DrugresistantEGFRmutations}, which uniformly samples 2–3 emitter cliques within a specified radius. However, naive uniform sampling increasingly favours inter-structure over intra-structure cliques as clique size grows (SI Sec. 2.2). Voidwalker addresses this by exploiting similarity in the structure of empty space in repeating but under sampled motifs in the point pattern to group emitters into structurally meaningful sets to facilitate uniform sampling of predominantly intra-structure cliques. Fig. \ref{fig:vw-g-res}(e) shows that Voidwalker improves the probability of sampling one such meaningful set, \(P_{\text{within}}\), substantially in comparison to radial uniform sampling of cliques of emitters, particularly when sampling larger cliques. Fig. \ref{fig:vw-g-res}(f) highlights the number of emitters assigned to each estimated centre across various labelling efficiencies, and offers some guidance on the selection of both the clique size to sample, and model size to estimate, using ASMBLR. At 1.0 labelling, for instance, a large concentration is observed at 8 points-per-centre, indicating a potential target structure of 8, and permitting reasonable clique sampling of size \(k\leq8\). In the 0.3 labelling case, no such spike at 8 points per centre exists - and as such no guidance on model size may be available in such under-sampled data - but a concentration of 2-3 points per centre is shown, suggesting that cliques of \(k\in\{2,3\}\) is a viable sampling strategy.

\subsection{Super-structure Discovery}
Most structures comprise multiple voids; a DNA-Origami \(3\times3\) grid has four voids, for example. Super-structure discovery identifies voids occurring in closer spatial proximity than expected under a CSR hard-shell null hypothesis; grouped voids form single structural units for clique sampling. Each structure in the simulated data has a small probability of extending into a connected pair of regular polygons. In such cases, the pair is considered a super-structure, with two component structures. In the DNA-Origami case, the overarching grid is considered a super-structure, and each of the four cells that make up this grid are considered a structure. 

\begin{figure}[ht]
    \centering
    \subfloat[]{\includegraphics[width=\linewidth]{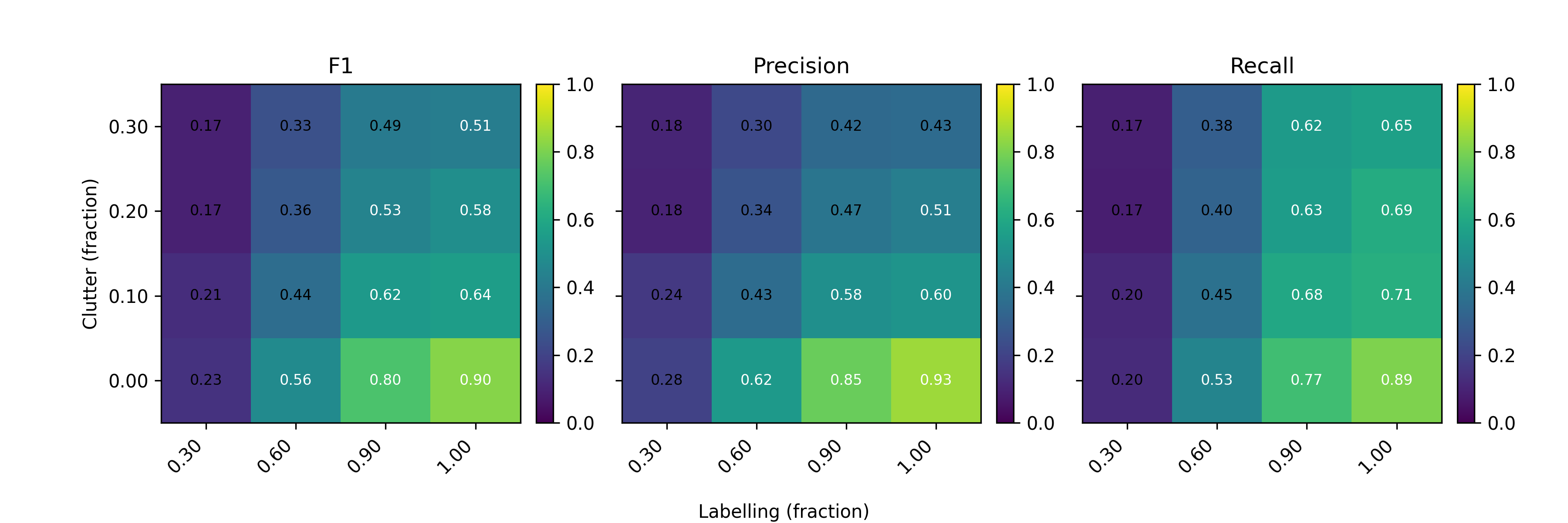}} \\
    \subfloat[]{\includegraphics[width=0.4\linewidth]{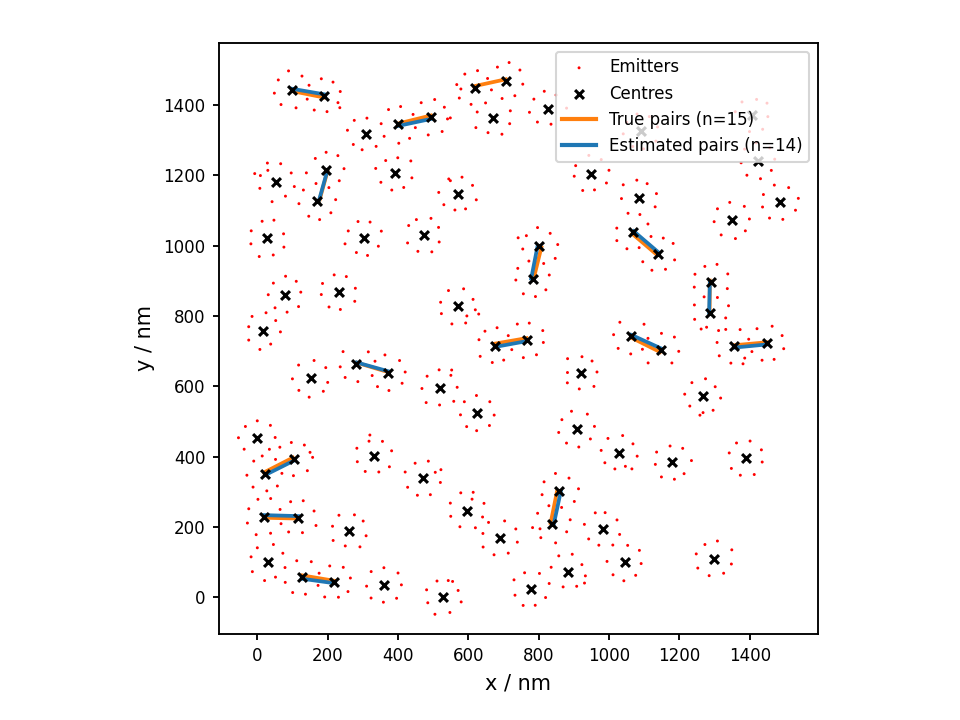}}
    \subfloat[]{\includegraphics[width=0.4\linewidth]{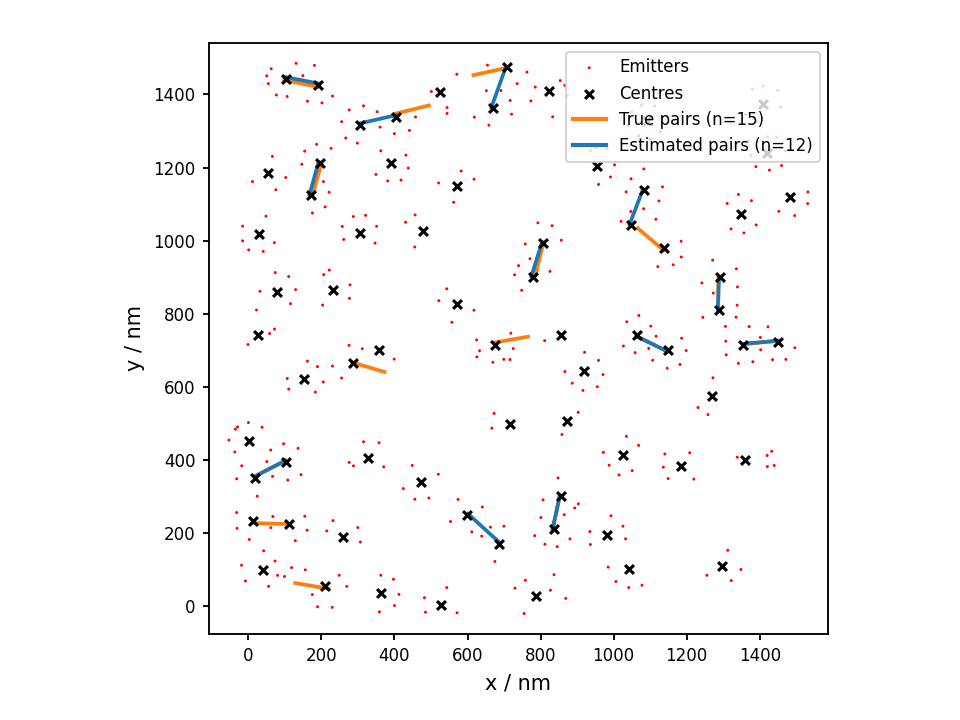}}
    \caption{Mark-based super-structure detection establishes data-quality thresholds distinct from centre detection. (a) Performance across labelling efficiencies (0.3, 0.6, 0.9, 1.0) and clutter levels ($0-30\%)$ on synthetic Nup96 dimer mixtures. Performance surfaces for F1 (left), precision (centre), and recall (right) over 100 replicates per condition. Numerical values indicate mean metrics. Super-structure discovery algorithm applied to labelling efficiencies of (b) 1.0 and (c) 0.6.}
    \label{fig:super-metrics}
\end{figure}

We develop a super-structure discovery algorithm, that links inferred centres into super-structures. We transform the emitter-centre assignment probabilities of Sec. \ref{sec:vw-g} into a per-centre probability distribution, itself considered a mark vector. These mark vectors undergo randomly labelled, permutative null simulations to distinguish between close spatial proximity by chance, and true super-structure (SI Sec. 1.4). We evaluate the edge prediction between estimated pairs against ground-truth pairs using precision, recall, and \(F1\) (Fig. \ref{fig:super-metrics}(a)).

Performance is heavily impacted by labelling (Fig. \ref{fig:super-metrics}(a). Under high labelling efficiency $(0.9-1.0)$ and low clutter (<0.1), the method achieves high \(F1\) \((\approx 0.8-0.9)\), indicating reliable discrimination of true super-structure edges. At intermediate labelling (0.6), performance drops to moderate levels ($F1$ $\approx 0.3-0.6$), reflecting reduced separability between true and false centre pairs as the marks become more sparse. At low labelling (0.3), performance collapses across all clutter regimes ($F1$ $\approx 0.1-0.2$), establishing an effective information threshold for this form of oligomer inference. Fig. \ref{fig:super-metrics}(b, c) highlight the effect that emitter sparsity has on the super-structure discovery algorithm as a result of both missing information at this stage, and the propagation of errors from previous steps in the pipeline.

This degradation reflects an informational bottleneck rather than a strictly algorithmic failure. At 0.3 labelling of 8-fold symmetrical structure, each centre is represented by, on average, $2-3$ emitters. The resulting posterior responsibility distributions have Shannon entropy approaching that of a uniform distribution, yielding Bhattacharyya distances between true constituent sub-structures that are statistically indistinguishable from random centre pairs (SI Sec. 2.3). Weighting the permutation testing by the posterior intensity of the field accounts for spatial heterogeneity, but cannot overcome signal collapse in information-poor marks obtained from under-labelled data. Thus, we establish a stricter data-quality requirement for super-structure inference, than for emitter sub-grouping.

\subsection{ASMBLR}
ASMBLR seeks to reconstruct complete structures (or repeating motifs) from a set of under-sampled cliques. This algorithm is required as clique populations are sampled from under-labelled and error prone measurement processes and to overcome limitations with template matching.

\begin{figure}[ht]
    \centering
    \subfloat[]{\includegraphics[width=0.22\linewidth]{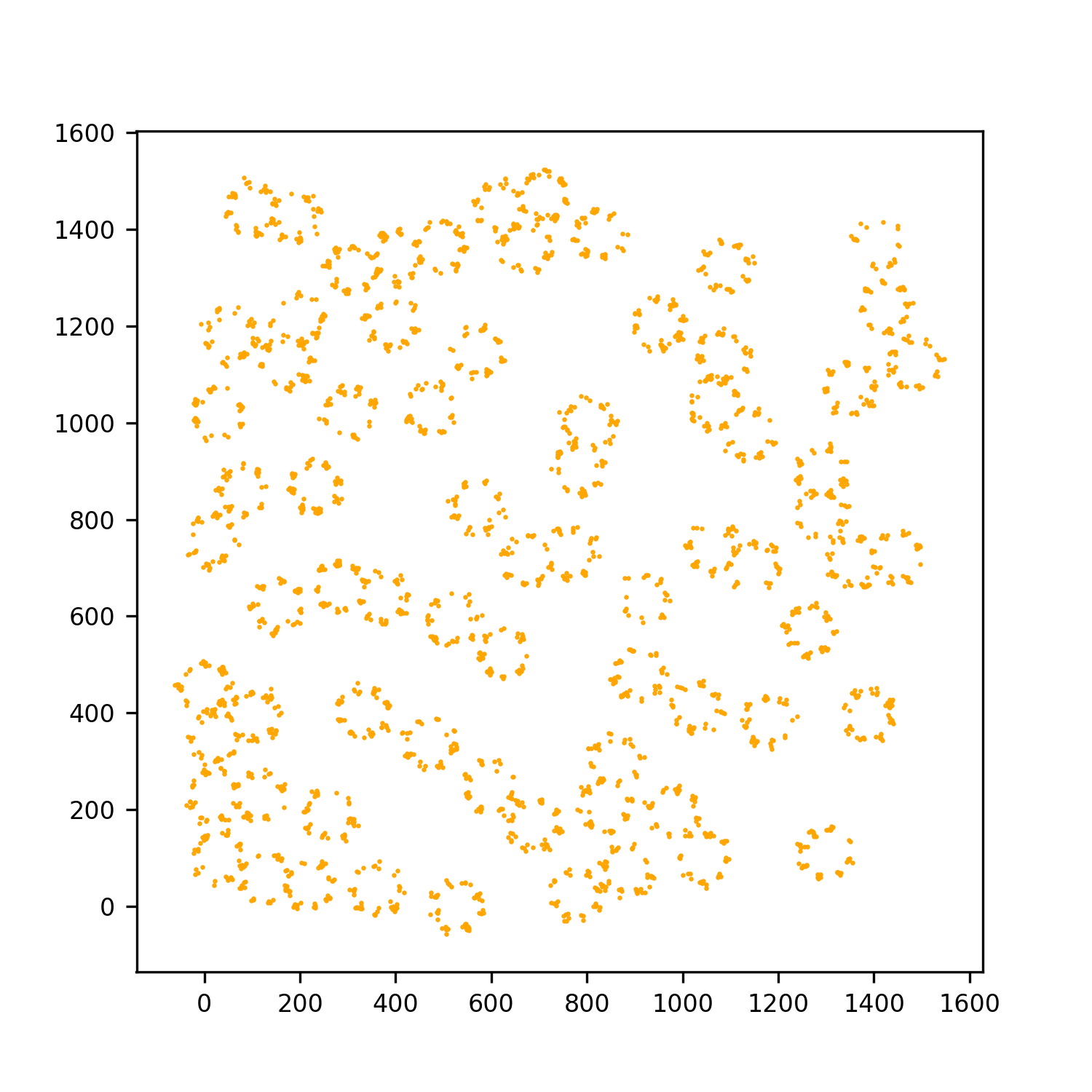}}
    \subfloat[]{\includegraphics[width=0.275\linewidth]{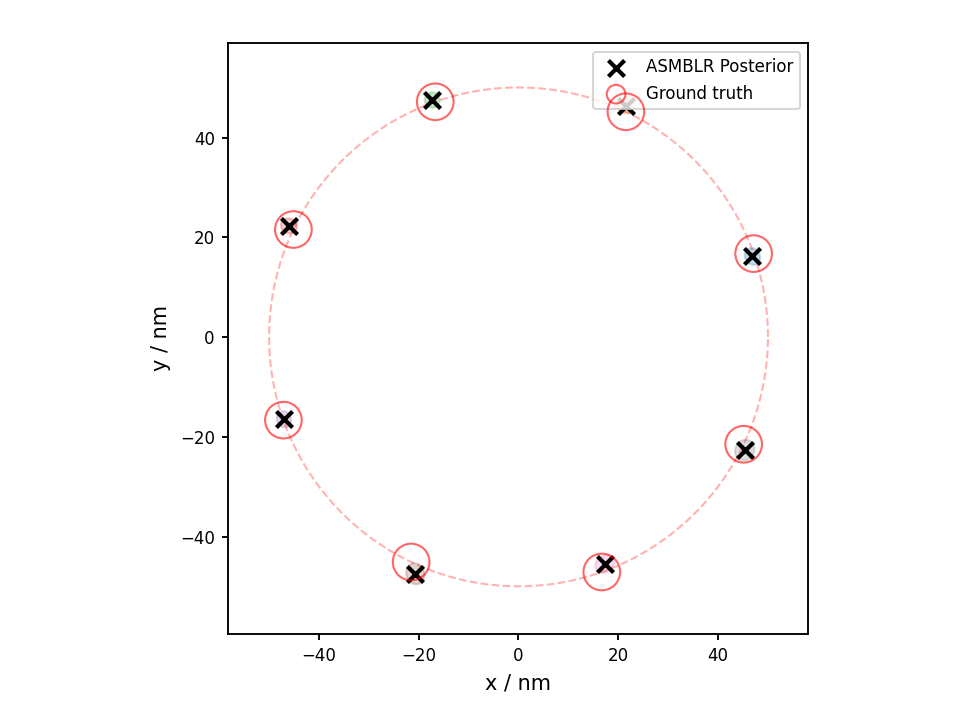}}
    \subfloat[]{\includegraphics[width=0.22\linewidth]{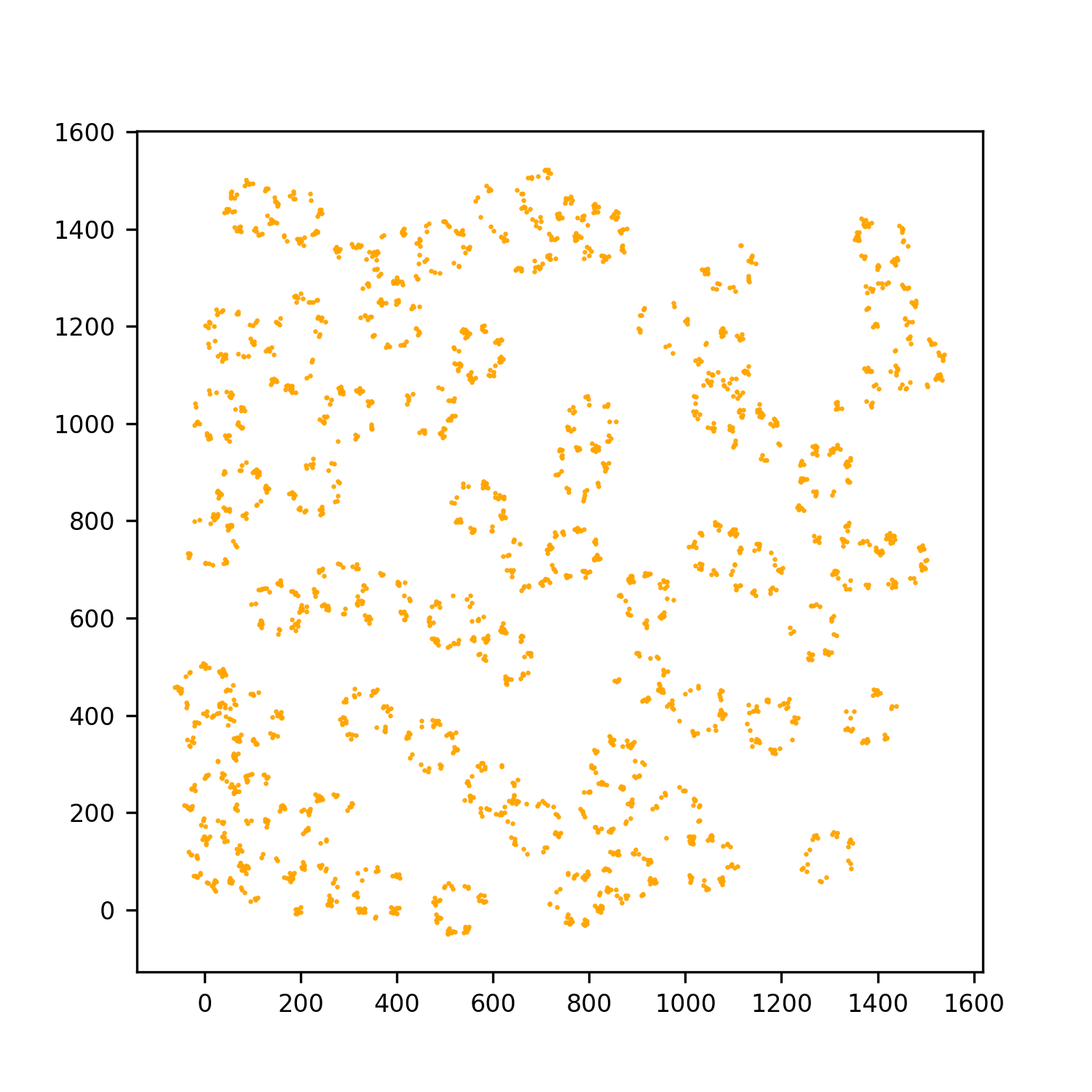}}
    \subfloat[]{\includegraphics[width=0.275\linewidth]{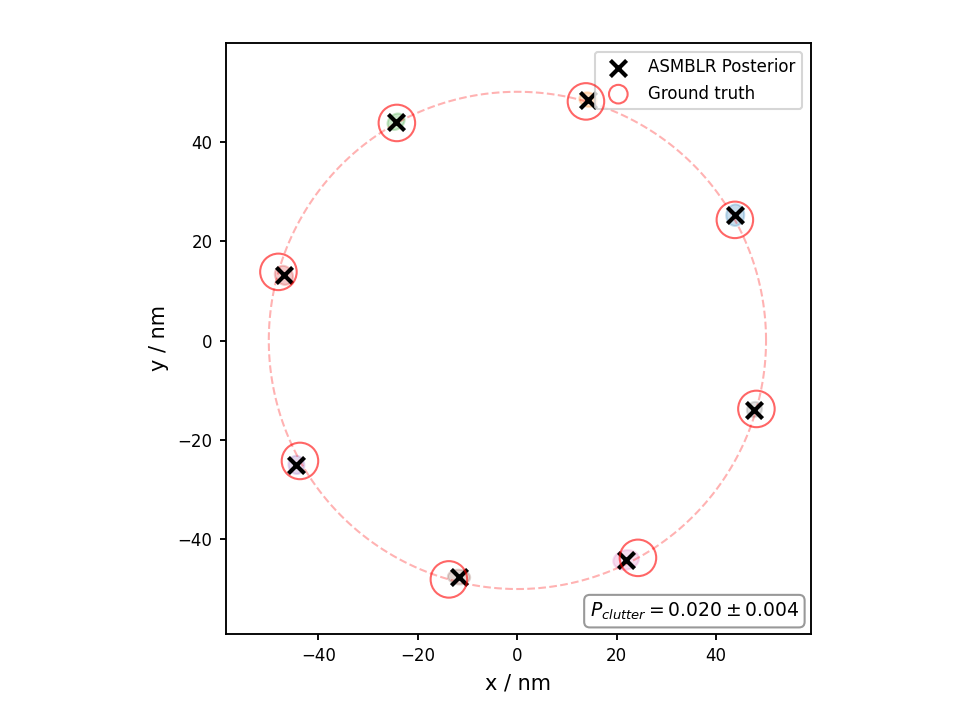}} \\
    \subfloat[]{\includegraphics[width=0.22\linewidth]{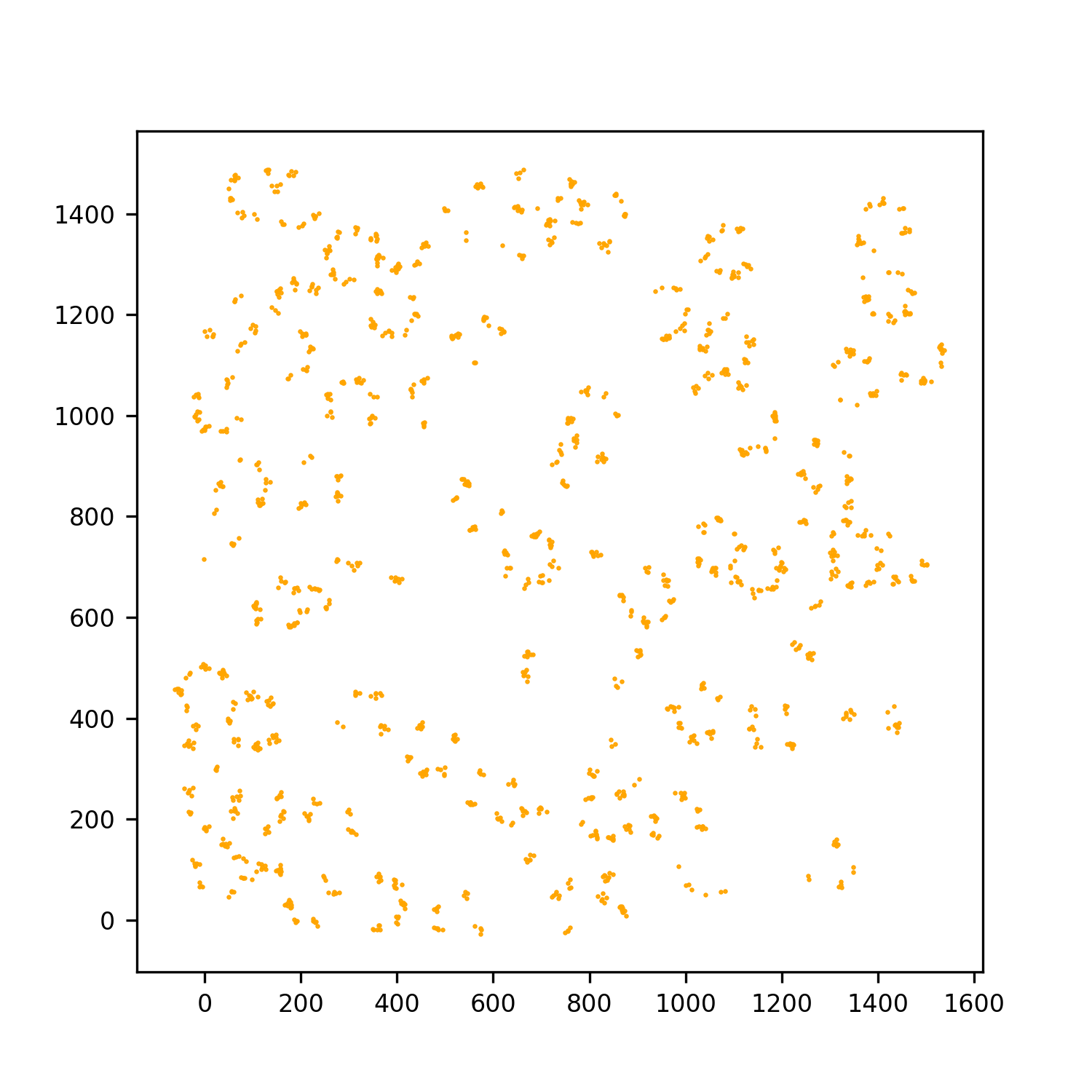}}
    \subfloat[]{\includegraphics[width=0.275\linewidth]{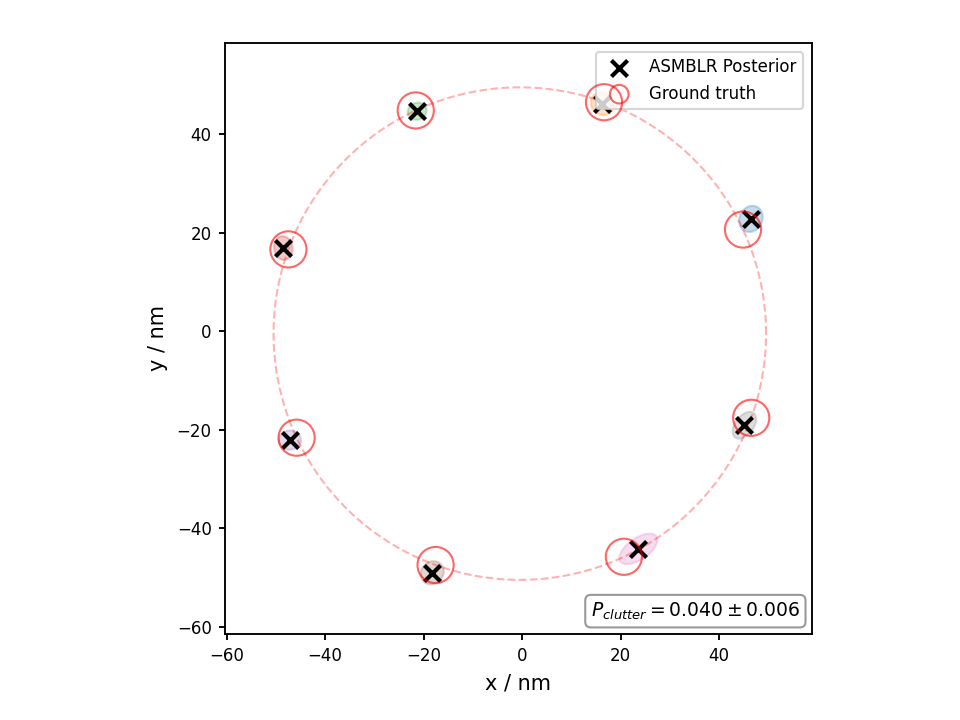}}
    \subfloat[]{\includegraphics[width=0.22\linewidth]{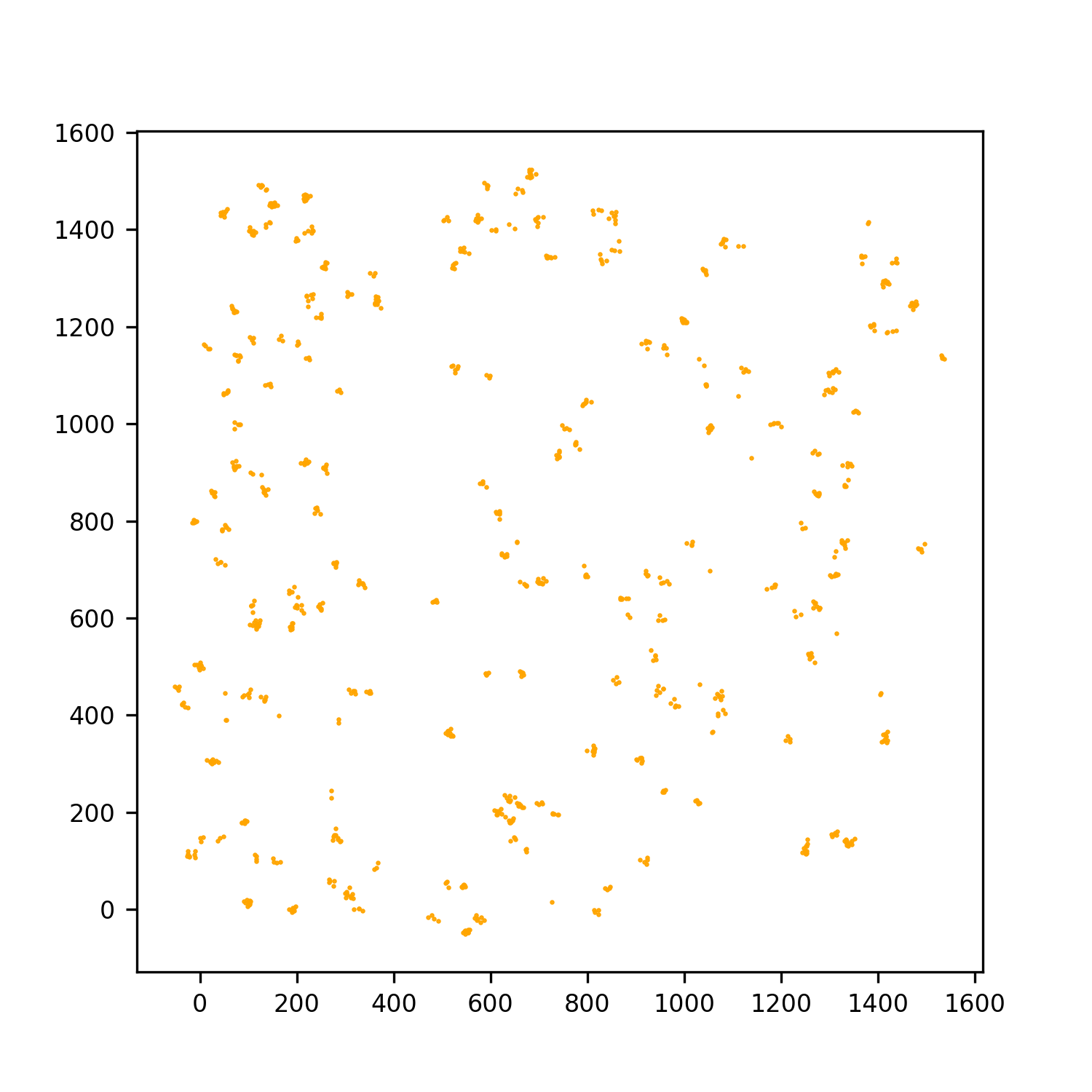}}
    \subfloat[]{\includegraphics[width=0.275\linewidth]{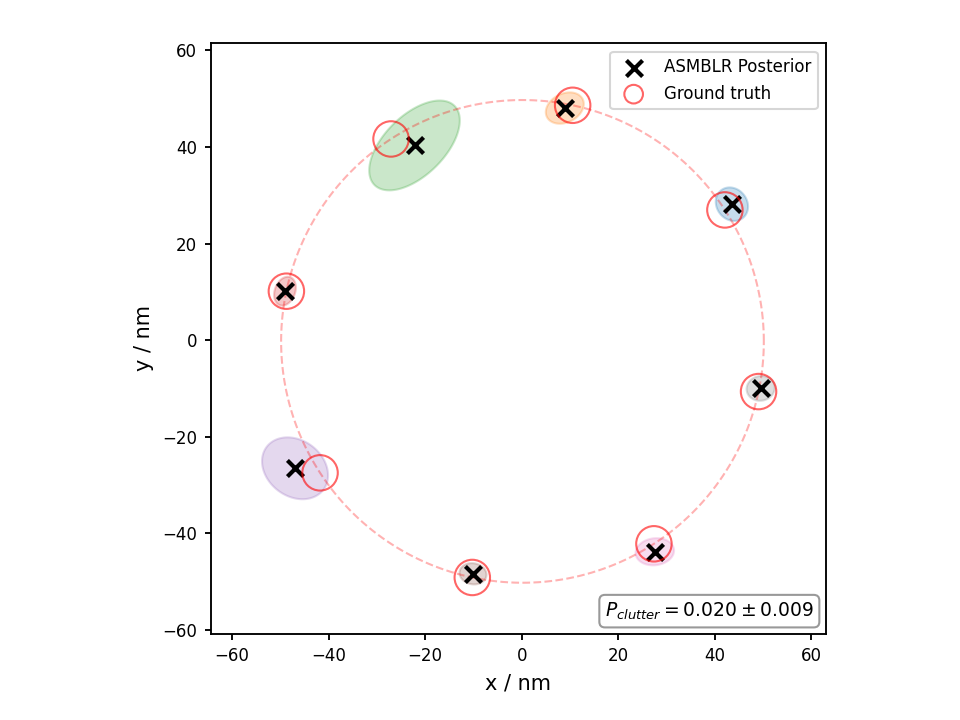}} \\
    \subfloat[]{\includegraphics[width=0.22\linewidth]{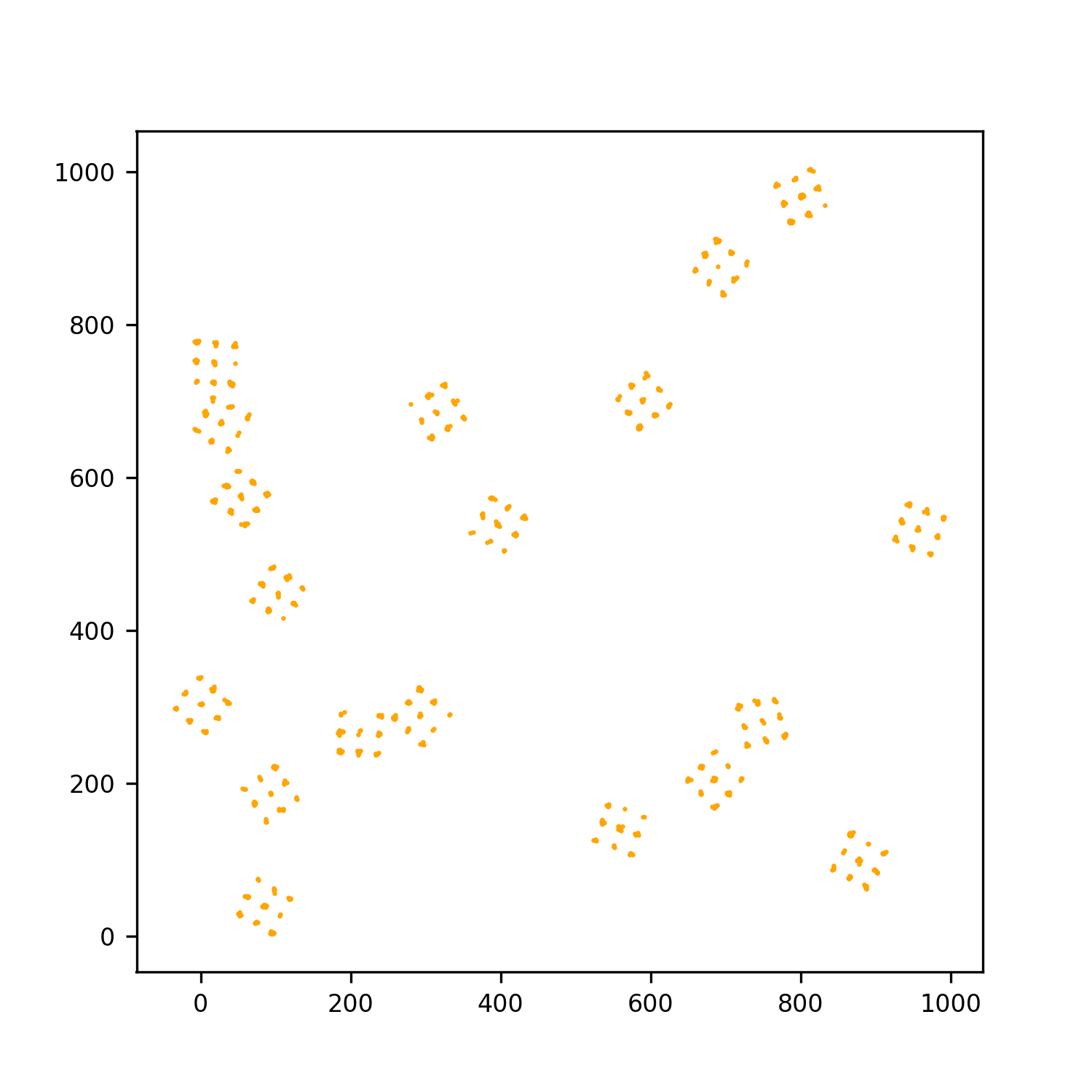}}
    \subfloat[]{\includegraphics[width=0.275\linewidth]{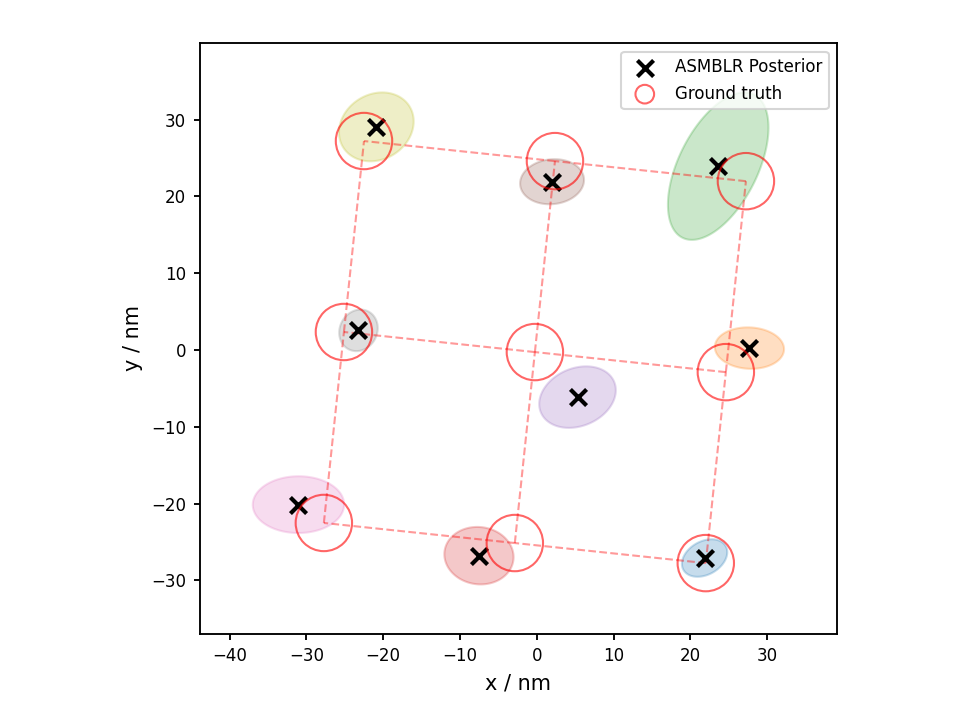}}
    \caption{ASMBLR reconstructs repeating molecular motifs across a range of data conditions. Observed measurements and model reconstruction with 67\% credible interval for labelling efficiencies of (a-b) 1.0, (c-d) 0.9, (e-f) 0.6, and (g-h) 0.3. 600 cliques of size 3 were used for the 8-fold model. Reconstruction of DNA-Origami 3x3 grids (i-j) also shown with 500 cliques of size 5.}
    \label{fig:asmblr-res}
\end{figure}

We demonstrate the efficacy of the ASMBLR algorithm under the variety of labelling conditions used thus far (Fig. \ref{fig:asmblr-res}). 

Fig. \ref{fig:asmblr-res} demonstrates ASMBLR's robustness to decreasing labelling efficiency. At full labelling (Fig. \ref{fig:asmblr-res}(a-b)) the algorithm clearly resolves the characteristic 8-fold symmetry and \(\approx50\)nm radius of the 2D Nup96 molecular structure. As labelling efficiency degrades to 0.3 (Fig. \ref{fig:asmblr-res}(g-h)), this characteristic symmetry and scale remains evident, successfully recovered despite the challenge in structural data availability, albeit with expanded uncertainty at select vertices. Critically, these reconstructions emerge solely from the internal consistency of the localisation data, supported by the Voidwalker-Gibbs structural inference, without reference to known Nup96 structural models. 

To assess 8-fold symmetry, we aligned an idealised ring to each posterior via Hungarian algorithm, then applied three tests: per-vertex Mahalanobis distance with Bonferroni correction (\(\alpha = 0.05/8\)), Fisher's combined statistic for joint consistency, and permutation testing against 10,000 random rotations to distinguish geometric regularity from chance.

This pattern of strong statistical support was maintained across all labelling conditions, though with appropriately increased uncertainty at lower labelling efficiencies as reflected in expanded credible regions (Fig. \ref{fig:asmblr-res}) and modestly elevated Mahalanobis distances. Across all labelling efficiencies, all criteria showed near-optimal performance: 8/8 vertices passing, Fisher \(p > 0.95\), permutation \(p < 0.05\), and mean Mahalanobis\(^2 < 2.0\). The inferred radius remained stable across all conditions, demonstrating that ASMBLR correctly recovers geometric scale even when individual vertex positions carry substantial uncertainty.

For DNA-Origami grids (Fig.~\ref{fig:asmblr-res}(i-j)) the framework correctly identified the lattice geometry under no grid structure assumption. Correct inference of emitter separation (25 nm) and statistical validation yielded strong evidence (Fisher $p > 0.95$, permutation $p < 0.05$, mean Mahalanobis$^2 < 2.0$) for agreement with ground truth. The central vertex shows modest displacement from the expected position; this likely reflects ASMBLR's treatment of pairwise separations as independent, whereas in practice the $O(K^2)$ separations derived from $K$ vertices are correlated. Additionally, the current implementation considers only axial localisation uncertainty, neglecting transverse contributions that become relevant when separation approaches the localisation precision. Despite these simplifications, all vertices remain statistically consistent with the ground truth model.

\subsection{MINFLUX Data}
\begin{figure}[ht]
    \centering
    \subfloat[]{\includegraphics[width=0.24\linewidth]{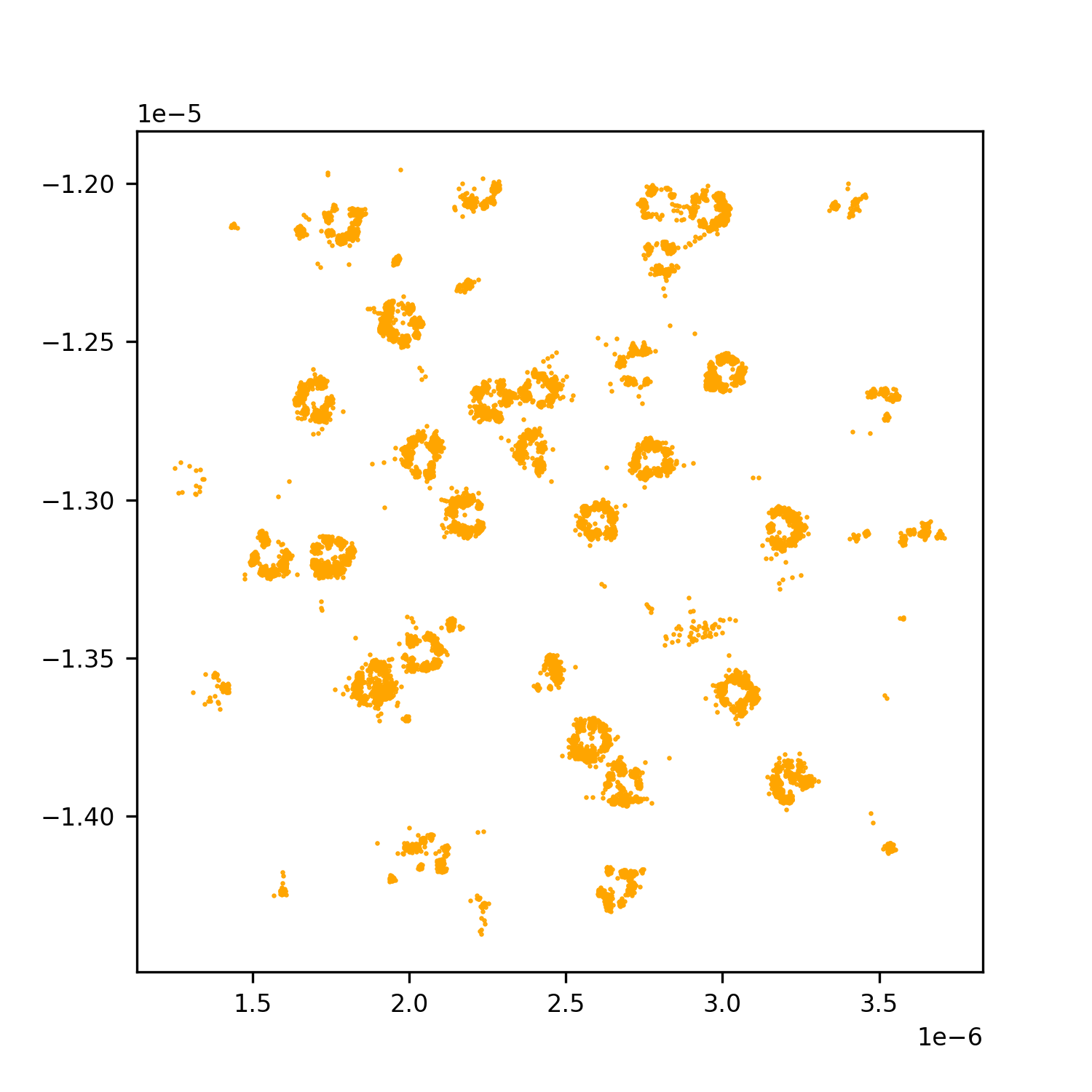}}
    \subfloat[]{\includegraphics[width=0.24\linewidth]{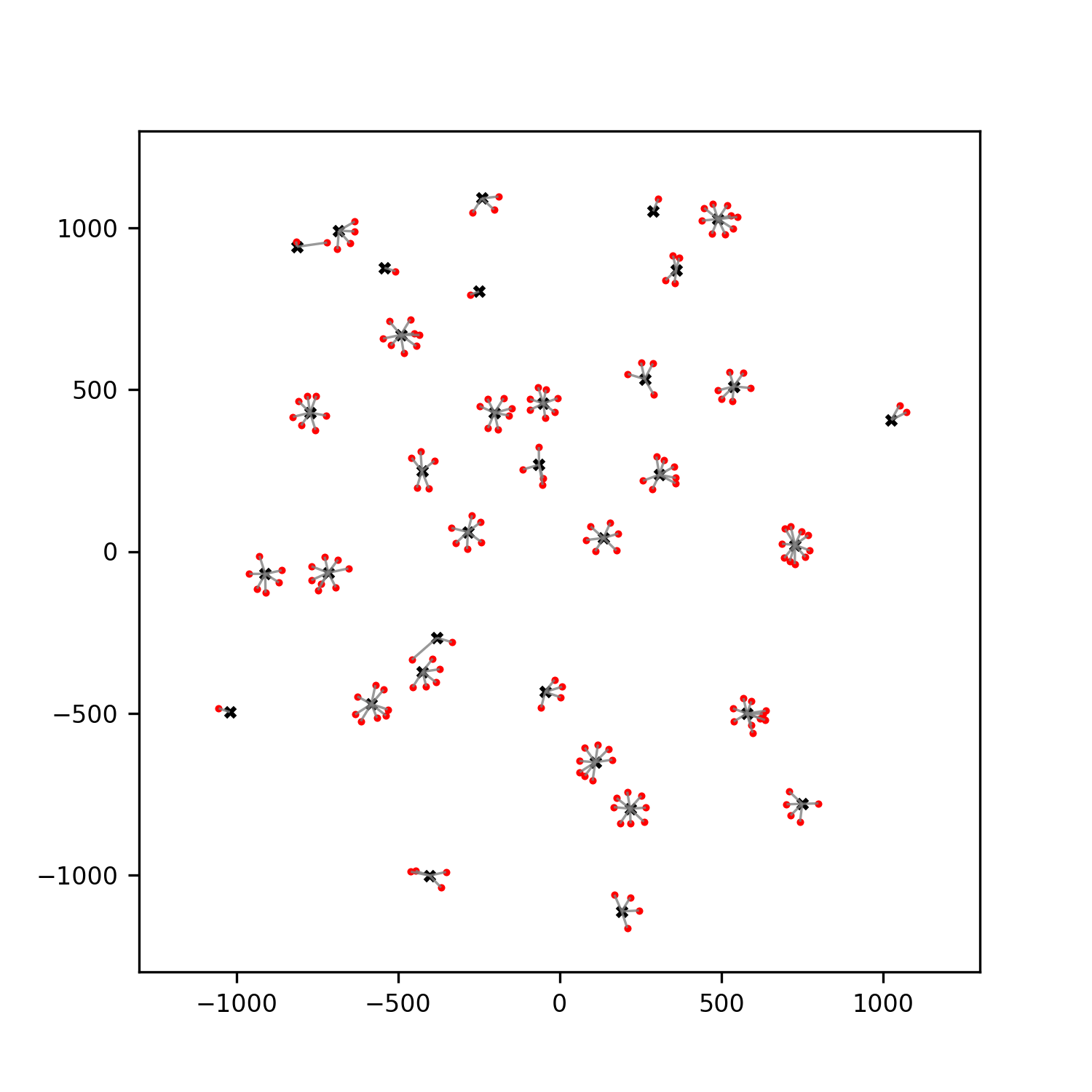}}
    \subfloat[]{\includegraphics[width=0.30\linewidth]{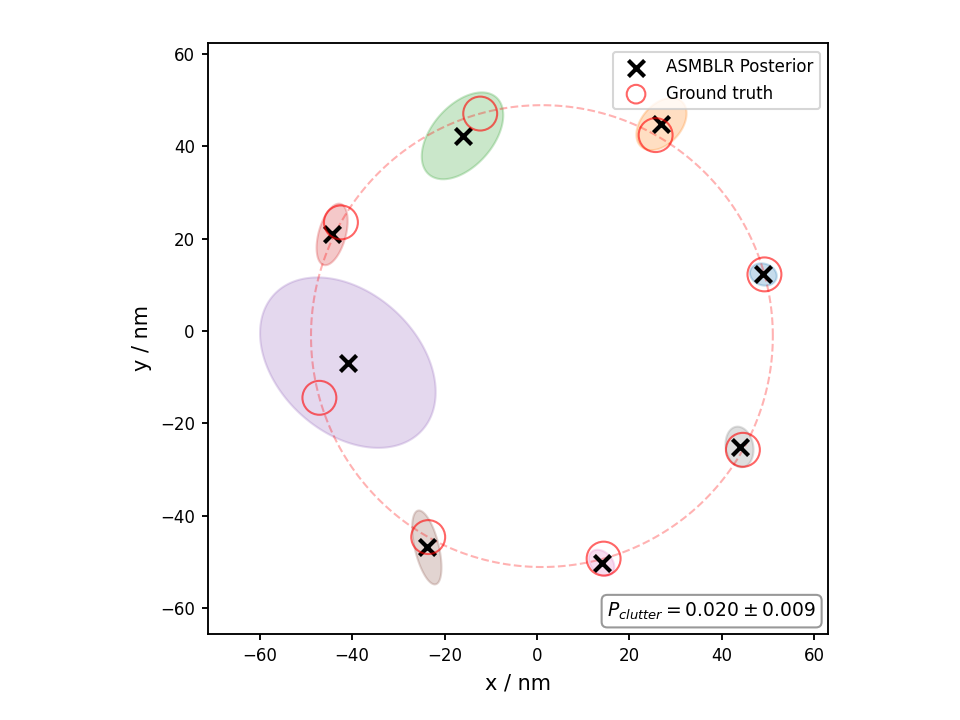}} \\ 

    \subfloat[]{\includegraphics[width=0.3\linewidth]{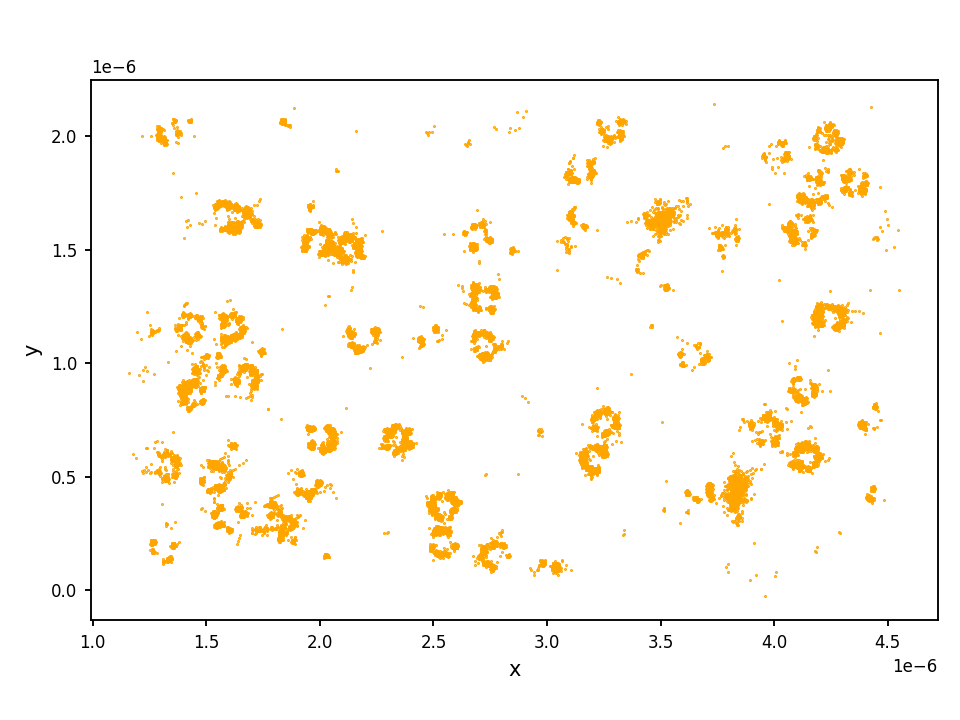}}
    \subfloat[]{\includegraphics[width=0.3\linewidth]{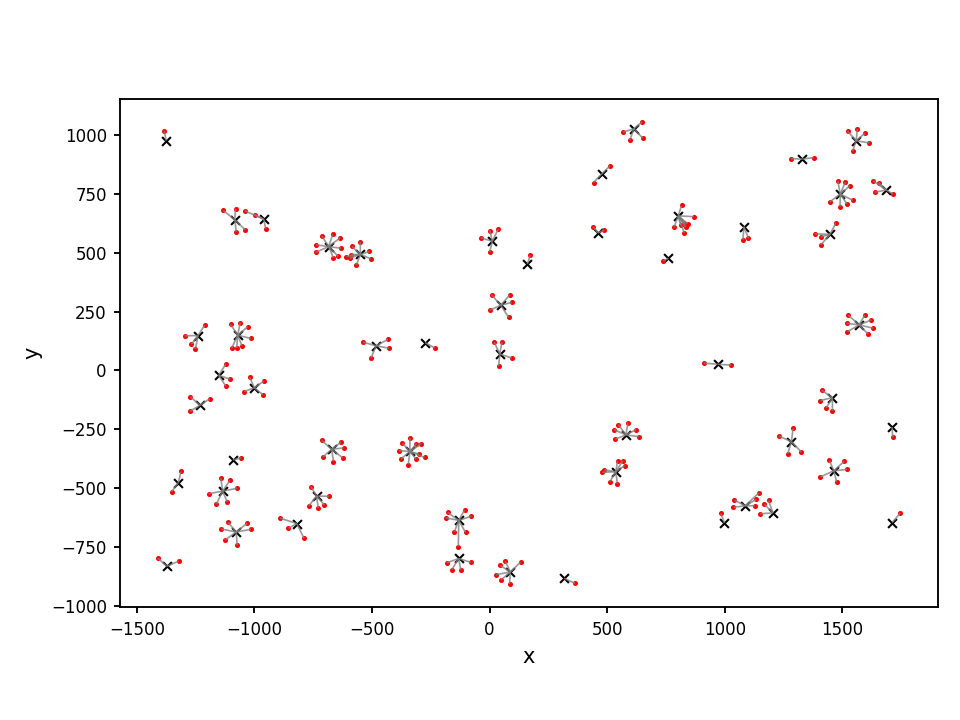}}
    \subfloat[]{\includegraphics[width=0.3\linewidth]{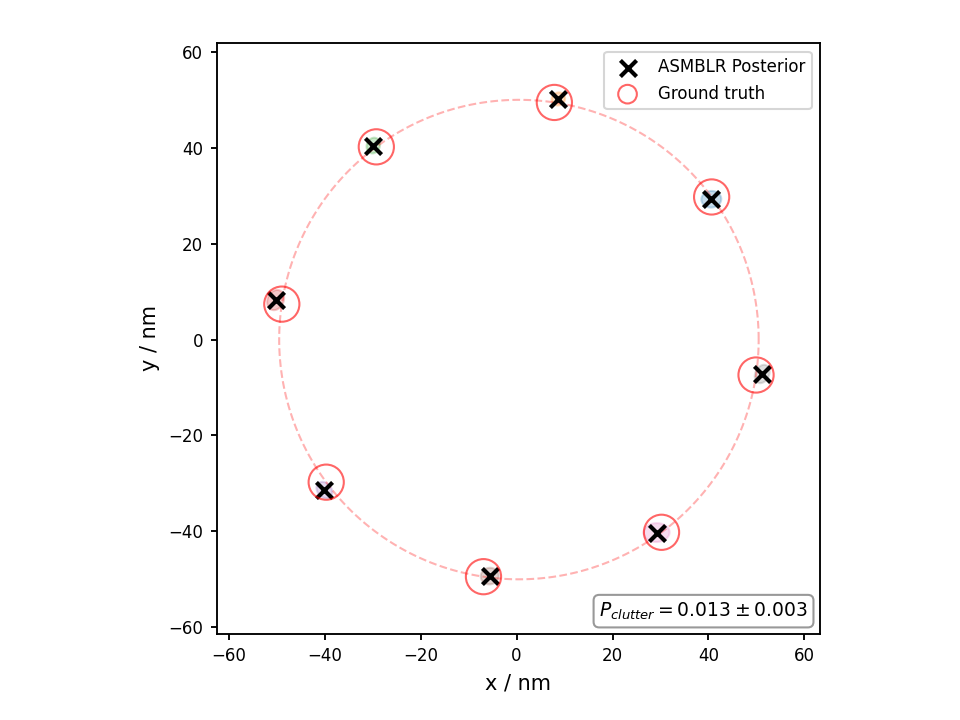}} \\

    \subfloat[]{\includegraphics[width=0.3\linewidth]{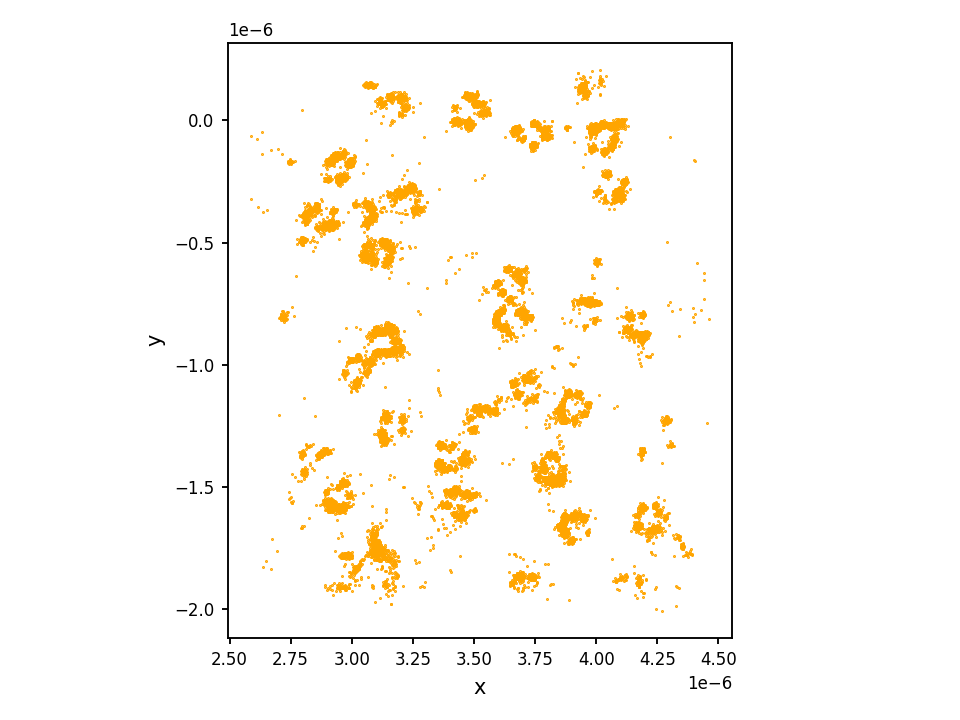}}
    \subfloat[]{\includegraphics[width=0.3\linewidth]{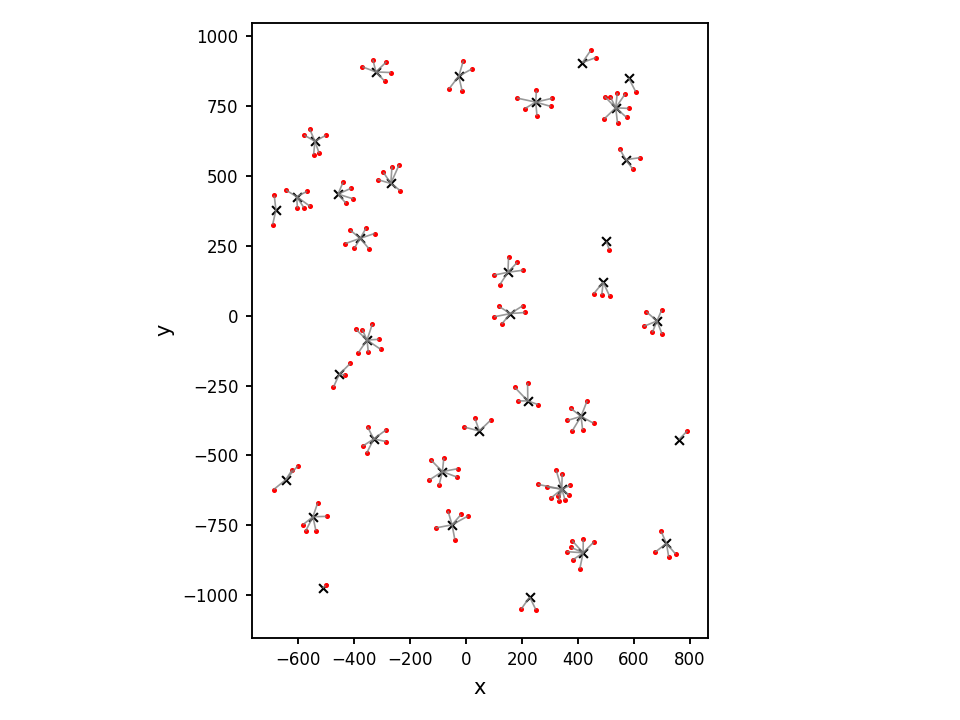}}
    \subfloat[]{\includegraphics[width=0.3\linewidth]{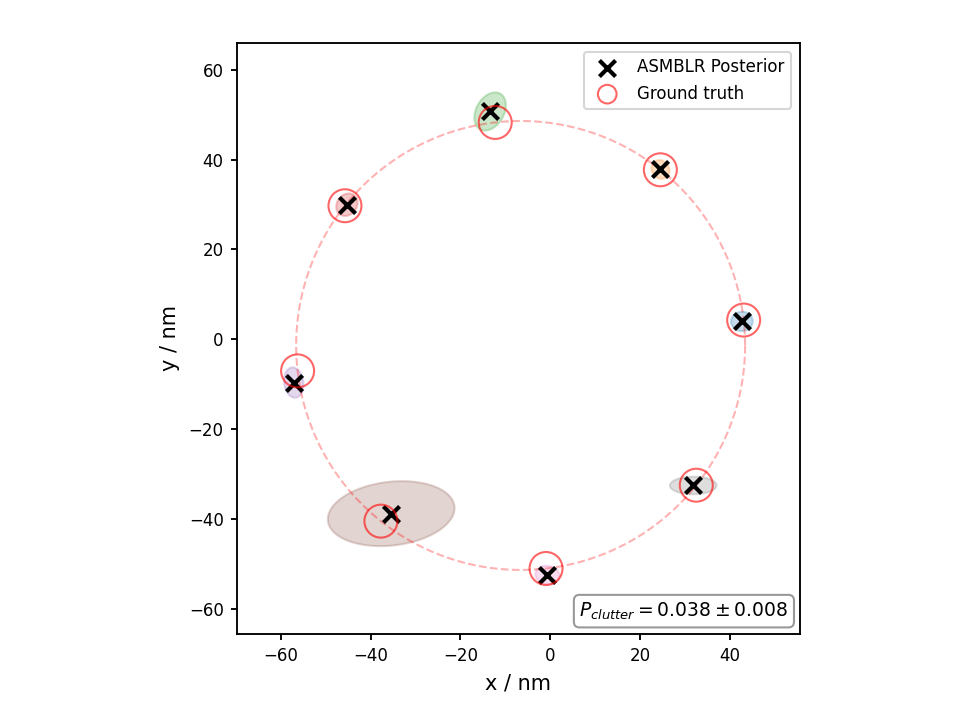}}
    
    \caption{The outlined framework successfully reconstructs the Nup96 characteristic structure from real MINFLUX localisation data. (a, d, g) Observed measurements, (b, e, h) GROUPA estimated emitters and Voidwalker assignments, (c, f, i) ASMBLR model reconstructions.}
    \label{fig:real-full-pipeline}
\end{figure}

The framework was validated on experimental MINFLUX data by reconstructing Nup96's characteristic 8-fold symmetry and $\sim 50$\,nm radius from real localisation data (Fig.\ref{fig:real-full-pipeline}). GROUPA requires no hyperparameter tuning on localisation uncertainty, operating solely on maximum measurement separation. Voidwalker-Gibbs then groups emitters via data-driven priors derived from empty-space statistics, without assumptions on measurement precision. The reconstructed structures (Fig.~\ref{fig:real-full-pipeline}(c,f,i)) recover the expected 8-fold symmetry and $\sim 50$\,nm radius, with quantitative validation consistent with synthetic benchmarks.

\FloatBarrier
\section{Discussion}
Extracting structural information from sparse, error prone and uncertain point patterns without strong \textit{a priori} information represents a fundamental challenge in single-molecule localisation microscopy. While demonstrated on MINFLUX data, the statistical foundations of pairwise hypothesis testing, inhomogeneous point processes and hard-core Gibbs processes presented here can be applied to any localisation-based single-molecule imaging technique (such as STORM) that produces point patterns with quantified uncertainty and incomplete labelling. The critical requirement is sufficient localisation precision such that the structural features of interest are resolved, encompassing most modern SMLM implementations.

Template-matching workflows conflate two distinct questions: ``Is this structure present?'' and ``Does the data support this structure?'' By first fitting known geometries and then evaluating goodness-of-fit, such approaches risk confirmation bias, discovering structures because one searched for them and not because the data independently support them. Our framework inverts this logic: we first ask what structures the data may support through purely data-driven clustering and centre detection, then enable hypothesis testing against known architectures if desired. This separation reduces the tendency toward false positives while maintaining compatibility with validation against crystallographic or cryo-EM references. Importantly, the uncertainty quantification at each stage provides an explicit measure of confidence, allowing researchers to distinguish between robust discoveries and findings contingent on favourable data quality. This proposed framework has three distinct use cases: determining structure of proteins with unknown or variable stoichiometry, validating template-matched results are statistically justifiable, and analysis of sparse labelling regimes where conventional methods struggle. For well-characterised structures imaged at high labelling efficiency and measurement precision, researchers may prefer familiar established tools. This framework produces more statistically defensible results than those of typical template matching approaches.

GROUPA reduces measurement clustering to pairwise 'sameness' evaluations, with higher-order structure inferred via community detection. This removes heuristic parameters and the need to estimate local emitter density a priori. GROUPA was found to outperform optimally tuned DBSCAN or Gaussian mixture models across measurement uncertainties 5 to 20 nm (Fig. \ref{fig:groupa-res}). A direct quantitative comparison to BaGoL \cite{fazel_high-precision_2022} was not undertaken, as the methods address fundamentally different analytical scales. BaGoL excels at sub-nanometre localisation refinement within manually defined ROIs and is well-suited for detailed analysis of individual structures where computational cost is secondary to precision. GROUPA prioritises scalability and automation, processing entire imaging fields without ROI pre-selection. For many experimental workflows, these approaches are complementary: researchers might apply GROUPA for field-wide partitioning and structural discovery, then reserve BaGoL for high-precision refinement of specific structures of interest. The comparison between measurement-uncertainty-aware Bayesian methods (GROUPA, BaGoL) and density-based heuristics (DBSCAN, HDBSCAN) reflects a broader trade-off between computational overhead and statistical rigour; future hybrid pipelines may leverage each method's strengths at appropriate scales.

Treating SMLM images as inhomogeneous point patterns, Voidwalker identifies statistically significant voids exceeding null-model expectations, providing data-driven priors on structure count and radius. This narrows the RJMCMC proposal space, improving convergence. The RJMCMC itself exploits core structural constraints inherent to SMLM: the expectation that structure interiors are devoid of fluorophores, and that sterically exclusive objects do not overlap. By encoding these constraints via a Gibbs point process with hard-core interaction, and leveraging Voidwalker-informed priors, the sampler constructs per-emitter probability distributions over centre assignments. Fig. \ref{fig:vw-g-res} demonstrates that this approach maintains, on average, >80\% assignment accuracy at labelling efficiencies as low as \(l=0.3\), indicating robust performance for emitter subset detection across the full range of data quality conditions examined. This Voidwalker step provides a vast improvement in downstream reconstruction of underlying molecular structure; radial uniform sampling of cliques yields small probabilities of structurally representative cliques, particularly at large clique sizes, where Voidwalker improves this probability substantially. As such, fewer cliques are clutter, and more cliques can meaningfully contribute to the reconstructed model.

The marked point-process approach to super-structure detection reveals a critical distinction between what is algorithmically feasible and what is information-theoretically possible (SI Sec. 2.3). While Voidwalker-Gibbs maintains >80\% assignment accuracy at 30\% labelling (Fig. \ref{fig:vw-g-res}(b)), super-structure inference requires $\geq90\%$ labelling to exceed F1$ > 0.8$ (Fig. \ref{fig:super-metrics}). This threshold difference is not an algorithmic shortcoming but a fundamental consequence of hierarchical inference: centre detection operates on local emitter-centre spatial relationships, whereas super-structure discovery requires discriminating between centre pairs based on global assignment patterns - a higher-order inference problem with compounded uncertainty. At 30\% labelling of 8-emitter rings the assignment posteriors flatten, each centre is represented by approximately $2-3$ emitters, yielding marks with Shannon information content below the minimum required to distinguish meaningful similarity from noise. Super-structure inference below this threshold requires introducing additional constraints (characteristic inter-centre distances, known stoichiometries) that risk the circular reasoning inherent to template-matching paradigms. The performance surfaces in Fig. \ref{fig:super-metrics} document this degradation explicitly. 

For experimental design, these results provide actionable thresholds: seek super-structure at high bio-labelling with low clutter, or defer super-structure inference at low labelling. Different molecular geometries exhibit distinct failure modes: rotationally symmetric structures (such as Nup96) display distributed criticality requiring cumulative damage, whereas structures with unique central elements (grid lattices) exhibit concentrated criticality. This suggests minimum detection thresholds are structure-specific, with symmetric architectures tolerating lower detection probabilities (SI Sec. 2.4), perhaps explaining Nup96's prevalence as a MINFLUX benchmark \cite{thevathasan_nuclear_2019}. A formal treatment of structural fragility will be developed subsequently.

These marked point-process distributions, when applicable, are further leveraged to partition centres into sub-structure and super-structure subsets, enabling template-free reconstruction via ASMBLR. By sampling fully connected subgraphs, or cliques, from emitter populations co-assigned to the same structural unit, ASMBLR exploits the internal geometry of these cliques to reconstruct repeating structural motifs without prior knowledge of molecular architecture, other than that the structures to be reassembled are fully-connected. The current framework assumes known motif size N, though both N and clique size may be guided by emitters per estimated centre (Fig. \ref{fig:vw-g-res}(f)). Moreover, we assume a \(\text{Beta}(\alpha=10, \beta=90)\) prior on clutter to reflect that optimally there is no clutter and 10\% average clutter is expected in typical datasets, and broadly isotropic measurement uncertainty. Operating on the inner space of the structural domain rather than external spatial context, molecular structures are resolved from their constituent emitter distributions alone. The statistical validation framework applied to ASMBLR reconstruction demonstrates that reference-free discovery is statistically distinguishable from template matching or overfitting. Comparing observed alignment quality to thousands of random rotations in the permutation test is particularly critical, in that it directly addresses the question of whether a discovered pattern reflects genuine geometric regularity or chance alignment to noise. Across all labelling conditions the observed 8-fold arrangements were significantly better aligned than random orientations \((p<0.01)\), providing unambiguous evidence that ASMBLR extracts biologically meaningful information. The high degree of symmetry present in the DNA-Origami grid, however, necessitates larger sizes of cliques to correctly constrain the model - explaining the slightly offset centre point. Using \(m=5\) or \(m=6\) would likely improve this model at the cost of additional computational overhead.

Graph-theoretic methods \cite{delaunay1934sphere, edelsbrunner_three-dimensional_1994} lack robustness to clutter; persistent homology \cite{edelsbrunner2002topological,carlsson2009topology} struggles with sparse data; deep learning \cite{speiser2021deep} requires annotated training sets that presuppose template knowledge. All rely on point distributions, suffering from incomplete labelling. The Bayesian framework adopted here is driven by the reproducibility of the empty space subdomains in the data, and provides principled uncertainty propagation at each stage: from measurement-to-emitter clustering (posterior over labels), through centre detection (posterior over positions, scale, and assignments), to super-structure inference (permutation-tested similarities with explicit p-values), and finally to structural validation (hypothesis tests with quantified significance). This end-to-end uncertainty quantification enables researchers to distinguish robust structural findings from those contingent on favourable data conditions.

The modular design enables parallelisation at natural boundaries: GROUPA processes fields in parallel; Voidwalker executes once per dataset; RJMCMC chains run independently; ASMBLR operates per oligomer. Wall-clock time is dominated by ASMBLR (hours for large datasets in poor-quality regimes), while the preceding pipeline requires <1hr. The tractability and VRAM memory requirements of the TopK framework as $N>14$ and $M>6$ increase by at least $P(N,M)$ (SI Sec. 1.5). While our problems are currently constrained to this space, solutions would be to replace the existing TopK scheme with a Plackett-Luce \cite{luce1959individual, plackett1975analysis} style sequential sampler that avoids the requirement to scan all potential assignments.

The current pipeline applies to 2D patterns of a single structure type. Extension to 3D is straightforward for pairwise testing and RJMCMC, though Voidwalker and super-structure detection require adaptation to volumetric fields. Multi-structure detection would require hierarchical intensity models or mixture-of-Gibbs processes, both tractable extensions.

The framework demonstrates that template-free structural biology at nanometre resolution is achievable without sacrificing statistical rigour. By explicitly testing multiple structural hypotheses and quantifying support, rather than single template assumption, we propose a path forward for SMLM analysis that addresses the critique raised by Prakash \cite{prakash_at_2022, prakash_assessment_2021} while maintaining compatibility with validation against known structures. The modular architecture enables incremental adoption of the individual methods; researchers may integrate any of GROUPA, Voidwalker, or ASMBLR in their own existing analysis pipelines, while the end-to-end pipeline aims to provide a complete solution for discovery-focussed applications. As SMLM moves toward discovering novel architectures in proteome-wide studies, template-free methods with rigorous uncertainty quantification will become essential rather than optional.

\paragraph{Data and code availability.} Code for synthetic data available on github \cite{simflux}. Pipeline code and experimental data available upon reasonable request. 

\paragraph{Supplementary information.} Supplementary information including end-to-end results for example datasets and full mathematical detail for the pipeline is available.

\paragraph{Acknowledgements.} Many thanks to Dr Evelyn Garlick of Aberrior Instruments for useful discussion.

\section{Methods}
\subsection{Data Availabilty}
Synthetic datasets are simulated via the Python-based SimFlux package \cite{simflux}. All datasets are simulated under a specified seed to analyse under varying labelling and clutter conditions. Real MINFLUX datasets acquired from Evelyn Garlick, Aberrior Instruments.

\subsection{GROUPA: measurement clustering}
GROUPA (SI Sec. 1.1) performs measurement-to-emitter clustering via pairwise Bayes factor tests followed by community detection. For each pair of measurements \(i,j\) with positions \(\mathbf{x}_i,\mathbf{x}_j\) and uncertainty \(\Sigma_i, \Sigma_j\), we compute the Bayes factor comparing \(H_0\) (same emitter) with \(H_1\) (distinct emitters). Under \(H_0\) both measurements arise from a shared distribution; under \(H_1\) two distinct distributions. We construct a weighted graph retaining edges of \(BF_{ij}>1.0\) and apply the Infomap algorithm \cite{Smiljanic2023MapEquation} to detect communities via the map equation \cite{rosvall_map_2009}.

\subsection{Voidwalker: intensity-guided void detection}
Voidwalker (SI Sec. 1.2) identifies statistically significant empty voids in the emitter point pattern to inform priors and proposals for downstream RJMCMC. We fit an inhomogeneous Log-Gaussian Cox Process (LGCP) \cite{Illian2008} to emitter locations using the SPDE method \cite{lindgren_explicit_2011}, constructing a finite-element mesh with resolution scaled to the bounding box \(W\), where inner and outer edges of \(W/40\) and \(W/10\) were used throughout the study. Intensity fit \(\hat{\lambda}(x)\) is validated by assessing \(\int\hat{\lambda}(\mathbf{x})d\mathbf{x}\) aligns with observed emitter count. Intensity is renormalised to constitute a probability distribution and voids are probabilistically seeded according to draws from this distribution, and subject to morphological growth and emitter-repulsive walks. 1500 voids are seeded, and their significance is determined via z-scores calculated from 1000 inhomogeneous Poisson process simulations on LGCP posterior-predictive draws under the null hypothesis of no structure. The number of voids seeded was chosen to be more than necessary to ensure the space is adequately searched; larger windows should thus seed a larger number of voids as there is no downside (other than computationally) to seeding many voids. Voids passing a \(p<0.05\) gate (testing Voidwalker on CSR simulations yielded few to no contributive voids under this gate) under this null are deemed active and contribute to the Poisson prior on structure count, Gaussian prior on structure radius, and proposal density for dimension changing moves in RJMCMC.

\subsection{RJMCMC: centre and super-structure inference}
We model structural centres as a Gibbs point process \cite{Illian2008} (SI Sec. 1.3) with hard-core interaction (minimum separation \(1.5r\) where $r$ is the structural radius) and Voidwalker-informed priors \(N_{\text{centres}}\sim\text{Poisson}(\lambda_{\text{void}})\) and \(r\sim\mathcal{N}(r_{\text{void}}, \sigma_{r_{\text{void}}})\). The posterior is explored via RJMCMC \cite{green_reversible_1995} with birth, death, split, merge, and shift moves; proposal scales are adaptively tuned using a Robbins-Monro \cite{RobbinsMonro1951} scheme targeting 23.4\% acceptance rate, an empirically stable rate. At each iteration, emitters are probabilistically assigned to the nearest centre under bivariate Gaussian density integrating the centre and radial uncertainty. We run 3 chains per dataset for 25-125k iterations depending on data quality (sparse labelling requires increased runtime), with 80\% burn-in, storing the top 20 centre assignments per emitter from post-burn-in samples. Gelman-Rubin \cite{brooks1998} \(\hat{R}<1.1\) observed for all datasets processed.

Each centre's mark is its posterior responsibility distribution over emitters. Centre pairs are scored via composite Bhattacharyya and harmonic-mean measures, scores are tested against an intensity-stratified permutation null under 1000 permutations; edges with similarity both exceeding a global 99th percentile of the null and individually passing a \(p<0.05\) gate are considered a dimeric pair (SI Sec. 1.4). 

\subsection{ASMBLR: molecular reconstruction}
For each identified structural unit, ASMBLR (SI Sec. 1.5) samples \(m\)-cliques - fully connected subgraphs of size \(m\) - using a centre assignment-conditioned Bron-Kerbosch algorithm. From datasets representing these sets of sampled \(m\)-cliques with positional uncertainty, ASMBLR reconstructs an n-fully connected graph with uncertainty and additional clutter component. The vertices and corresponding uncertainties of this graph constitute the model posterior. These posteriors, and their joint model, are validated against idealised models of the ground truth structure using per-vertex Bonferroni-corrected \(\chi^2\) tests, Fisher's combined statistic, and 1000 rotation permutation tests (see SI Sec. 1.5 for full mathematical detail and validation methodology.)

\bibliographystyle{unsrt}
\bibliography{references}{}

@book{Illian2008,
  author    = {Janine Illian and Antti Penttinen and Helga Stoyan and Dietrich Stoyan},
  title     = {Statistical Analysis and Modelling of Spatial Point Patterns},
  publisher = {John Wiley \& Sons},
  year      = {2008},
  address   = {Chichester, UK},
  isbn      = {978-0-470-01491-2}
}

@article{gwosch_minflux_2020,
    title = {{MINFLUX} nanoscopy delivers {3D} multicolor nanometer resolution in cells},
    volume = {17},
    issn = {1548-7091, 1548-7105},
    url = {https://www.nature.com/articles/s41592-019-0688-0},
    doi = {10.1038/s41592-019-0688-0},
    language = {en},
    number = {2},
    urldate = {2024-12-02},
    journal = {Nature Methods},
    author = {Gwosch, Klaus C. and Pape, Jasmin K. and Balzarotti, Francisco and Hoess, Philipp and Ellenberg, Jan and Ries, Jonas and Hell, Stefan W.},
    month = feb,
    year = {2020},
    pages = {217--224},
}

@article{balzarotti_nanometer_2017,
    title = {Nanometer resolution imaging and tracking of fluorescent molecules with minimal photon fluxes},
    volume = {355},
    copyright = {http://www.sciencemag.org/about/science-licenses-journal-article-reuse},
    issn = {0036-8075, 1095-9203},
    url = {https://www.science.org/doi/10.1126/science.aak9913},
    doi = {10.1126/science.aak9913},
    language = {en},
    number = {6325},
    urldate = {2024-12-02},
    journal = {Science},
    author = {Balzarotti, Francisco and Eilers, Yvan and Gwosch, Klaus C. and Gynnå, Arvid H. and Westphal, Volker and Stefani, Fernando D. and Elf, Johan and Hell, Stefan W.},
    month = feb,
    year = {2017},
    pages = {606--612},
}

@article{prakash_at_2022,
    title = {At the molecular resolution with {MINFLUX}?},
    volume = {380},
    issn = {1364-503X, 1471-2962},
    url = {https://royalsocietypublishing.org/doi/10.1098/rsta.2020.0145},
    doi = {10.1098/rsta.2020.0145},
    language = {en},
    number = {2220},
    urldate = {2024-12-02},
    journal = {Philosophical Transactions of the Royal Society A: Mathematical, Physical and Engineering Sciences},
    author = {Prakash, Kirti},
    month = apr,
    year = {2022},
    pages = {20200145},
}

@article{prakash_assessment_2021,
    title = {Assessment of {3D} {MINFLUX} data for quantitative structural biology in cells},
    volume = {20},
    issn = {1548-7091, 1548-7105},
    url = {https://www.nature.com/articles/s41592-022-01694-x},
    doi = {10.1038/s41592-022-01694-x},
    language = {en},
    number = {1},
    urldate = {2024-12-02},
    journal = {Nature Methods},
    author = {Prakash, Kirti and Curd, Alistair P.},
    month = jan,
    year = {2023},
    pages = {48--51},
}

@misc{gwosch_assessment_2022,
    title = {Assessment of {3D} {MINFLUX} data for quantitative structural biology in cells revisited},
    url = {http://biorxiv.org/lookup/doi/10.1101/2022.05.13.491065},
    doi = {10.1101/2022.05.13.491065},
    language = {en},
    urldate = {2024-12-02},
    publisher = {Biophysics},
    author = {Gwosch, Klaus C. and Balzarotti, Francisco and Pape, Jasmin K. and Hoess, Philipp and Ellenberg, Jan and Ries, Jonas and Matti, Ulf and Schmidt, Roman and Sahl, Steffen J. and Hell, Stefan W.},
    month = may,
    year = {2022},
}

@article{lelek_single-molecule_2021,
    title = {Single-molecule localization microscopy},
    volume = {1},
    issn = {2662-8449},
    url = {https://www.nature.com/articles/s43586-021-00038-x},
    doi = {10.1038/s43586-021-00038-x},
    language = {en},
    number = {1},
    urldate = {2024-12-02},
    journal = {Nature Reviews Methods Primers},
    author = {Lelek, Mickaël and Gyparaki, Melina T. and Beliu, Gerti and Schueder, Florian and Griffié, Juliette and Manley, Suliana and Jungmann, Ralf and Sauer, Markus and Lakadamyali, Melike and Zimmer, Christophe},
    month = jun,
    year = {2021},
    pages = {39},
}

@misc{hammer_density-based_2024,
    title = {Density-based optimization for unbiased, reproducible clustering applied to single molecule localization microscopy},
    copyright = {© 2024, Posted by Cold Spring Harbor Laboratory. The copyright holder for this pre-print is the author. All rights reserved. The material may not be redistributed, re-used or adapted without the author's permission.},
    url = {https://www.biorxiv.org/content/10.1101/2024.11.01.621498v1},
    doi = {10.1101/2024.11.01.621498},
    language = {en},
    urldate = {2025-04-02},
    publisher = {bioRxiv},
    author = {Hammer, Joseph L. and Devanny, Alexander J. and Kaufman, Laura J.},
    month = nov,
    year = {2024},
    note = {Pages: 2024.11.01.621498
Section: New Results},
}

@article{wu_facam_2023,
    title = {{FACAM}: {A} {Fast} and {Accurate} {Clustering} {Analysis} {Method} for {Protein} {Complex} {Quantification} in {Single} {Molecule} {Localization} {Microscopy}},
    volume = {10},
    copyright = {http://creativecommons.org/licenses/by/3.0/},
    issn = {2304-6732},
    shorttitle = {{FACAM}},
    url = {https://www.mdpi.com/2304-6732/10/4/427},
    doi = {10.3390/photonics10040427},
    language = {en},
    number = {4},
    urldate = {2025-08-20},
    journal = {Photonics},
    author = {Wu, Cheng and Kuang, Weibing and Zhou, Zhiwei and Zhang, Yingjun and Huang, Zhen-Li},
    month = apr,
    year = {2023},
    note = {Publisher: Multidisciplinary Digital Publishing Institute},
    pages = {427},
}

@incollection{campello2013density,
  author    = {Ricardo J. G. B. Campello and Davoud Moulavi and Joerg Sander},
  title     = {Density-based Clustering based on Hierarchical Density Estimates},
  booktitle = {Advances in Knowledge Discovery and Data Mining},
  publisher = {Springer Berlin Heidelberg},
  year      = {2013},
  pages     = {160--177},
}

@inproceedings{mcinnes2017accelerated,
  author    = {Leland McInnes and John Healy},
  title     = {Accelerated Hierarchical Density Based Clustering},
  booktitle = {2017 IEEE International Conference on Data Mining Workshops (ICDMW)},
  month     = nov,
  year      = {2017},
  pages     = {33--42},
}

@inproceedings{malzer2020hybrid,
  author    = {Claudia Malzer and Marcus Baum},
  title     = {A Hybrid Approach To Hierarchical Density-based Cluster Selection},
  booktitle = {2020 IEEE International Conference on Multisensor Fusion and Integration for Intelligent Systems (MFI)},
  year      = {2020},
  address   = {Karlsruhe, Germany},
  pages     = {223--228},
}

@inproceedings{10.5555/3001460.3001507,
author = {Ester, Martin and Kriegel, Hans-Peter and Sander, J\"{o}rg and Xu, Xiaowei},
title = {A density-based algorithm for discovering clusters in large spatial databases with noise},
year = {1996},
publisher = {AAAI Press},
booktitle = {Proceedings of the Second International Conference on Knowledge Discovery and Data Mining},
pages = {226–231},
numpages = {6},
location = {Portland, Oregon},
series = {KDD'96}
}

@article{khater_review_2020,
    title = {A {Review} of {Super}-{Resolution} {Single}-{Molecule} {Localization} {Microscopy} {Cluster} {Analysis} and {Quantification} {Methods}},
    volume = {1},
    issn = {2666-3899},
    url = {https://www.ncbi.nlm.nih.gov/pmc/articles/PMC7660399/},
    doi = {10.1016/j.patter.2020.100038},
    number = {3},
    urldate = {2025-04-02},
    journal = {Patterns},
    author = {Khater, Ismail M. and Nabi, Ivan Robert and Hamarneh, Ghassan},
    month = jun,
    year = {2020},
    pmid = {33205106},
    pmcid = {PMC7660399},
    pages = {100038},
}

@article{fazel_bayesian_2019,
    title = {Bayesian {Multiple} {Emitter} {Fitting} using {Reversible} {Jump} {Markov} {Chain} {Monte} {Carlo}},
    volume = {9},
    copyright = {2019 The Author(s)},
    issn = {2045-2322},
    url = {https://www.nature.com/articles/s41598-019-50232-x},
    doi = {10.1038/s41598-019-50232-x},
    language = {en},
    number = {1},
    urldate = {2025-06-11},
    journal = {Scientific Reports},
    author = {Fazel, Mohamadreza and Wester, Michael J. and Mazloom-Farsibaf, Hanieh and Meddens, Marjolein B. M. and Eklund, Alexandra S. and Schlichthaerle, Thomas and Schueder, Florian and Jungmann, Ralf and Lidke, Keith A.},
    month = sep,
    year = {2019},
    note = {Publisher: Nature Publishing Group},
    pages = {13791},
}

@article{fazel_high-precision_2022,
    title = {High-precision estimation of emitter positions using {Bayesian} grouping of localizations},
    volume = {13},
    copyright = {2022 The Author(s)},
    issn = {2041-1723},
    url = {https://www.nature.com/articles/s41467-022-34894-2},
    doi = {10.1038/s41467-022-34894-2},
    language = {en},
    number = {1},
    urldate = {2025-05-01},
    journal = {Nature Communications},
    author = {Fazel, Mohamadreza and Wester, Michael J. and Schodt, David J. and Cruz, Sebastian Restrepo and Strauss, Sebastian and Schueder, Florian and Schlichthaerle, Thomas and Gillette, Jennifer M. and Lidke, Diane S. and Rieger, Bernd and Jungmann, Ralf and Lidke, Keith A.},
    month = nov,
    year = {2022},
    note = {Publisher: Nature Publishing Group},
    pages = {7152},
}

@article{green_reversible_1995,
    title = {Reversible jump {Markov} chain {Monte} {Carlo} computation and {Bayesian} model determination},
    volume = {82},
    issn = {0006-3444},
    url = {https://doi.org/10.1093/biomet/82.4.711},
    doi = {10.1093/biomet/82.4.711},
    number = {4},
    urldate = {2025-06-11},
    journal = {Biometrika},
    author = {Green, Peter J},
    month = dec,
    year = {1995},
    pages = {711--732},
}

@article{milenkoviae_uncovering_2008,
    title = {Uncovering {Biological} {Network} {Function} via {Graphlet} {Degree} {Signatures}},
    volume = {6},
    issn = {1176-9351},
    url = {https://www.ncbi.nlm.nih.gov/pmc/articles/PMC2623288/},
    urldate = {2025-02-12},
    journal = {Cancer Informatics},
    author = {Milenkoviæ, Tijana and Pržulj, Nataša},
    month = apr,
    year = {2008},
    pmid = {19259413},
    pmcid = {PMC2623288},
    pages = {257--273},
}

@misc{pineda_spatial_2024,
    title = {Spatial {Clustering} of {Molecular} {Localizations} with {Graph} {Neural} {Networks}},
    url = {http://arxiv.org/abs/2412.00173},
    doi = {10.48550/arXiv.2412.00173},
    urldate = {2025-03-28},
    author = {Pineda, Jesús and Masó-Orriols, Sergi and Bertran, Joan and Goksör, Mattias and Volpe, Giovanni and Manzo, Carlo},
    month = nov,
    year = {2024},
    note = {arXiv:2412.00173 [cs]},
}

@article{khater_super_2018,
    title = {Super {Resolution} {Network} {Analysis} {Defines} the {Molecular} {Architecture} of {Caveolae} and {Caveolin}-1 {Scaffolds}},
    volume = {8},
    copyright = {2018 The Author(s)},
    issn = {2045-2322},
    url = {https://www.nature.com/articles/s41598-018-27216-4},
    doi = {10.1038/s41598-018-27216-4},
    language = {en},
    number = {1},
    urldate = {2025-08-21},
    journal = {Scientific Reports},
    author = {Khater, Ismail M. and Meng, Fanrui and Wong, Timothy H. and Nabi, Ivan Robert and Hamarneh, Ghassan},
    month = jun,
    year = {2018},
    note = {Publisher: Nature Publishing Group},
    pages = {9009},
}

@article{iyer2024DrugresistantEGFRmutations,
  title = {Drug-Resistant {{EGFR}} Mutations Promote Lung Cancer by Stabilizing Interfaces in Ligand-Free Kinase-Active {{EGFR}} Oligomers},
  author = {Iyer, R. Sumanth and Needham, Sarah R. and Galdadas, Ioannis and Davis, Benjamin M. and Roberts, Selene K. and Man, Rico C. H. and {Zanetti-Domingues}, Laura C. and Clarke, David T. and Fruhwirth, Gilbert O. and Parker, Peter J. and Rolfe, Daniel J. and Gervasio, Francesco L. and {Martin-Fernandez}, Marisa L.},
  year = {2024},
  month = mar,
  journal = {Nature Communications},
  volume = {15},
  number = {1},
  pages = {2130},
  issn = {2041-1723},
  doi = {10.1038/s41467-024-46284-x},
  urldate = {2024-12-02},
  langid = {english}
}

@misc{dereudre_introduction_2018,
    title = {Introduction to the theory of {Gibbs} point processes},
    url = {http://arxiv.org/abs/1701.08105},
    doi = {10.48550/arXiv.1701.08105},
    urldate = {2025-07-08},
    publisher = {arXiv},
    author = {Dereudre, David},
    month = apr,
    year = {2018},
    note = {arXiv:1701.08105 [math]},
}

@article{birge_gaussian_2001,
    title = {Gaussian model selection},
    volume = {3},
    issn = {1435-9855, 1435-9863},
    url = {https://ems.press/doi/10.1007/s100970100031},
    doi = {10.1007/s100970100031},
    language = {en},
    number = {3},
    urldate = {2025-06-30},
    journal = {Journal of the European Mathematical Society},
    author = {Birgé, Lucien and Massart, Pascal},
    month = sep,
    year = {2001},
    pages = {203--268},
}

@article{Smiljanic2023MapEquation,
  title     = {Community Detection with the Map Equation and Infomap: Theory and Applications},
  author    = {Smiljani{\'c}, Jelena and Bl{\"o}cker, Christopher and Holmgren, Anton and Edler, Daniel and Neuman, Magnus and Rosvall, Martin},
  journal   = {arXiv preprint arXiv:2311.04036},
  year      = {2023},
  doi       = {10.48550/arXiv.2311.04036},
  url       = {https://arxiv.org/abs/2311.04036}
}

@article{rosvall_map_2009,
    title = {The map equation},
    volume = {178},
    copyright = {http://www.springer.com/tdm},
    issn = {1951-6355, 1951-6401},
    url = {http://link.springer.com/10.1140/epjst/e2010-01179-1},
    doi = {10.1140/epjst/e2010-01179-1},
    language = {en},
    number = {1},
    urldate = {2025-03-12},
    journal = {The European Physical Journal Special Topics},
    author = {Rosvall, M. and Axelsson, D. and Bergstrom, C. T.},
    month = nov,
    year = {2009},
    pages = {13--23},
}

@article{lindgren_explicit_2011,
    title = {An explicit link between {Gaussian} fields and {Gaussian} {Markov} random fields: the stochastic partial differential equation approach},
    volume = {73},
    copyright = {© 2011 Royal Statistical Society},
    issn = {1467-9868},
    shorttitle = {An explicit link between {Gaussian} fields and {Gaussian} {Markov} random fields},
    url = {https://onlinelibrary.wiley.com/doi/abs/10.1111/j.1467-9868.2011.00777.x},
    doi = {10.1111/j.1467-9868.2011.00777.x},
    language = {en},
    number = {4},
    urldate = {2025-09-18},
    journal = {Journal of the Royal Statistical Society: Series B (Statistical Methodology)},
    author = {Lindgren, Finn and Rue, Håvard and Lindström, Johan},
    year = {2011},
    note = {\_eprint: https://rss.onlinelibrary.wiley.com/doi/pdf/10.1111/j.1467-9868.2011.00777.x},
    pages = {423--498},
}

@book{santalo2004integral,
  author    = {Santaló, Luis A.},
  title     = {Integral Geometry and Geometric Probability},
  edition   = {2nd},
  publisher = {Cambridge University Press},
  year      = {2004},
  address   = {Cambridge, UK}
}

@article{leone_folded_normal,
 ISSN = {00401706},
 URL = {http://www.jstor.org/stable/1266560},
 author = {F. C. Leone and L. S. Nelson and R. B. Nottingham},
 journal = {Technometrics},
 number = {4},
 pages = {543--550},
 publisher = {[Taylor & Francis, Ltd., American Statistical Association, American Society for Quality]},
 title = {The Folded Normal Distribution},
 urldate = {2025-03-03},
 volume = {3},
 year = {1961}
}

@article{roberts1997weak,
  title={Weak convergence and optimal scaling of random walk Metropolis algorithms},
  author={Roberts, Gareth O. and Gelman, Andrew and Gilks, Walter R.},
  journal={The Annals of Applied Probability},
  volume={7},
  number={1},
  pages={110--120},
  year={1997},
  publisher={Institute of Mathematical Statistics}
}

@inproceedings{koren2008factorization,
  title={Factorization meets the neighborhood: a multifaceted collaborative filtering model},
  author={Koren, Yehuda},
  booktitle={Proceedings of the 14th ACM SIGKDD international conference on Knowledge discovery and data mining},
  pages={426--434},
  year={2008}
}

@article{liu2000multiple,
  title={The multiple-try method and local optimization in Metropolis sampling},
  author={Liu, Jun S. and Liang, Faming and Wong, Wing Hung},
  journal={Journal of the American Statistical Association},
  volume={95},
  number={449},
  pages={121--134},
  year={2000},
  publisher={Taylor \& Francis},
  doi={10.1080/01621459.2000.10473908}
}

@article{RobbinsMonro1951,
  author  = {Robbins, Herbert and Monro, Sutton},
  title   = {A Stochastic Approximation Method},
  journal = {The Annals of Mathematical Statistics},
  year    = {1951},
  volume  = {22},
  number  = {3},
  pages   = {400--407},
  doi     = {10.1214/aoms/1177729586}
}

@article{balanov2024einstein,
  title={Einstein from Noise: Statistical Analysis},
  author={Balanov, Amnon and Huleihel, Wasim and Bendory, Tamir},
  journal={arXiv preprint arXiv:2407.05277},
  year={2024},
  eprint={2407.05277},
  archivePrefix={arXiv},
  primaryClass={eess.SP}
}

@article{edelsbrunner_three-dimensional_1994,
    title = {Three-dimensional alpha shapes},
    volume = {13},
    issn = {0730-0301, 1557-7368},
    url = {https://dl.acm.org/doi/10.1145/174462.156635},
    doi = {10.1145/174462.156635},
    language = {en},
    number = {1},
    urldate = {2026-01-20},
    journal = {ACM Transactions on Graphics},
    author = {Edelsbrunner, Herbert and Mücke, Ernst P.},
    month = jan,
    year = {1994},
    pages = {43--72},
}

@article{hubert1985comparing,
  title={Comparing partitions},
  author={Hubert, Lawrence and Arabie, Phipps},
  journal={Journal of Classification},
  volume={2},
  number={1},
  pages={193--218},
  year={1985},
  publisher={Springer}
}

@article{fowlkes1983method,
  title={A method for comparing two hierarchical clusterings},
  author={Fowlkes, Edward B and Mallows, Colin L},
  journal={Journal of the American Statistical Association},
  volume={78},
  number={383},
  pages={553--569},
  year={1983},
  publisher={Taylor \& Francis}
}

@article{vinh2010information,
  title={Information theoretic measures for clusterings comparison: Variants, properties, normalization and correction for chance},
  author={Vinh, Nguyen Xuan and Epps, Julien and Bailey, James},
  journal={Journal of Machine Learning Research},
  volume={11},
  pages={2837--2854},
  year={2010}
}

@article{kuhn1955hungarian,
  title={The Hungarian method for the assignment problem},
  author={Kuhn, Harold W},
  journal={Naval Research Logistics Quarterly},
  volume={2},
  number={1-2},
  pages={83--97},
  year={1955},
  publisher={Wiley}
}

@article{kingwell2018antibacterial,
  title={Antibacterial agents: New antibiotic hits {Gram}-negative bacteria},
  author={Kingwell, Katie},
  journal={Nature Reviews Drug Discovery},
  volume={17},
  number={11},
  pages={785},
  year={2018},
  publisher={Nature Publishing Group},
  doi={10.1038/nrd.2018.182},
  pmid={30337723}
}

@article{needham2016egfr,
  title={{EGFR} oligomerization organizes kinase-active dimers into competent signalling platforms},
  author={Needham, Sarah R and Roberts, Selene K and Arkhipov, Anton and Mysore, Venkatesh P and Tynan, Christopher J and Zanetti-Domingues, Laura C and Kim, Eric T and Losasso, Valeria and Korovesis, Dimitrios and Hirsch, Michael and Rolfe, Daniel J and Clarke, David T and Winn, Martyn D and Lajevardipour, Alireza and Clayton, Andrew H A and Pike, Linda J and Perani, Michela and Parker, Peter J and Shan, Yibing and Shaw, David E and Martin-Fernandez, Marisa L},
  journal={Nature Communications},
  volume={7},
  pages={13307},
  year={2016},
  publisher={Nature Publishing Group},
  doi={10.1038/ncomms13307},
  pmid={27796308}
}

@article{speiser2021deep,
  author    = {Artur Speiser and Lucas-Raphael M{\"u}ller and Philipp Hoess and 
               Ulf Matti and Christopher J. Obara and Wesley R. Legant and 
               Anna Kreshuk and Jakob H. Macke and Jonas Ries and Srinivas C. Turaga},
  title     = {Deep learning enables fast and dense single-molecule localization 
               with high accuracy},
  journal   = {Nature Methods},
  volume    = {18},
  number    = {9},
  pages     = {1082--1090},
  year      = {2021},
  publisher = {Nature Publishing Group},
  doi       = {10.1038/s41592-021-01236-x}
}

@article{carlsson2009topology,
  author    = {Gunnar Carlsson},
  title     = {Topology and data},
  journal   = {Bulletin of the American Mathematical Society},
  volume    = {46},
  number    = {2},
  pages     = {255--308},
  year      = {2009},
  doi       = {10.1090/S0273-0979-09-01249-X}
}

@article{delaunay1934sphere,
  author    = {Boris Delaunay},
  title     = {Sur la sph{\`e}re vide},
  journal   = {Bulletin de l'Acad{\'e}mie des Sciences de l'URSS, Classe des Sciences Math{\'e}matiques et Naturelles},
  volume    = {6},
  pages     = {793--800},
  year      = {1934}
}

@article{edelsbrunner2002topological,
  author    = {Herbert Edelsbrunner and David Letscher and Afra Zomorodian},
  title     = {Topological persistence and simplification},
  journal   = {Discrete \& Computational Geometry},
  volume    = {28},
  number    = {4},
  pages     = {511--533},
  year      = {2002},
  publisher = {Springer},
  doi       = {10.1007/s00454-002-2885-2}
}

@book{luce1959individual,
  author    = {R. Duncan Luce},
  title     = {Individual Choice Behavior: A Theoretical Analysis},
  publisher = {John Wiley \& Sons},
  address   = {New York},
  year      = {1959}
}

@article{plackett1975analysis,
  author    = {Robin L. Plackett},
  title     = {The analysis of permutations},
  journal   = {Journal of the Royal Statistical Society: Series C (Applied Statistics)},
  volume    = {24},
  number    = {2},
  pages     = {193--202},
  year      = {1975},
  publisher = {Wiley},
  doi       = {10.2307/2346567}
}

@misc{simflux,
  author = {Jack Peyton},
  title = {simflux},
  version = {0.1.2},
  year = {2025},
  howpublished={\url{https://github.com/j-peyton/SimFlux}}
}

@article{thevathasan_nuclear_2019,
    title = {Nuclear pores as versatile reference standards for quantitative superresolution microscopy},
    volume = {16},
    copyright = {2019 The Author(s), under exclusive licence to Springer Nature America, Inc.},
    issn = {1548-7105},
    url = {https://www.nature.com/articles/s41592-019-0574-9},
    doi = {10.1038/s41592-019-0574-9},
    language = {en},
    number = {10},
    urldate = {2025-02-19},
    journal = {Nature Methods},
    publisher = {Nature Publishing Group},
    author = {Thevathasan, Jervis Vermal and Kahnwald, Maurice and Cieśliński, Konstanty and Hoess, Philipp and Peneti, Sudheer Kumar and Reitberger, Manuel and Heid, Daniel and Kasuba, Krishna Chaitanya and Hoerner, Sarah Janice and Li, Yiming and Wu, Yu-Le and Mund, Markus and Matti, Ulf and Pereira, Pedro Matos and Henriques, Ricardo and Nijmeijer, Bianca and Kueblbeck, Moritz and Sabinina, Vilma Jimenez and Ellenberg, Jan and Ries, Jonas},
    month = oct,
    year = {2019},
    pages = {1045--1053},
}

@article{brooks1998,
  author  = {Brooks, Stephen P. and Gelman, Andrew},
  title   = {General Methods for Monitoring Convergence of Iterative Simulations},
  journal = {Journal of Computational and Graphical Statistics},
  volume  = {7},
  number  = {4},
  pages   = {434--455},
  year    = {1998},
  doi     = {10.1080/10618600.1998.10474787}
}

@article{hastings1970,
  author = {Hastings, W. K.},
  title = {Monte Carlo Sampling Methods Using Markov Chains and Their Applications},
  journal = {Biometrika},
  volume = {57},
  number = {1},
  pages = {97--109},
  year = {1970},
  doi = {10.1093/biomet/57.1.97}
}

@article{smiljanic2020,
  author = {Smiljani\'{c}, Jelena and Edler, Daniel and Rosvall, Martin},
  title = {Mapping flows on sparse networks with missing links},
  journal = {Physical Review E},
  volume = {102},
  number = {1},
  pages = {012302},
  year = {2020},
  doi = {10.1103/PhysRevE.102.012302}
}

@article{haario_adaptive_2001,
    title = {An adaptive {Metropolis} algorithm},
    volume = {7},
    issn = {1350-7265},
    url = {https://projecteuclid.org/journals/bernoulli/volume-7/issue-2/An-adaptive-Metropolis-algorithm/bj/1080222083.full},
    number = {2},
    urldate = {2025-08-05},
    journal = {Bernoulli},
    publisher = {Bernoulli Society for Mathematical Statistics and Probability},
    author = {Haario, Heikki and Saksman, Eero and Tamminen, Johanna},
    month = apr,
    year = {2001},
    pages = {223--242},
}

@article{roberts2001,
  author = {Roberts, Gareth O. and Rosenthal, Jeffrey S.},
  title = {Optimal Scaling for Various {M}etropolis-{H}astings Algorithms},
  journal = {Statistical Science},
  volume = {16},
  number = {4},
  pages = {351--367},
  year = {2001},
  doi = {10.1214/ss/1015346320}
}

@article{moller1998log,
  author    = {M{\o}ller, Jesper and Syversveen, Anne Randi and Waagepetersen, Rasmus Plenge},
  title     = {Log {G}aussian Cox Processes},
  journal   = {Scandinavian Journal of Statistics},
  volume    = {25},
  number    = {3},
  pages     = {451--482},
  year      = {1998},
  publisher = {Wiley}
}

@book{boyd2004convex,
  author    = {Boyd, Stephen and Vandenberghe, Lieven},
  title     = {Convex Optimization},
  publisher = {Cambridge University Press},
  year      = {2004},
  address   = {Cambridge, UK}
}

\end{document}

\renewcommand{\thefigure}{S\arabic{figure}}
\renewcommand{\thetable}{S\arabic{table}}
\renewcommand{\theequation}{S\arabic{equation}}

\maketitle

\tableofcontents

\newpage
\section{Methods}\label{sec:methods}
Let the measurement set \(Y=\{y_i\}_{i=1}^{n_y}\) denote the raw observed data MINFLUX provides.

\begin{figure}[ht]
    \centering
    \includegraphics[width=0.5\linewidth]{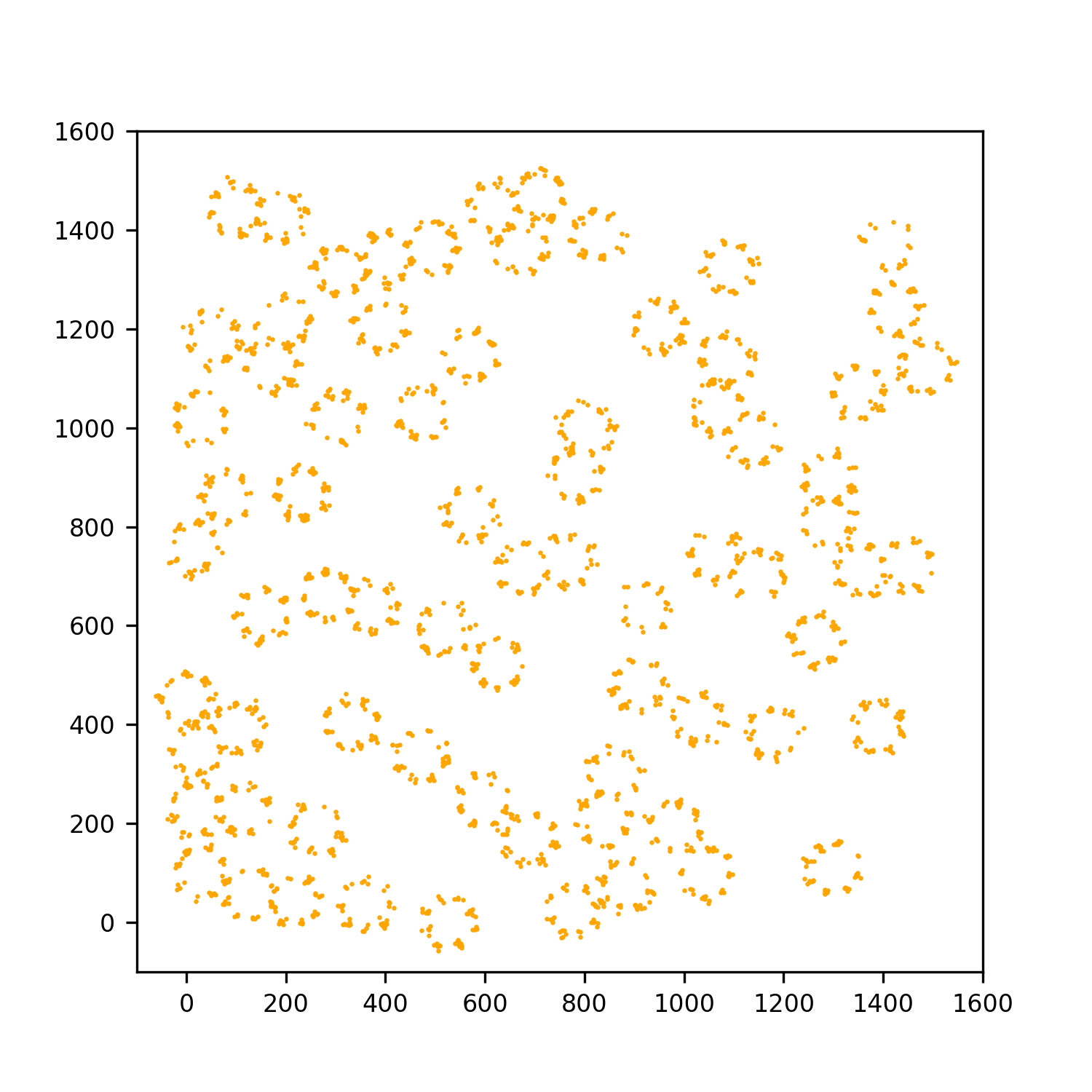}
    \caption{A synthetic point pattern representing homogeneously distributed regular-octagonal structures with emitters at each vertex that have been subject to repeated (but varying) numbers of measurements. These datasets were generated by SIMFLUX \cite{simflux} and are similar to point patterns observed when imaging Nup96 using MINFLUX. This dataset was further modified to contain both monomeric and dimeric octagon structures.}
    \label{fig:methods-observed}
\end{figure}

Fig. \ref{fig:methods-observed} depicts such a measurement set, and will be carried forward to illustrate each aspect of the method.

\subsection{GROUPA: Grouping Observations Under Pairwise Associations}\label{subsec:groupa}
GROUPA is an uncertainty aware algorithm for clustering measurements into groups whose centre represents the most likely position of an unseen parent point locations (an emitter). This process is re-framed from attempting to fit a global model to instead evaluating a series of more constrained models; the relative evidence that each pair of localisations represent measurements from:

\begin{align*}
    H_0&: \text{a common emitter}, \\
    H_1&: \text{two distinct emitters}.
\end{align*}

This pairwise formulation offers several advantages to those discussed prior:

\begin{itemize}
  \item Pairwise comparisons reduce problem complexity to a series of binary hypothesis tests, tractable through overlap integrals of probability densities, i.e. permits efficient sparse matrix and Kdtree representations.
  \item Only localisation pairs within a reasonable cut-off distance are considered, avoiding combinatorial explosion.
  \item Strong global priors on emitter counts or their density in the dataset are not required.
\end{itemize}

Let each localisation \(y_i\in Y\) be represented by a Gaussian mixture model (GMM)\footnote{The GMM model could be weighted and components are not necessarily restricted to space – for example molecular dipoles.},
\[
p(\mathbf{y}) = \sum_{k} w_k \, \mathcal{N}(\mathbf{y}\mid \mu_k, \Sigma_k),
\qquad
q(\mathbf{y}) = \sum_{\ell} v_\ell \, \mathcal{N}(\mathbf{y}\mid \nu_\ell, \Lambda_\ell).
\]

The overlap integral between two such mixtures is
\[
I \;=\; \int p(\mathbf{y})\,q(\mathbf{y})\,d\mathbf{y}
     \;=\; \sum_{k}\sum_{\ell} w_k v_\ell \int 
          \mathcal{N}(\mathbf{y}\mid \mu_k, \Sigma_k)\
          \mathcal{N}(\mathbf{y}\mid \nu_\ell, \Lambda_\ell)\,d\mathbf{y}.
\]

As each inner integral has the closed form
\[
\int \mathcal{N}(\mathbf{y}\mid \mu_k, \Sigma_k)\
     \mathcal{N}(\mathbf{y}\mid \nu_\ell, \Lambda_\ell)\,d\mathbf{y}
   \;=\; \mathcal{N}(\mu_k \mid \nu_\ell, \Sigma_k + \Lambda_\ell),
\]
\(I\) reduces to a weighted sum of Gaussian–Gaussian overlaps. This quantity measures the raw probability mass assigned jointly by both uncertainty models.

Under the single-emitter hypothesis \(H_0\), two independent measurements should fall within the emitter’s effective uncertainty region.  
This region is approximated by the \(99.5\%\) confidence ellipsoid defined by the Gaussian covariance in \(d\) dimensional space:
\[
r^2 = \chi^2_{d,\alpha}, \quad \alpha = 0.995.
\]
The effective volume is then estimated as the union of the two uncertainty regions
\[
V_{\mathrm{eff}} \;\approx\; \mathrm{Vol}\!\left( \mathcal{E}(\mu_p, \Sigma_p, r_p) \;\cup\; \mathcal{E}(\mu_q, \Sigma_q, r_q) \right),
\]
which can be readily and rapidly estimated using Monte Carlo sampling \cite{hastings1970} providing a baseline scale for coincidental overlap under \(H_0\).

The pairwise Bayes factor is defined as
\[
\mathrm{BF}_{ij} \;=\; I \cdot V_{\mathrm{eff}},
\]
where \(I\) captures the local overlap of two Gaussian mixture models, and \(V_{\mathrm{eff}}\) accounts for the prior-volume penalty associated with two independent emitter locations under \(H_1\).

Assume a uniform prior over a feasible region \(R\) of volume \(V_{\mathrm{eff}}\). Then

- Under \(H_0\):
\[
p(\text{data}\mid H_0) \;=\; \int_R \frac{1}{V_{\mathrm{eff}}}\, p(z)\,q(z)\,dz 
\;=\; \frac{I}{V_{\mathrm{eff}}},
\]

- Under \(H_1\):
\[
p(\text{data}\mid H_1) \;=\; \int_R\!\!\int_R \frac{1}{V_{\mathrm{eff}}^2}\,p(z_1)\,q(z_2)\,dz_1dz_2
\;=\; \frac{1}{V_{\mathrm{eff}}^2}.
\]

As \(p\) and \(q\) are normalised, the resulting Bayes factor is therefore
\[
\mathrm{BF}_{ij} \;=\; \frac{p(\text{data}\mid H_0)}{p(\text{data}\mid H_1)}
\;=\; I \cdot V_{\mathrm{eff}}.
\]

Thus, the evidence for a common emitter naturally factorises into a local overlap term (\(I\)) and a prior-volume term (\(V_{\mathrm{eff}}\)), which penalises the extra latent parameter in the two-emitter hypothesis.

Pairwise Bayes factors above a threshold (typically BF \(>\)1) are retained as weighted edges in an undirected graph where nodes represent localisations and edges represent evidence of common origin. As these edges provide only first-order evidence of association and do not directly encode higher-order dependencies, techniques such as modularity and other spectrum-based approaches that evaluate partitions from static measures of pairwise connectivity are unsuitable for this clustering problem. Instead, the Infomap algorithm \cite{Smiljanic2023MapEquation} was used, which exploits flow-based dynamics to minimize the map equation (an information-theoretic objective corresponding to the Shannon entropy of a random walk’s description length under a given partition \cite{rosvall_map_2009}) and therefore captures both second-order and higher-order structure emerging from local evidences \cite{Smiljanic2023MapEquation}. This technique provides an efficient alternative to fully Bayesian network models that attempt to infer higher order interactions directly, yielding comparable explanatory power at substantially lower computational cost. 

GROUPA can be used to cluster millions of measurements locally (connectivity dependent), and is amenable to chunking. In dense systems, connectivity can be sampled by considering mutual-KNN or setting a global upper limit on max separation between points. Mutual-KNN will avoid over-relying more uncertain points that will have more neighbours otherwise. In a sparse, under-sampled network, a poor selection of KNN may lead to overfitting \cite{smiljanic2020}.

\begin{figure}[ht]
    \centering
    \includegraphics[width=0.5\linewidth]{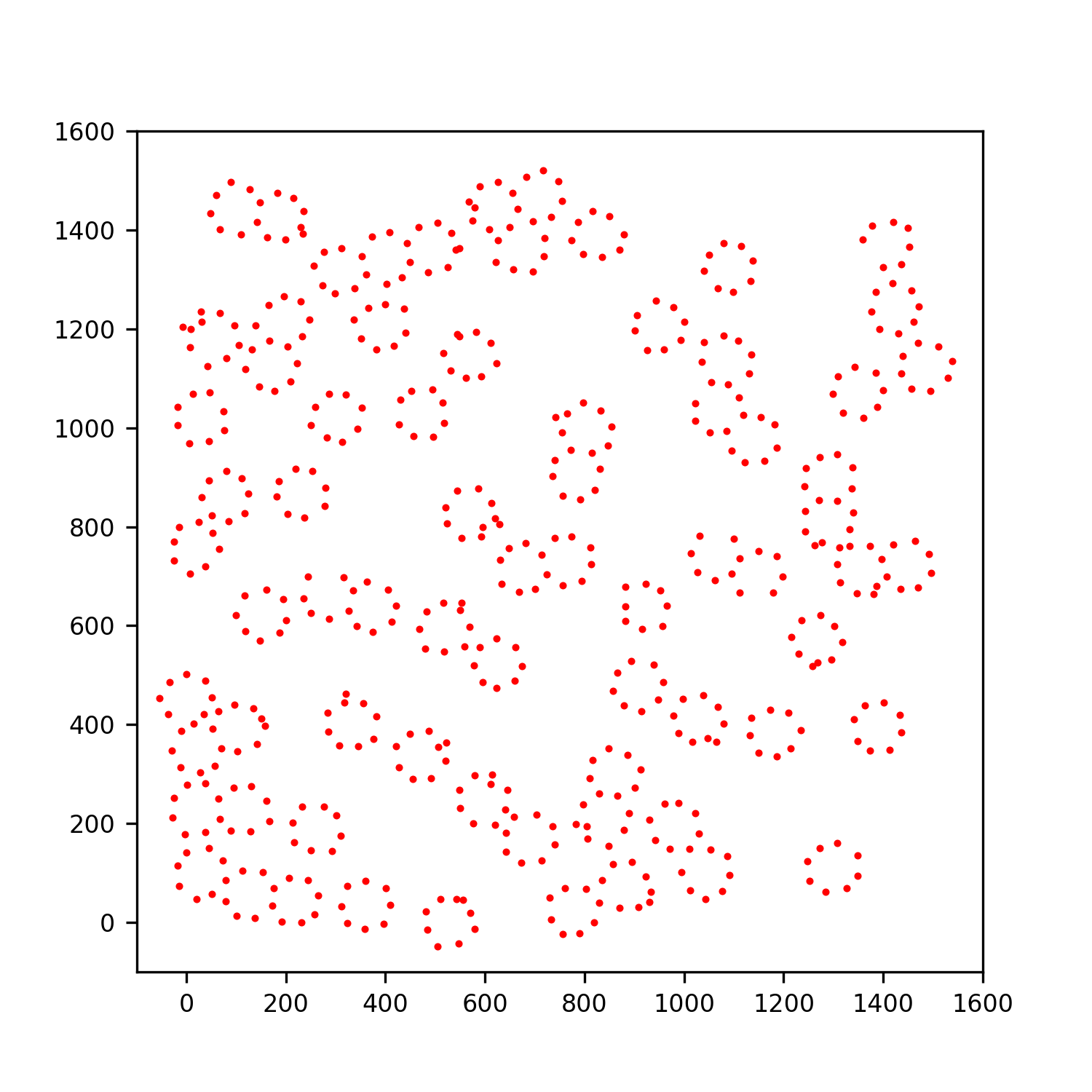}
    \caption{The result of the emitter estimation method applied to the measurement set in Fig. \ref{fig:methods-observed}}
    \label{fig:methods-emitters}
\end{figure}

For localisations modelled as single Gaussian, evaluating \(I\) requires a single Gaussian-Gaussian overlap and is constant time per pair. In the more general case of Gaussian mixtures with \(K_p\) and \(K_q\) components, the overlap integral expands into \(O(K_p K_q)\) closed-form terms. Since clustering only considers pairs within a spatial cutoff, the total graph construction scales approximately as \(O(N \cdot k)\) for \(N\) localisations and average neighbourhood size \(k\), rather than \(O(N^2)\).

\subsection{Voidwalker}\label{subsec:voidmethods}
Once emitter positions have been estimated from the measurement set, Voidwalker is the first step taken in structural inference. Voidwalker uses the posterior of the intensity of the point pattern to seed voids probabilistically, then grows and walks these voids to fill their local empty space, and tests all voids to determine which empty space is of statistically significant interest, providing priors and a proposal space for future RJMCMC.

Let \(W \subset \mathbb{R}^{2}\) be an observation window with boundary \(\partial W\) and Lebesgue measure \(|W|\). Observe the spatial point pattern of the estimated emitters \(X=\{x_{n}\}_{n=1}^{N}\subset W\). For any \(u\in W\), define distances to the point set and to the boundary, 

\[
d_{X}(u):=\min_{1\leq n \leq N}\|u-x_{n}\|, \quad d_{\partial W}(u):=\inf _{v\in\partial W}\|u-v\|.
\]

Define the clearance field, \(r(u)\), as the pointwise radius of the largest empty disc centred at \(u\):

\begin{equation}\label{void-emptiness}
r(u):=\min\{d_{X}(u), d_{\partial W}(u)\}, \quad B(u, r(u))\cap X = \emptyset,  \quad B(u,r(u))\subset W,
\end{equation}

where the empty disc \(B(u,r(u))\) is referred to as a void. 

The inhomogeneous intensity is modelled as
\begin{equation}\label{eq:lgcp}
\lambda(u)\;=\;\exp\{\eta(u)\},\qquad \eta(u)\;=\;\beta_0 + \omega(u),\qquad u\in W,
\end{equation}
where $\beta_0$ is an intercept and $\omega$ is a mean-zero Matérn Gaussian random field represented by a Stochastic Partial Differential Equation (SPDE) \cite{lindgren_explicit_2011} on a triangular mesh built on \( W\). Penalised complexity priors are placed on the Matérn hyperparameters, with range and marginal standard deviation scaled to the window span. The likelihood is the standard Log-Gaussian Cox Process (LGCP) \cite{moller1998log} with the observation domain passed as a named polygonal sampler.

This fit provides posterior draws of the intensity field

\begin{equation}\label{eq:emitterintensitydraws}
    \lambda^{(s)}(u) = \exp\left( \eta^{(s)}(u)\right), \quad s=1, \dots, S,
\end{equation}

and consequently a variance corrected posterior mean intensity 

\begin{equation}\label{eq:emittermeanintensity}
    \hat{\lambda}(u)\simeq\exp\left( \mathbb{E}[\eta (u)|X] + \frac{1}{2}\text{Var} [\eta (u) | X] \right).
\end{equation}

\begin{figure}[ht]
    \centering
    \includegraphics[width=0.6\linewidth]{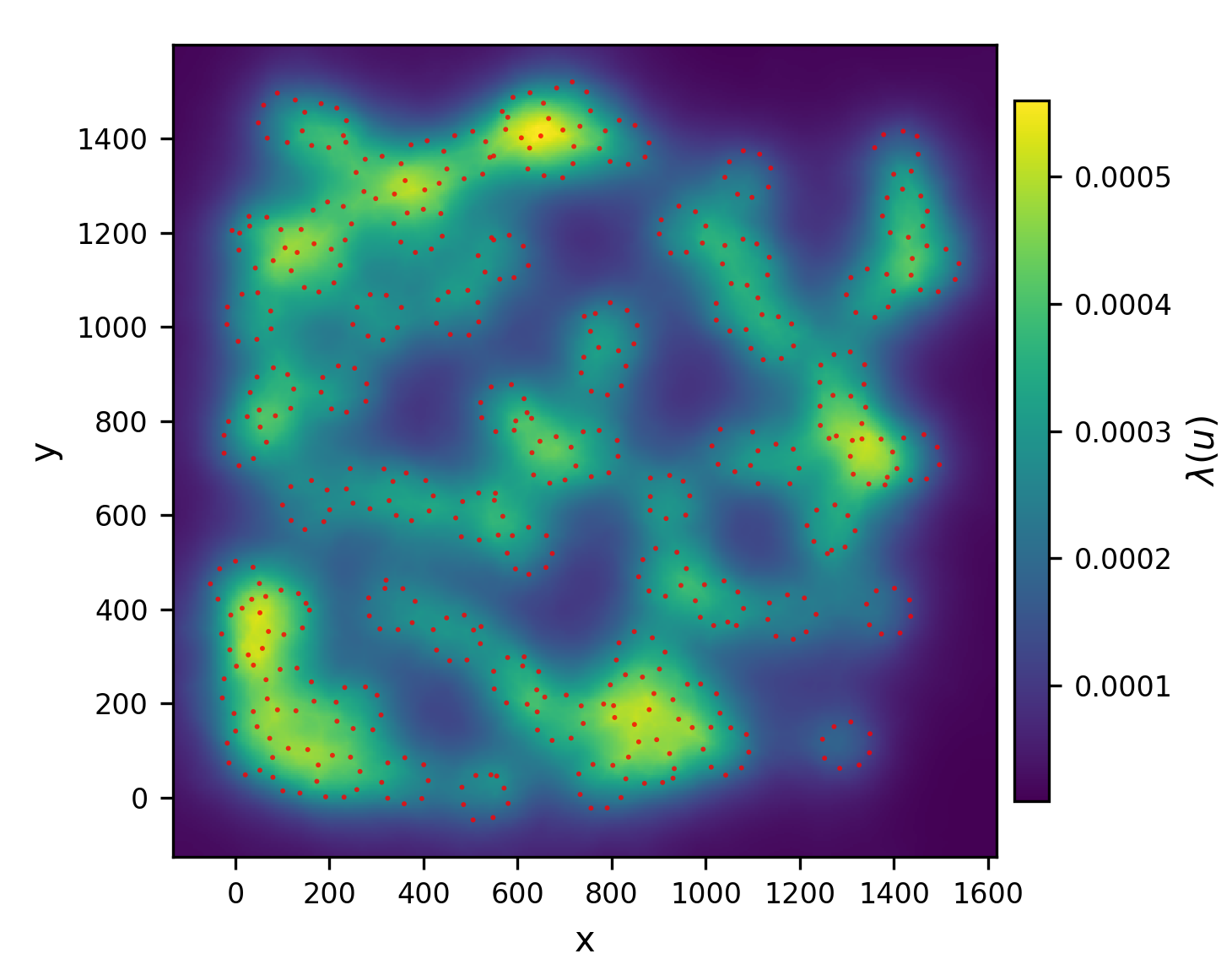}
    \caption{Posterior mean intensity estimate of emitter point pattern Fig. \ref{fig:methods-emitters}.}
    \label{fig:methods-intensity}
\end{figure}

Discretise the observation window $W$ into a regular grid of $|\mathcal{G}|$ square cells, each with side length $\delta$ and area $\Delta A = \delta^2$. Let $\{\lambda_g\}_{g\in\mathcal{G}}$ denote the posterior mean intensity evaluated at each cell centroid. The expected count over a region $\mathcal{R}\subseteq W$ is then approximated by summing over cells whose centroids fall within $\mathcal{R}$:
\begin{equation}\label{eq:exp-count-grid}
    \mu(\mathcal{R})\;=\;\int_{\mathcal{R}}\lambda(u)\,du
    \;\approx\;
    \sum_{g\in \mathcal{G}(\mathcal{R})}\lambda_g\,\Delta A,
\end{equation}
where $\mathcal{G}(\mathcal{R}) = \{g \in \mathcal{G} : c_g \in \mathcal{R}\}$ and $c_g$ denotes the centroid of cell $g$.

To avoid trivial seeding of voids, a guard function is implemented to enforce minimum distance criteria between both \(X\) and \(\partial W\). Set two distance criteria, \(\rho_X>0\) and \(\rho_{\partial W}>0\), and define

\[
G(u):=\mathds{1} \{d_{X}(u)\geq\rho_X\}\mathds{1} \{d_{\partial W}(u)\geq\rho_{\partial W}\}.
\]

Draw a finite set of seeds\(\{\iota_{k}\}_{k=1}^{\mathcal{I} _{0}}\subset W\) from a guarded, intensity-weighted proposal \(q(u)\propto \hat{\lambda}(u)G(u)\), \(u\in W\), implemented on a raster \(\{x_p, y_q\}\) by normalising \(\hat{\lambda}_{pq}G_{pq}\).

For each seed \(s_k\), consider the empty-ball radius function, \(r(u)\). Voidwalker seeks the local maximiser \(c^*_k\) of \(r(u)\) in the basin of attraction of \(\iota_k\). At a local maximum \(c^*\) there is a contact set \(H(c^*)\subset X\cup \partial W\) such that

\[
r(c^*) = \min_{h\in H(c^* )}\|c^* - h\|, \quad \text{and }\sum_{h\in H(c^* )}\alpha_h\frac{c^* - h}{\|c^* - h\|}=0,
\]

for some \(\alpha_h\geq0\) with \(\sum_h \alpha_h=1\). This first order Karush-Kuhn-Tucker (KKT) \cite{boyd2004convex} condition is not directly solved, but is approximated by point repulsive stochastic growths and random walks of each seeded void. At each step, void radii are increased. If any of the  point set are violating a voids emptiness condition (Eq. \ref{void-emptiness}), the void centre undergoes a random walk step in the direction of the resultant force vector from all violating points. Growth terminates when either (i) a pre-defined maximum radius is reached, chosen to exceed the spatial scale of interest (75\,nm for the synthetic Nup96 data), or (ii) the void attains a local maximum of the clearance field, i.e.\ $r(c_k)$ cannot increase under any small perturbation of $c_k$.

Let this set of grown voids be denoted \(v_k:=(c_k, r_k)\), where \(c_k=c_k^*\) and \(r_k=r(c_k^*)\). This set is thinned by a user-defined minimum radius gate, and by non-maximum suppression (NMS). In \((x, y, r)\), NMS sorts voids by descending \(r\), greedily keeping a candidate and suppressing any similar in location and in size.

\begin{figure}[ht]
    \centering
    \includegraphics[width=0.6\linewidth]{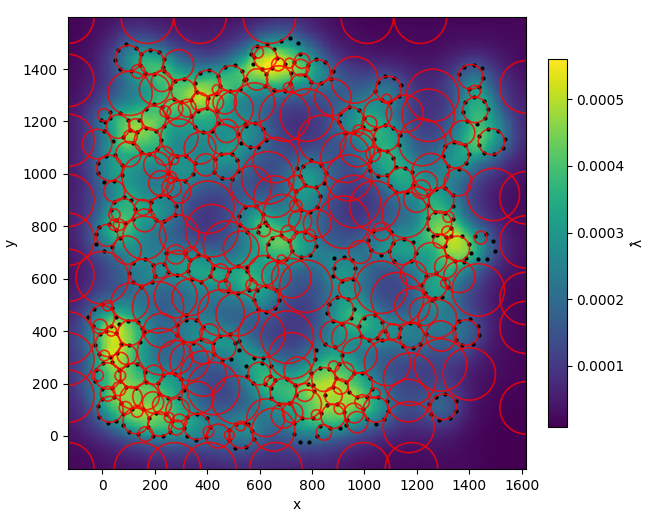}
    \caption{All seeded voids after the growth and walk process.}
    \label{fig:voids_all}
\end{figure}

Voids are calibrated against the posterior predictive draw (PPD) null computed from the draws described in Eq. \ref{eq:emitterintensitydraws}. Throughout PPD null simulations, two characteristic statistics are computed on the already existing candidate void geometry, and an additional statistic is computed from inhomogeneous simulations on these draws. For a candidate void \(i\), with location and radius \(c_i,\) \(r_i\), define the ball and annulus respectively

\[
B_i:=B(c_i, r_i), \quad R_i:=\{x\in W: \|x-c_i\|\leq r_i(1+f_r )\},
\]

with \(f_r \in (0, 1)\). For simulation $m=1,\dots,M$, and candidate voids \(i=1,\dots,n\):
\begin{enumerate}
  \item Pick a posterior draw $\lambda^{(s_m)}$.
  \item Simulate $X^{(m)}\mid \lambda^{(s_m)}$ as an inhomogeneous Poisson process on $W$.
  \item For each candidate $i$ with centre $c_i$ and radius $r_i$, define the annulus
        $R_i=\{x\in W: r_i<\|x-c_i\|\le r_i(1+f_r)\}$ with $f_r\in(0,1)$, and compute
        \[
        \mu_i^{(m)}=\int_{R_i}\lambda^{(s_m)}(u)\,du,\qquad
        N_i^{(m)}=\#\{x\in X^{(m)}:x\in R_i\},\qquad
        Z_i^{(m)}=\frac{N_i^{(m)}-\mu_i^{(m)}}{\sqrt{\mu_i^{(m)}}}\,.
        \]
\end{enumerate}

Pool all simulated scores to form the empirical null CDF
\[
\hat F_0(z)=\frac{1}{Mn}\sum_{m=1}^{M}\sum_{i=1}^{n}\mathds{1}\!\left[\,Z_i^{(m)}\le z\,\right],
\qquad
t_Z(\alpha)=\hat F_0^{-1}(1-\alpha).
\]

On the existing void candidates, compute
\[
\mu_i=\int_{R_i}\hat\lambda(u)\,du,\quad
N_i=\#\{x\in X:x\in R_i\},\quad
Z_i=\frac{N_i-\mu_i}{\sqrt{\mu_i}},
\quad
p_i=1-\hat F_0(Z_i).
\]
Activate candidate $i$ iff $p_i\le\alpha$ and $Z_i\ge t_Z(\alpha)$, to form the set of active voids

\[
\mathcal{A}=\{i:(p_i\leq\alpha)\cap (Z_i\geq t_Z(\alpha))\}. 
\]

Active voids inform the Poisson prior on structure, the Gaussian prior on radius, and the birth proposal for use downstream in RJMCMC. 

Assuming the radii of active voids to follow a normal distribution, compute the mean and variance \((\hat{\mu}_r, \hat{\sigma}_r)\), and apply a band 

\[
\mathcal{A}^*=\left\{i\in \mathcal{A}:|r_i - \hat{\mu}_r|\leq2\hat{\sigma}_r\right\}
\]

to discard outliers.

\begin{figure}[ht]
    \centering
    \includegraphics[width=0.6\linewidth]{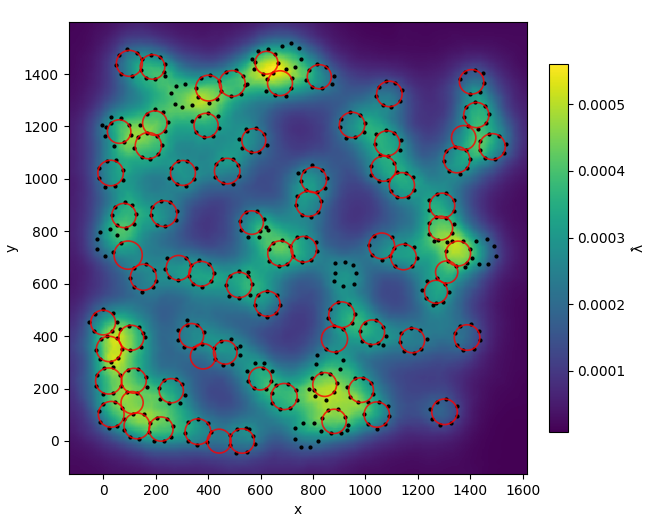}
    \caption{Set of thinned activated voids, \(\mathcal{A}^*\).}
    \label{fig:voids_active}
\end{figure}

Let the cardinality $|\mathcal{A}^*|$ define the rate parameter of a Poisson prior on the number of structures, and let the radii $\{r_i : i \in \mathcal{A}^*\}$ define a Gaussian prior on structure radius with mean $\hat{\mu}_r$ and variance $\hat{\sigma}_r^2$ estimated from active voids. Let $C = \{c_i : i \in \mathcal{A}^*\}$ denote the centres of active voids.

The birth proposal distribution is a two-component mixture:
\begin{equation}\label{eq:birthproposal}
    q_b(x) \;\propto\; \omega_\lambda \frac{\lambda(x)}{\int_W \lambda(u)\,du} 
    \;+\; \omega_{\mathcal{A}^*} \frac{1}{|\mathcal{A}^*|} 
    \sum_{i \in \mathcal{A}^*} \phi(x;\, c_i,\, \hat{\sigma}_b I),
\end{equation}
where:
\begin{itemize}
    \item $\omega_\lambda, \omega_{\mathcal{A}^*} \in [0,1]$ are mixture 
    weights satisfying $\omega_\lambda + \omega_{\mathcal{A}^*} = 1$, 
    controlling the relative contribution of each component;
    \item the first component samples proportionally to the normalised 
    posterior intensity $\frac{\lambda(x)}{\int_W \lambda(u)\,du}$, encouraging 
    births in regions of high emitter density;
    \item $\phi(x;\, c_i,\, \hat{\sigma}_b I)$ denotes the bivariate 
    Gaussian density with mean $c_i$ and isotropic covariance 
    $\hat{\sigma}_b^2 I_2$, where $\hat{\sigma}_b$ is a bandwidth 
    parameter controlling dispersion around each void centre;
    \item the second component places Gaussian kernels at each active 
    void centre, encouraging births near regions already identified 
    as structurally significant.
\end{itemize}
This mixture proposal concentrates birth attempts in promising regions of the domain while retaining coverage of the full observation window through the intensity component.

\begin{figure}[ht]
    \centering
    \includegraphics[width=0.6\linewidth]{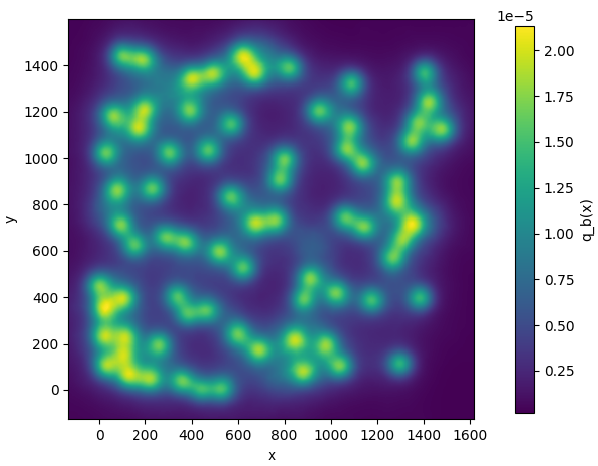}
    \caption{Birth proposal \(q_b\) calculated from intensity and \(\mathcal{A}^*\) coverage}
    \label{fig:birth_proposal}
\end{figure}

\subsection{Gibbs Process Reversible Jump MCMC}\label{subsec:gibbsmethods}
The structural centres of the biological subjects are modelled as a realisation of a Gibbs Point Process\cite{Illian2008, dereudre_introduction_2018}. This approach enables spatial regularisation through radial and repulsion energies, while maintaining no geometric assumptions as to the true shape of the underlying structure.
Let $X=\{x_n\}_{n=1}^N \subset \mathbb{R}^d$ denote estimated emitters. Let \(C=\{c_i\}_{i=1}^{k}\) be the latent structural centres of these emitter configurations.

The probability density of a general point configuration \(\mathbf{x}\) = \(\{ x_{1}, \dots, x_{n}\}\) in a domain \(W\subset\mathbb{R}^{d}\), under a Gibbs point process, is given by 

\[
p(\mathbf{x})=\frac{1}{Z}\text{exp}\left( -U(\mathbf{x}) \right),
\]

where \(U(\mathbf{x})\) is the energy function (the form of which encodes the interaction between points) and \(Z\) is the normalising constant, or partition function:

\[
Z = \int\text{exp}(-U(\mathbf{x}))d\mathbf{x},
\]

where \(Z\) is typically intractable \cite{Illian2008}.

Biological structures imaged via SMLM, such as protein oligomers, have physical size and are therefore sterically inhibitive. This serves to enforce an overlap constraint: two distinct structures cannot occupy the same physical space. Thus, protein oligomeric arrangements can be represented as a series of voids (occupied by unlabelled protein mass) surrounded by a `ring' of fluorescence labels, within which there cannot be any emitter in the interior of the structure. While the strictness of this paradigm depends on the presence of spurious localisations within void interiors, such a construction applies to the majority of protein oligomer structures, including dimers. These key facts permit the formation of a biologically grounded, Gibbs point process likelihood defining voids and their structural centres:
\begin{equation}\label{eq:gibbslikelihood}
    P(X \mid C) = \exp\left( -\sum_{i=1}^{k}\left[ \sum_{x_n:a_n=i} 
    \frac{(\|x_n - c_i\| - r)^2}{2\sigma_r^2} \right] 
    + \sum_{i<j}\bigl(\max(0,\, d_{\min} - \|c_i - c_j\|)\bigr)^2 \right),
\end{equation}
where the first term represents the radial energy penalising emitters that deviate from a radius $r$ around their assigned structural centre (with tolerance $\sigma_r$), and the second term penalises overlap between structural centres.

The parameter $d_{\min}$ defines a soft minimum separation, imposing a quadratic penalty rather than a hard constraint. This relaxation is essential for MCMC mixing: strict non-overlap enforcement would reject nearly all shift proposals for spatially proximate structures, preventing positional refinement even when centres are correctly identified.

The Gibbs process has a tendency to overfit structural centres to explain clutter, and as such a penalised Poisson prior is set on the on the number of structures \(k\) and number of emitters $N$:

\begin{equation}\label{eq:gibbspriork}
    \pi(k)=\frac{\lambda_C^k}{k!}e^{-\lambda_C}\cdot\exp\left(  -k^2\log\left(  \frac{N}{k+1}  \right)  \right),
\end{equation}

where \(\lambda_C\) is the Poisson mean derived from the Voidwalker process discussed in Sec. \ref{subsec:voidmethods}. The penalty term \(-k^2\log\left(  \frac{N}{k+1}  \right)\), derived in Birg\'e \& Massart \cite{birge_gaussian_2001}, is intended to prevent an explosion of structural centres explaining singular points of clutter\footnote{The inclusion of \(k+1\) in the denominator as opposed to \(k\) as in Birg\'e \& Massart is to ensure numerical stability in downstream work in the rare case that \(k=0\).}.

This penalty, originally proposed in the context of model selection for Gaussian sequences, translates naturally to spatial inference in SMLM. In the context of SMLM, each structural centre acts analogously to a changepoint: a latent spatial origin from which nearby emitter configurations arise. Just as overfitting in changepoint detection can model noise instead of signal variation, inferring too many structural centres risks modelling clutter rather than true emitter configurations. The penalty discourages unnecessary proliferation of structural centres and improves identifiability in the trans-dimensional inference setting, providing a spatial analogue to the penalty used in temporal segmentation problems.

We place a hard-core prior on the structural centres $C=\{c_i\}_{i=1}^{k}$ with intensity $\lambda_C>0$ and hard-core distance $d_{\min}>0$. Its density with respect to a unit-rate Poisson process on $W$ is

\begin{equation}\label{eq:gibbspriorc}
    \pi(C \mid \lambda_C,d_{\min})
    \;=\;
    \frac{1}{Z_{\mathrm{hc}}(\lambda_C,d_{\min};W)}\;
    \lambda_C^{\,k}\;
    \prod_{i<j}\mathbf{1}\!\left\{\|c_i-c_j\|\ge d_{\min}\right\},
\end{equation}
where $k=|C|$ and $Z_{\mathrm{hc}}(\lambda_C,d_{\min};W)$ is the normalising constant.

Following from Eq.'s \ref{eq:gibbslikelihood}, \ref{eq:gibbspriork}, and \ref{eq:gibbspriorc}, the model posterior for this Gibbs point process is 

\begin{align}
    \notag P(C,k\mid X) \propto &\exp\left(  -\sum_{i=1}^{k}\left[  \sum_{x_{n}\in c_{i}}\frac{(\|x_{n}-c_{i}\|-r)^{2}}{2\sigma_{r}^2}  \right] +\sum_{i<j}(\max(0, d_{\min}-\|c_{i}-c_{j}\|))^2 \right) \\
    \notag & \cdot \frac{\lambda_C^k}{k!}e^{-\lambda_C}\cdot\exp\left(  -k^2\log\left(  \frac{N}{k+1}  \right)  \right) \\
    \label{eq:gibbsposterior}& \cdot \frac{1}{Z_{\mathrm{hc}}(\lambda_C,d_{\min};W)}\;
    \lambda_C^{\,k}\;
    \prod_{i<j}\mathbf{1}\!\left\{\|c_i-c_j\|\ge d_{\min}\right\}
\end{align}

The Voidwalker–Gibbs method estimates both the number of structural centres, $k$, their radius \(r\), and their locations $C=\{c_i\}_{i=1}^k\subset W\subset\mathbb{R}^d$ The target posterior is
\begin{equation}
\pi(k,C\mid X)\;\propto\; L(X\mid C)\;\pi(C\mid \lambda_C,d_{\min})\;\pi(k),
\end{equation}
where $L(X\mid C)$ is the Gibbs likelihood in Eq. \ref{eq:gibbslikelihood}, $\pi(C\mid \lambda_C,d_{\min})$ is the hard-core prior in Eq. \ref{eq:gibbspriorc} , and $\pi(k)$ is the penalised Poisson prior on the number of structures, Eq. \ref{eq:gibbspriork}. As the dimensionality of the model is now a parameter to be fit, standard MCMC methods are insufficient. Thus, a Reversible Jump MCMC (RJMCMC)\cite{green_reversible_1995} approach is taken.

An RJMCMC transition proposes to move from $(k,C)$ to $(k',C')$ by first choosing a move type $m\in\mathcal{M}$ with probability $j_m(k,C)$, then drawing auxiliary variables $u\sim q_m(u\mid k,C)$ and applying a dimension-matching diffeomorphism
\begin{equation}
(k',C',u') \;=\; g_m\bigl(k,C,u\bigr),\qquad 
\dim(C)+\dim(u)\;=\;\dim(C')+\dim(u').
\end{equation}
Let $m'$ denote the reverse move and $J_m$ the Jacobian determinant of the mapping $(C,u)\mapsto (C',u')$. The Metropolis–Hastings acceptance probability in the notation of Green~\cite{green_reversible_1995} is
\begin{equation}
\alpha_m\bigl((k,C)\to(k',C')\bigr)\;=\;\min\!\left\{1,\;
\frac{\pi(k',C'\mid X)\, j_{m'}(k',C')\, q_{m'}(u'\mid k',C')}{\pi(k,C\mid X)\, j_m(k,C)\, q_m(u\mid k,C)}\;
\bigl|J_m\bigr| \right\}.
\label{eq:rjmcmc-accept}
\end{equation}
Moves that violate the hard-core constraint ($\min_{i\neq j}\|c_i-c_j\|<d_{\min}$) are rejected. Available moves for this model are similar to those of Fazel \textit{et al}\cite{fazel_bayesian_2019, fazel_high-precision_2022}:
\begin{itemize}
    \item \textbf{Shift:} Uniformly select and translate an existing centre.
    \item \textbf{Birth:} Add a new centre to the model, drawing from the Voidwalker-informed proposal space.
    \item \textbf{Death:} Remove an existing centre from the model, selected uniformly.
    \item \textbf{Split:} Uniformly select a centre and split it along the axis of its emitters.
    \item \textbf{Merge:} Uniformly select a centre and a sister centre within a neighbourhood around it, merging these at their arithmetic mean.
\end{itemize}

The probabilities of these moves are selected such that the probabilities of birth and death \(j_b\) and \(j_d\) are equal, and the probabilities of split and merge \(j_s\) and \(j_m\) are equal. The probability of a shift move is then \(j_{\text{shift}}=1-(j_b+j_d+j_s+j_m)\). Additionally, at every iteration, there is a chance of a radius move \(j_R\), in which the radius is adjusted according to a randomly drawn value from a Gaussian distribution, prior to the selection and undertaking of the typical model moves.

\begin{figure}[ht]
    \centering
    \includegraphics[width=0.5\linewidth]{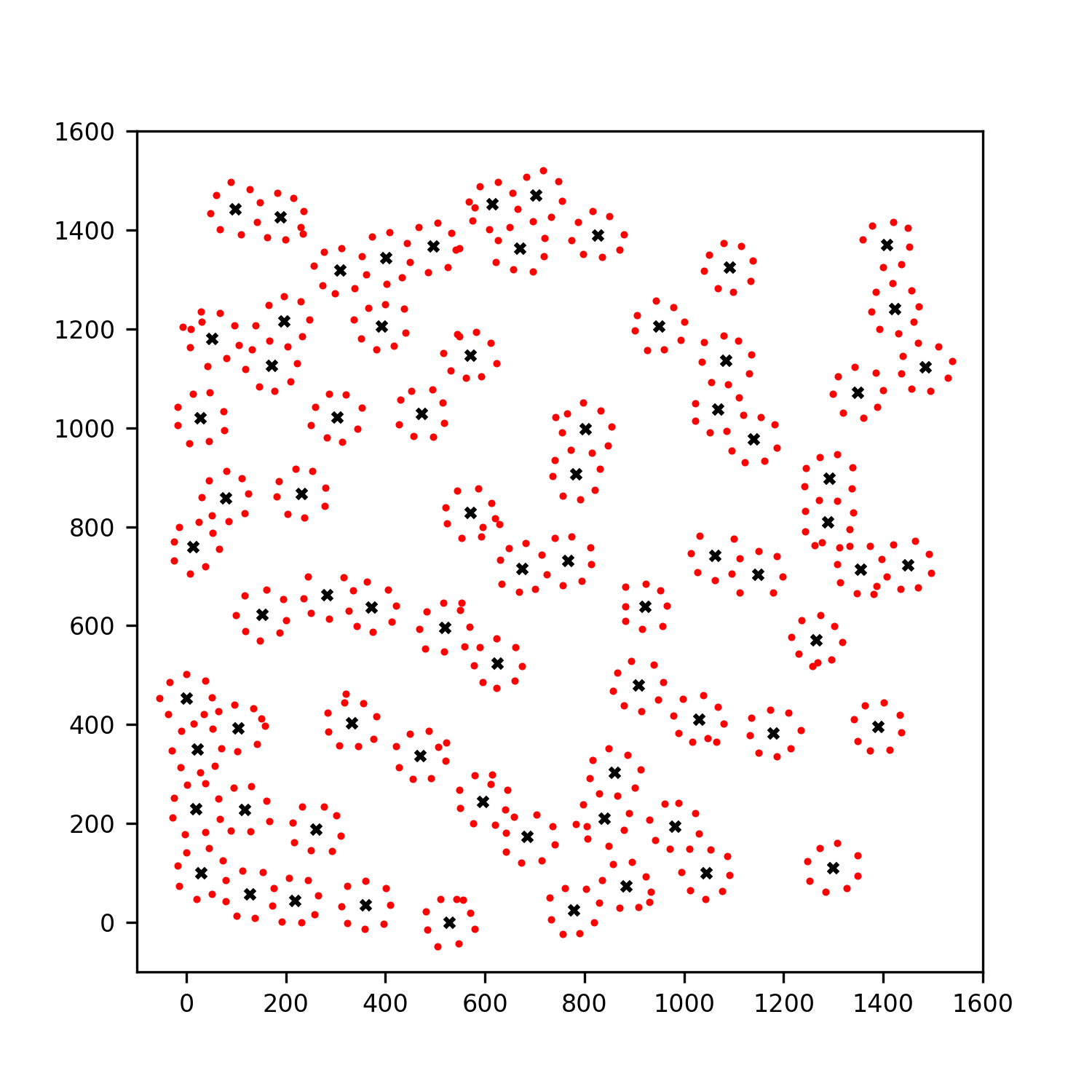}
    \caption{RJMCMC inferred, latent structural centres \(C\), of the emitter set \(X\).}
    \label{fig:methods-centres}
\end{figure}

\paragraph{Move probabilities and proposal families.}
Let $j_{\mathrm{b}}=j_{\mathrm{d}}$ and $j_{\mathrm{s}}=j_{\mathrm{m}}$ be the birth/death and split/merge move probabilities, respectively, and set $j_{\text{shift}}=1-(j_{\mathrm{b}}+j_{\mathrm{d}}+j_{\mathrm{s}}+j_{\mathrm{m}})-j_r$ where $j_r$ is the probability of attempting a (within-model) radius update; cf.\ the specific acceptance forms in the subsequent paragraphs. 
For the shift move we use a \emph{mixture} proposal: (i) a \emph{small} Gaussian translation with scale $\sigma_{\text{sml}}$, (ii) a \emph{large} Gaussian translation with scale $\sigma_{\text{lrg}}$, and (iii) with a small probability $p_{\text{unif}}\ll 1$, a uniform draw in the window $W$ with reflection at the boundary. 
To keep the scales commensurate with the current structure size, we parameterise
\[
\sigma_{\text{sml}} = f_{\text{sml}}\,r, 
\qquad
\sigma_{\text{lrg}} = f_{\text{lrg}}\,r,
\]
with fractions $f_{\text{sml}},f_{\text{lrg}}>0$ clamped to user-chosen bounds (e.g.\ $f_{\text{sml}}\in[f_{\min}^{\text{sml}},f_{\max}^{\text{sml}}]$, $f_{\text{lrg}}\in[f_{\min}^{\text{lrg}},f_{\max}^{\text{lrg}}]$). 
The radius update uses a Gaussian random walk with reflection and hard clamping to weakly informative bounds $[r_{\min},r_{\max}]$:
\[
r' \;=\; \mathrm{clip}\!\left(r+\epsilon,\;r_{\min},r_{\max}\right),\qquad \epsilon\sim\mathcal{N}(0,\sigma_r^2).
\]

\paragraph{Adaptive Robbins-Monro.}
Proposal scales for radius and shift moves are adapted by a Robbins-Monro \cite{haario_adaptive_2001, roberts2001} scheme towards target acceptance rates $a^\star$ (we use $a_r^\star\approx 0.234$ for scalar $r$ and $a_{\text{shift}}^\star\approx 0.468$ for random-walk shifts in $\mathbb{R}^2$). 
Let $a_t\in\{0,1\}$ be the accept indicator for the most recent proposal of a given type (radius or a particular shift component). 
Maintain an exponentially weighted moving average (EWMA)
\[
\widehat a_t \;=\; (1-\rho_t)\,\widehat a_{t-1} \;+\; \rho_t\,a_t,\qquad 
\rho_t \;=\; t^{-\beta},\ \ \beta\in(0.5,1],
\]
and update the log-scale of the proposal by
\[
\log s_{t+1} \;=\; \log s_t \;+\; \gamma_t\,(\widehat a_t - a^\star),
\qquad 
\gamma_t \;=\; t^{-\alpha},\ \ \alpha\in(0.5,1].
\]
Instantiating $s_t$ by the radius standard deviation $\sigma_r$ yields the radius adaptation; instantiating $s_t$ by the step fractions $f_{\text{sml}}$ and $f_{\text{lrg}}$ yields
\[
f_{\bullet,t+1} \;=\; f_{\bullet,t}\,\exp\left(\gamma_t(\widehat a_t-a^\star)\right),
\qquad \bullet\in\{\text{sml},\text{lrg}\},
\]
after which $f_{\bullet,t+1}$ is clamped to its admissible interval. 
For numerical stability and to prevent runaway adaptation after trans-dimensional moves, we (i) reflect at the window boundary in all random-walk proposals, (ii) hard clamp $r$ to $[r_{\min},r_{\max}]$, and (iii) after birth/death/split/merge we dampen the adaptation by halving the effective counts $t$ in $\rho_t,\gamma_t$ and recentering $\widehat a_t$ partway to $a^\star$. 

Acceptance probabilities remain those in the fixed-$k$ ``Radius update'' and ``Within-model Shift'' paragraphs and in the trans-dimensional formulas~\eqref{eq:birth-accept}-\eqref{eq:death-accept} and~\eqref{eq:split-accept}. As proposal rates remain unchanged by the Robbins-Monro scheme, detailed balance remains intact.

\paragraph{Radius update ($k$ fixed).}
Augment the state with the radius $r>0$ so the target becomes
\[
\pi(k,C,r\mid X)\;\propto\;L(X\mid C,r)\;\pi(C\mid \lambda_C,d_{\min})\;\pi(k)\;\pi(r),
\]
where the radial term of the Gibbs likelihood contributes
\[
L(X\mid C,r)\;\propto\;\exp\!\left(-\frac{1}{2\sigma_r^2}\sum_{i=1}^{k}\ \sum_{x_n:a_n=i}\bigl(\|x_n-c_i\|-r\bigr)^{2}\right).
\]
With probability $j_r$ at each iteration, propose a radius move 

\[
r' \;=\; r + \epsilon,\qquad \epsilon \sim \mathcal{N}(0,\tau_r^2),
\]
with reflection at the boundary $r'>0$ to maintain proposal symmetry. The scale of the radius shift is chosen from a small or large move  with user-specified standard deviations, or a uniform move clamped to a weakly informative \([r_{\min}, r_{\max}]\).
The acceptance probability is
\[
\alpha_r \;=\; \min\!\left\{1,\;
\frac{L(X\mid C,r')}{L(X\mid C,r)}\;
\frac{\pi(r')}{\pi(r)}\right\}
\;=\;
\min\!\left\{1,\;
\exp\!\Bigl(-\frac{\Delta U_r}{2\sigma_r^2}\Bigr)\,
\frac{\pi(r')}{\pi(r)}\right\},
\]
where the radial energy difference is
\[
\Delta U_r \;=\; \sum_{i=1}^{k}\ \sum_{x_n\in c_i}\Bigl[\bigl(\|x_n-c_i\|-r'\bigr)^2 - \bigl(\|x_n-c_i\|-r\bigr)^2\Bigr].
\]

\paragraph{Within-model Shift ($k$ fixed).}
When a shift move is selected, uniformly select an index \(J\in \{1,\dots,k\}\) and propose a local translation of the corresponding centre. Let $\eta\sim q_{\text{shift}}(\eta)$. To encourage exploration of the space, \(q_{\text{shift}}\) is chosen from a small or large Gaussian draw with standard deviations tied to the current R - such that a small move aims to refine a position and a large moves aims to move from a structures boundary to its interior - or a uniform move clamped to the Lebesgue measure of the point set \(|B|\). Set
\[
c_J' = \mathrm{Reflect}_W\!\bigl(c_J+\eta\bigr),\qquad
C'=\bigl\{c_1,\dots,c_{J-1},c_J',c_{J+1},\dots,c_k\bigr\}.
\]
If the hard-core constraint is violated, $\min_{i\neq J}\|c_J'-c_i\|<d_{\min}$, reject immediately; otherwise accept with probability
\[
\alpha_{\text{shift}}
=\min\!\left\{1,\;
\frac{L(X\mid C')\,\pi(C'\mid \lambda_C,d_{\min})}{L(X\mid C)\,\pi(C\mid \lambda_C,d_{\min})}
\;\times\;
\frac{\tfrac{1}{k}\,q_{\text{shift}}(-\eta)}{\tfrac{1}{k}\,q_{\text{shift}}(\eta)}
\right\}.
\]
With a symmetric proposal and reflecting boundary (so $q_{\text{shift}}(-\eta)=q_{\text{shift}}(\eta)$), the proposal factors cancel and the ratio reduces to the target-density ratio.

\paragraph{Birth/Death ($k\leftrightarrow k\pm 1$).}
For a birth move, choose the move with probability $j_{\mathrm{b}}(k,C)$ and draw a proposed new centre $c_\star\sim q_{\mathrm{b}}(\cdot\mid k,C)$, from the proposal space discovered via Voidwalker in Sec. \ref{subsec:voidmethods}. Set
\[
k'=k+1,\qquad C'=C\cup\{c_\star\},\qquad u'=\varnothing,\qquad J_{\mathrm{b}}=1.
\]

Birth proposals violating the hard-core constraint are immediately rejected.

The reverse move, death, chooses one of the $k'$ centres uniformly to delete, such that $q_{\mathrm{d}}(u'\mid k',C')=1/(k')$ and $j_{\mathrm{d}}(k',C')$ is the probability of proposing a death at $(k',C')$. The birth acceptance probability from \eqref{eq:rjmcmc-accept} is
\begin{equation}
\alpha_{\mathrm{b}}=\min\!\left\{1,\;
\frac{L(X\mid C\cup\{c_\star\})\,\pi(C\cup\{c_\star\}\mid \lambda_C,d_{\min})\,\pi(k+1)}{L(X\mid C)\,\pi(C\mid \lambda_C,d_{\min})\,\pi(k)}\;
\frac{j_{\mathrm{d}}(k+1,C\cup\{c_\star\})}{j_{\mathrm{b}}(k,C)}\;
\frac{1/(k+1)}{q_{\mathrm{b}}(c_\star\mid k,C)} \right\}.
\label{eq:birth-accept}
\end{equation}

The death move $(k,C)\to(k-1,C\setminus\{c_j\})$ selects an index $J$ uniformly from $\{1,\dots,k\}$, so $q_{\mathrm{d}}(J\mid k,C)=1/k$, and has acceptance probability

\begin{equation}
\begin{split}
\alpha_{\mathrm{d}}
= \min\left\{1,\,
\frac{L\!\bigl(X\mid C\setminus\{c_J\}\bigr)\,
       \pi\!\bigl(C\setminus\{c_J\}\mid \lambda_C,d_{\min}\bigr)\,
       \pi(k-1)}
      {L(X\mid C)\,\pi(C\mid \lambda_C,d_{\min})\,\pi(k)}
\right.\\[2pt]
\left.\times\hspace{0.1cm}\frac{j_{\mathrm{b}}\!\bigl(k-1,\,C\setminus\{c_J\}\bigr)}
                 {j_{\mathrm{d}}(k,C)}
      \times\frac{q_{\mathrm{b}}\!\bigl(c_J\mid k-1,\,C\setminus\{c_J\}\bigr)}
                 {1/k}
\right\}.
\end{split}
\label{eq:death-accept}
\end{equation}

\paragraph{Split/Merge ($k\leftrightarrow k\pm 1$).}
To encourage model exploration in cases where birth/death may be insufficient, such as a pair of centres attempting to explain the same emitter configuration, or a single centre explaining two configurations, split and merge moves are implemented. For a split move with probability \(j_s\), uniformly select an index $J\in\{1,\dots,k\}$, draw auxiliary $u=(r,\omega)$ with density $q_{\mathrm{s}}(r,\omega)$ (radius and angle), and set in $d=2$
\[
\begin{split}
c^{(1)} = &c_J + r(\cos\omega,\sin\omega),\qquad
c^{(2)} = c_J - r(\cos\omega,\sin\omega),\qquad \\
&C'=\bigl(C\setminus\{c_J\}\bigr)\cup\{c^{(1)},c^{(2)}\}, \quad k'=k+1.
\end{split}
\]
The reverse move with probability \(j_m\) merges a selected pair $\{i,j\}$ - where the members of the pair are within \(3R\) of one another to encourage contextually sensible merging - to their midpoint $c=(c_i+c_j)/2$ and recovers $(r,\omega)$. The dimension-matching map $(c_J,r,\omega)\mapsto(c^{(1)},c^{(2)})$ has Jacobian determinant $\lvert J_{\mathrm{s}}\rvert = 4r$ in $d=2$. The split acceptance probability is
\begin{equation}
\alpha_{\mathrm{s}}=\min\!\left\{1,\;
\frac{L(X\mid C')\,\pi(C'\mid \lambda_C,d_{\min})\,\pi(k+1)}{L(X\mid C)\,\pi(C\mid \lambda_C,d_{\min})\,\pi(k)}\;
\frac{j_{\mathrm{m}}(k+1,C')\, q_{\mathrm{m}}(i^\star,j^\star\mid C')}{j_{\mathrm{s}}(k,C)\, p_{\mathrm{sel}}(J\mid C)\, q_{\mathrm{s}}(r,\omega)}\;
\lvert J_{\mathrm{s}}\rvert
\right\},
\label{eq:split-accept}
\end{equation}
where $q_{\mathrm{m}}(i^\star,j^\star\mid C')$ is the probability of choosing the reverse merge pair (typically the two newly created centres). The merge acceptance is the reciprocal form with $\lvert J_{\mathrm{m}}\rvert=\lvert J_{\mathrm{s}}\rvert^{-1}$ and the corresponding selection and proposal factors.

\subsection{Emitter Assignments and Super-Structure Discovery}\label{subsec:assignments}
A persistent identifier for each centre is maintained during RJMCMC, and updated directly as moves are accepted. At the beginning of sampling, the existing centres are assigned consecutive IDs \(\{1,\dots,k\}\). A shift move translates an existing centre but leaves its ID unchanged. A birth move creates a new centre and assigns it a fresh ID, a death move retires the chosen centre’s ID. A split replaces one ID by two new IDs; a merge replaces two IDs by a single new ID. This deterministic index bookkeeping removes the need for per-iteration assignments, saving on runtime and avoiding potential labelling errors that accompany spatial implementations of the Hungarian algorithm.

Let \(t\leq T\) index accepted states and let \(b\) denote the burn-in length. For each persistent ID \(j\) that exists at iteration \(t>b\), the current centre position \(c_j^{(t)}\) and (if sampled) the current radius \(r^{(t)}\) are appended to that ID’s history. From the retained draws of ID \(j\), \(\{c_j^{(s)}\}_{s\in\mathcal{S}_j}\) where \(\mathcal{S}_j\subset\{b{+}1,\dots,T\}\), posterior uncertainty is summarised via empirical moments:
\begin{equation}\label{eq:track-moments}
    \mu_{j}=\frac{1}{S_{j}}\sum_{s\in\mathcal{S}_j} c_j^{(s)}, 
    \qquad 
    \Sigma_{j}=\frac{1}{S_{j}-1}\sum_{s\in\mathcal{S}_j}\bigl(c_j^{(s)}-\mu_j\bigr)\bigl(c_j^{(s)}-\mu_j\bigr)^{\!\top},
\end{equation}
with \(S_j=\lvert\mathcal{S}_j\rvert\). A running estimate of the squared radius is maintained to capture radial spread in the predictive model:
\begin{equation}\label{eq:r2-mean}
    \widehat{\mathbb{E}[r^2]}\;=\;\frac{1}{S_r}\sum_{s=b+1}^{T} \bigl(r^{(s)}\bigr)^2.
\end{equation}

Assignments are computed both online (periodically during sampling for diagnostics) and offline (once, post burn-in, for final labelling). In both cases, the same Gaussian predictive form is used. Let \(\sigma_{\text{loc}}^2\) denote the global localisation variance (assumed isotropic) and \(\epsilon>0\) a small numerical jitter. Assuming random radial orientation, the radial contribution adds \(\tfrac{1}{2}\widehat{\mathbb{E}[r^2]}\,I_2\) to the covariance, yielding
\begin{equation}\label{eq:predictive-mvn}
    x\mid j \;\sim\; \mathcal{N}\!\left(\mu_j,\;
    \Sigma_j \;+\; \Bigl(\sigma^{2}_{\mathrm{loc}}+\tfrac{1}{2}\widehat{\mathbb{E}[R^2]}\Bigr) I_2 \;+\; \epsilon I_2\right).
\end{equation}

Given the set of currently active IDs \(\mathcal{J}\), define the unnormalised likelihoods
\begin{equation}\label{eq:assignment-like}
    \ell_{nj}\;=\;\phi\!\left(x_n;\,\mu_j,\ \Sigma_j+\Bigl(\sigma^{2}_{\mathrm{loc}}+\tfrac{1}{2}\widehat{\mathbb{E}[r^2]}\Bigr) I_2+\epsilon I_2\right),
    \qquad n=1,\dots,N,\ \ j\in\mathcal{J},
\end{equation}
where \(\phi(\cdot;\mu,\Sigma)\) is the bivariate Gaussian density. Row-normalising gives per-emitter assignment probabilities
\begin{equation}\label{eq:assignment-probs}
    p_{nj}=\frac{\ell_{nj}}{\sum_{j'\in\mathcal{J}}\ell_{nj'}}, 
    \qquad \sum_{j\in\mathcal{J}}p_{nj}=1\ \ \text{for all }n.
\end{equation}
Emitters are assigned stochastically, via drawing \(j\sim\mathrm{Categorical}(p_{n\cdot})\).

For online diagnostics the same formulas are reused with \((\mu_j,\Sigma_j)\) replaced by running moments of each active ID up to the current iteration \(t\), and with \(\widehat{\mathbb{E}[r^2]}\) replaced by its running average to date. This yields stable, low-latency estimates without any cross-iteration matching. The offline pass simply fixes \((\mu_j,\Sigma_j)\) and \(\widehat{\mathbb{E}[r^2]}\) to their post burn-in values in \eqref{eq:track-moments}–\eqref{eq:r2-mean} and evaluates \eqref{eq:assignment-like}–\eqref{eq:assignment-probs} once.

IDs created near the end of sampling may have small \(S_j\); these are not eliminated, but instead their \(\Sigma_j\) are treated with extra jitter \(\epsilon I_2\). As IDs are persistent, no gating parameter or assignment solver is required; births and splits create new IDs and deaths and merges retire or replace IDs deterministically.

With the only assumption that emitter micro-structure forms ring-like arrangements, the process thus far is geared toward finding single structures. In the case of super-structure chains, or a DNA-Origami grid, these single structures are reflective of the individual components of the overall macro-structure. Thus, a method of distinguishing between single structures as isolated or paired is developed.

In a purely spatial meaning, two close single structures are indistinguishable from a super-structure formed by two sub-structures. Thus, the assignment information is utilised. For each emitter \(n\), up to \(K_{\max}\) candidate centres are stored with weights, determined in Sec. \ref{subsec:assignments}. Assemble the sparse matrix 

\[
P\in\mathbb{R}_{\geq 0}^{N \times |C|}, \quad P_{n,i}\geq 0,
\]

by placing each stored weight for emitter \(n\) and centre \(i\) at \((n, i)\), and zero elsewhere. Let \(m_n^{\text{row}}\) denote the stored probability mass for emitter \(n\). Define the row normalised matrix

\[
\Tilde{P}_{n,i}=\frac{P_{n,i}}{m_n^{\text{row}}}\quad \text{s.t.} \quad \forall n, \hspace{0.2cm}\sum_{i=1}^{|C|}\Tilde{P}_{n,i}=1.
\]

Define column sums \(\sum_{n=1}^N\Tilde{P}_{n,i}\), and the mark matrix 

\[
M\in \mathbb{R}^{|C|\times N}_{\geq 0}, \quad M_{i,n}=m_i(n)=\frac{\Tilde{P}_{n,i}}{\sum_{n=1}^N\Tilde{P}_{n,i}},
\]

where \(m_i( \cdot )\equiv0\) if \(\sum_{n=1}^N\Tilde{P}_{n,i}=0\). Each row \(m_i=\left( m_i(1),\dots ,m_i(N) \right)\) lies in the \((N-1)-\)simplex

\[
\Delta^{N-1}=\left\{  u\in \mathbb{R}^{N}_{\geq 0}:\sum_{n=1}^{N}u_n=1  \right\}.
\]

Define a bounded similarity \(S_{ij}\in[0,1]\) to assess the shared support of emitters \(i,j\). Let

\[
k_B(i,j)= \sum_{n=1}^N\sqrt{m_i(n)m_j(n)},
\]

where \(k_B\in [0,1]\) is the Bhattacharyya distance. Let

\[
T(i,j)=\sum_{n=1}^{N}\frac{m_i(n)m_j(n)}{m_i(n)+m_j(n)},
\]

where \(T(i,j)=0\) if \(m_i(n)+m_j(n)=0\). Generally, \(T(i,j)\leq\frac{1}{2}\). As such, it is counted twice in the final \(S_{ij}\) to avoid silently weighting toward \(k_B\). Let

\[
S_{ij}=\frac{1}{2}\left(  k_B(i,j)+2T(i,j)  \right), \quad S_{ii}=0.
\]

High \(S_{ij}\) suggests centres with similar emitter distributions, where \(S_{ij}=1\) implies an identical distribution of emitters.

Each \(S_{ij}\) is tested against a randomly labelled (RL) null to determine the significance of the shared support, under the hypothesis

\begin{align*}
    H_0: i, j &\text{  are dissimilar / monomeric structures},\\
    H_1: i,j &\text{ are similar / a dimer pair}.
\end{align*}

Let \(\hat{\lambda}:\mathbb{R}^2\rightarrow\mathbb{R}_{\geq0}\) be the LGCP-fit intensity field of the centres, estimated by the same SPDE process as in Sec. \ref{subsec:voidmethods}. Form \(Q\) quantile bins of \(\{\hat{\lambda}_i\}\). In each bin, permute labels independently; preserving inhomogeneity. Repeat permutation for \(b=1,\dots, B\), collecting upper-triangular entries \(\left\{  S_{ij}^{(b)}: i<j \right\}\). Calculate pairwise RL p-values 

\[
p_{ij}=\frac{1+\#\left\{ b:S_{ij}^{(b)}\geq S_{ij}\right\}}{B+1}.
\]

To stabilise pairs across heterogeneous regions, pool all permuted scores 

\[
S_{\text{pool}}=\bigcup_{b=1}^B\left\{ S_{ij}^{(b)}:i<j\right\},
\]

setting a single global barrier at a high quantile, \(q\). If both \(p_{ij}\leq\alpha\) and \(S_{ij}\geq S_{\text{pool}[0.99]}\) hold, centres \(i,j,\) are considered a pair, and emitter assignments are relabelled such that any emitters belonging to a grouped centre share the same label as that group.

\begin{figure}[ht]
    \centering
    \subfloat[]{\includegraphics[width=0.457\linewidth]{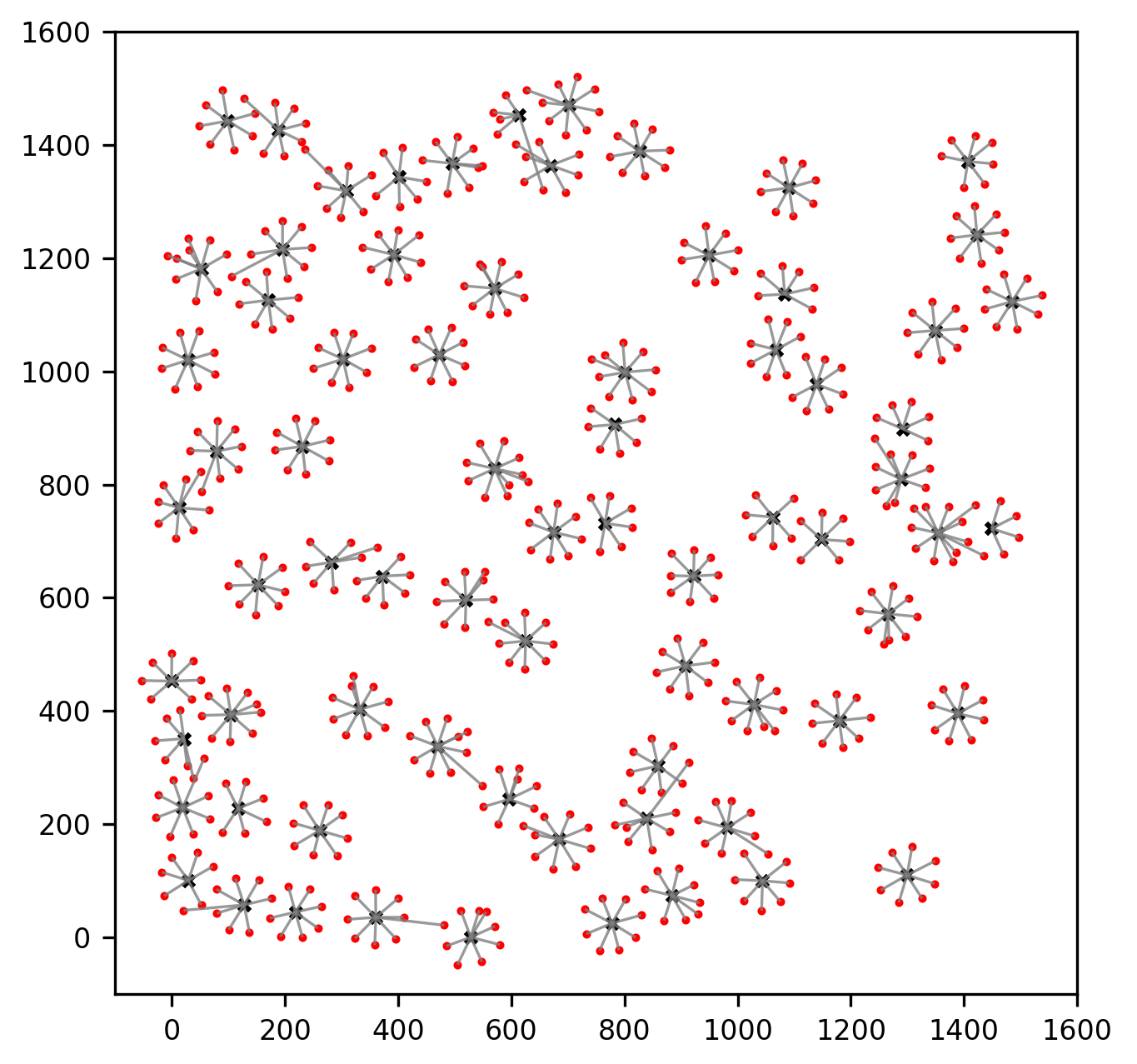}}
    \subfloat[]{\includegraphics[width=0.45\linewidth]{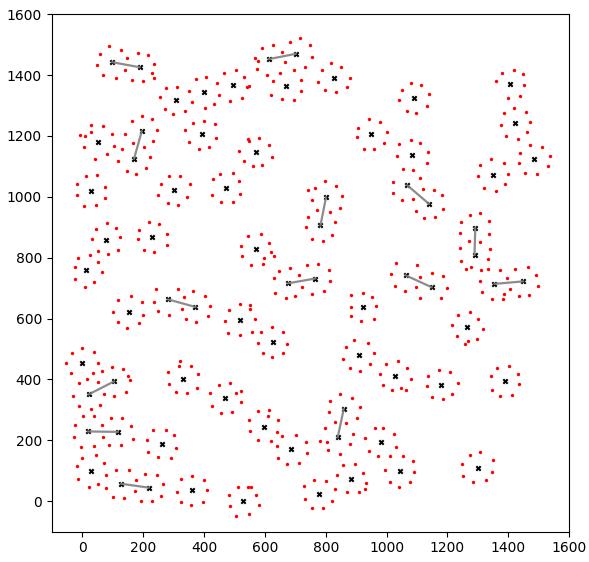}}
    \caption{(a) Sampled emitter assignments. (b) Mark-relabelling-based super-structuring of inferred centres.}
    \label{fig:methods-assign-super}
\end{figure}

Following super-structuring, the data are separated based on the number of connected structures. The data in Fig. \ref{fig:methods-assign-super}, for example, is separated into two datasets; single and paired.

\subsection{ASMBLR: Assembling Structured Molecular Building blocks from Localisation Reconstructions}\label{subsec:asmblr}
Given the sampled and finalised emitter-to-centre assignments per dataset, discussed in Sec. \ref{subsec:assignments}, let
\[
a_n \in \{-1, 0, 1,\dots ,J-1\}, \qquad n=1, \dots ,N,
\]

denote the per-emitter centre labels, where \(-1\) represents emitters that were unassigned. Define the index of assigned emitters \(I=\{n:\, a_n\ge 0\}\) and, for each centre \(j\), the group

\[
G_j\;=\;\{\,n\in I:\ a_n=j\,\}, \qquad j=0, \dots ,J-1.
\]

Within each \(G_j\) consider the induced complete graph on its vertices, so every \(k\)-subset of \(G_J\) is a \(k\)-clique. Let \(k\ge 2\) be the desired clique size and 

\[
    \Omega\;=\;\bigl\{\, j:\ |G_j|\ge k \,\bigr\},\qquad
    \mathcal{K}\;=\;\bigcup_{j\in\Omega}\binom{G_j}{k}
\]

be the centres with enough members and the universe of admissible \(k\)-cliques, respectively.

Draw \(n_{\text{samples}}\) distinct cliques via repeated two-stage sampling with de-duplication:

\begin{enumerate}
    \item \textbf{Filter:} Discard all unassigned emitters \(a_n=-1\). Build \(\Omega\), and a map $j\mapsto G_j$.
    \item \textbf{Propose:} Choose \(J^\star\sim\mathrm{Unif}(\Omega)\), and choose $S^\star\sim\mathrm{Unif}\!\left(\binom{G_{J^\star}}{k}\right)$, without replacement, within $G_{J^\star}$.
    \item \textbf{De-duplicate:} If \(S^\star\) has not been sampled before, accept it; otherwise reject and resample.
\end{enumerate}

The procedure ends when \(n_\text{samples}\) have been accepted\footnote{A necessary condition is thus \(\sum_{j\in\Omega}\binom{|G_j|}{k} \;\ge\; n_{\text{samples}}\),
otherwise fewer than $n_{\text{samples}}$ distinct cliques exist after removing unassigned emitters.}, or a fixed number of attempts have been made to sample distinct cliques. The output is

\[
\{\, X_{S^{(r)}} \,\}_{r=1}^{n_{\text{samples}}}\in\mathbb{R}^{n_{\text{samples}}\times k\times 2},
\]

the set of emitter coordinates of each selected clique.

On a single proposal, the probability of a particular clique \(S\in\binom{G_j}{k}\) being proposed is

\[
\mathbb{P}\{S\ \text{proposed}\} \;=\; \frac{1}{\mid\Omega\mid}\cdot\frac{1}{\binom{\mid G_j\mid}{k}},
\]

so the sampler is uniform within groups, treating groups equally. Consequently, when groups have varying sizes, the overall law on \(\mathcal{K}\) is not uniform across all cliques; cliques from larger groups have smaller per-proposal probability. If exact global uniformity on \(\mathcal{K}\) is required, one should choose the group with weights proportional to \(\binom{\mid G_j\mid}{k}\) and draw \(S\sim\mathrm{Unif}\!\bigl(\binom{G_j}{k}\bigr)\). To quantify the probability that a sampled clique is structurally coherent, define \(P_{\mathrm{within}}(k)\) as the probability that all \(k\) members of a sampled \(k\)-clique originate from the same underlying structure.

\paragraph{Uniform radial sampling.}
Consider a radial neighbourhood (e.g.\ within a fixed cut-off) containing \(s\) emitters
from a single true structure and \(c\) additional emitters arising from clutter or
neighbouring structures, so that the neighbourhood size is \(s+c\).
If a \(k\)-clique is sampled uniformly without replacement from this neighbourhood, then
\begin{equation}
P_{\mathrm{within}}^{\mathrm{rad}}(k; s,c)
\;=\;
\frac{\binom{s}{k}}{\binom{s+c}{k}}.
\label{eq:pwithin-radial}
\end{equation}

\paragraph{Voidwalker/assignment-guided sampling.}
Let \(a\in[0,1]\) denote the per-emitter assignment accuracy, i.e.\ the probability that
an emitter is assigned to its true centre. Under an independence approximation for
assignment correctness across clique members, the probability that all \(k\) members of a
clique drawn within an inferred group are correctly assigned is
\begin{equation}
P_{\mathrm{within}}^{\mathrm{VW}}(k; a)
\;=\;
a^{k}.
\label{eq:pwithin-vw}
\end{equation}
This provides an analytical upper bound on the rate of structurally coherent clique draws
achievable by assignment-guided sampling at a given accuracy \(a\).

Following the identification of candidate molecular features through the sampling of cliques from co-assigned emitters, the model can be reduced to a set of spatial relationships between connected emitters. Let \(\mathbf{X}_{a_n}=\{\mathbf{x}_{a_n1},\dots, \mathbf{x}_{a_nN_a}\}\) be the set of \(N_a\) emitters in \(\mathbb{R}^2\) that share assignment label \(a\), where each \(\mathbf{x}_{a_ni}=(x_{a_ni_{1}},x_{a_ni_{2}})\) is a two dimensional vector representing its location in \(W\). This point set is then limited to the inner-space of \(W\); the \(N_a\times N_a\) Euclidean distance matrix of separations of co-assigned emitters, denoted \(R\). Each entry \(R_{ij}\) represents the Euclidean distance between emitters \(i\) and \(j\), given by:

\[
R_{ij}=\sqrt{\sum_{d=1}^{2}(x_{a_ni{_d}}-x_{a_nj_{d}})^2},
\]

where \(a_n\) represents the identical assignment label of both emitters i and j.

The data for the reconstruction algorithm consists of a series of point sets formed from the sampled cliques, where \(N_{\text{cliques}}\) is the number of cliques. Each clique contains a measurement set of size $M\leq N$ representing a localised set of detections $\mathbf{L}_n = \{\mathbf{l}_1, \dots, \mathbf{l}_M\}$ for each $n=1,\dots,N_\text{cliques}$. Each $\mathbf{l}_i$ is a $d$-dimensional vector representing a point in the domain for $i=1,\dots, M$, analogous to the model locations with associated uncertainty. This ground truth uncertainty is represented by an additional Gaussian distribution centred on the true positions, which is not available from measurements and so is provided \textit{a priori}, based on knowledge of the problem area, i.e. DNA origami reproducibility or protein complex stability. Each measurement set has rotational and translational freedom relative to the model, addressed by considering only the internal spatial relationships within $\mathbf{L}_n$. Each measurement set can therefore also be represented as a Euclidean distance matrix, $S_n$, where the full set of data can therefore be represented as the vector $\mathbf{S}=\left(S_1,\dots,S_{N_\text{cliques}}\right)$.

\begin{figure}[ht!]
    \centering
    \includegraphics[width=0.75\linewidth]{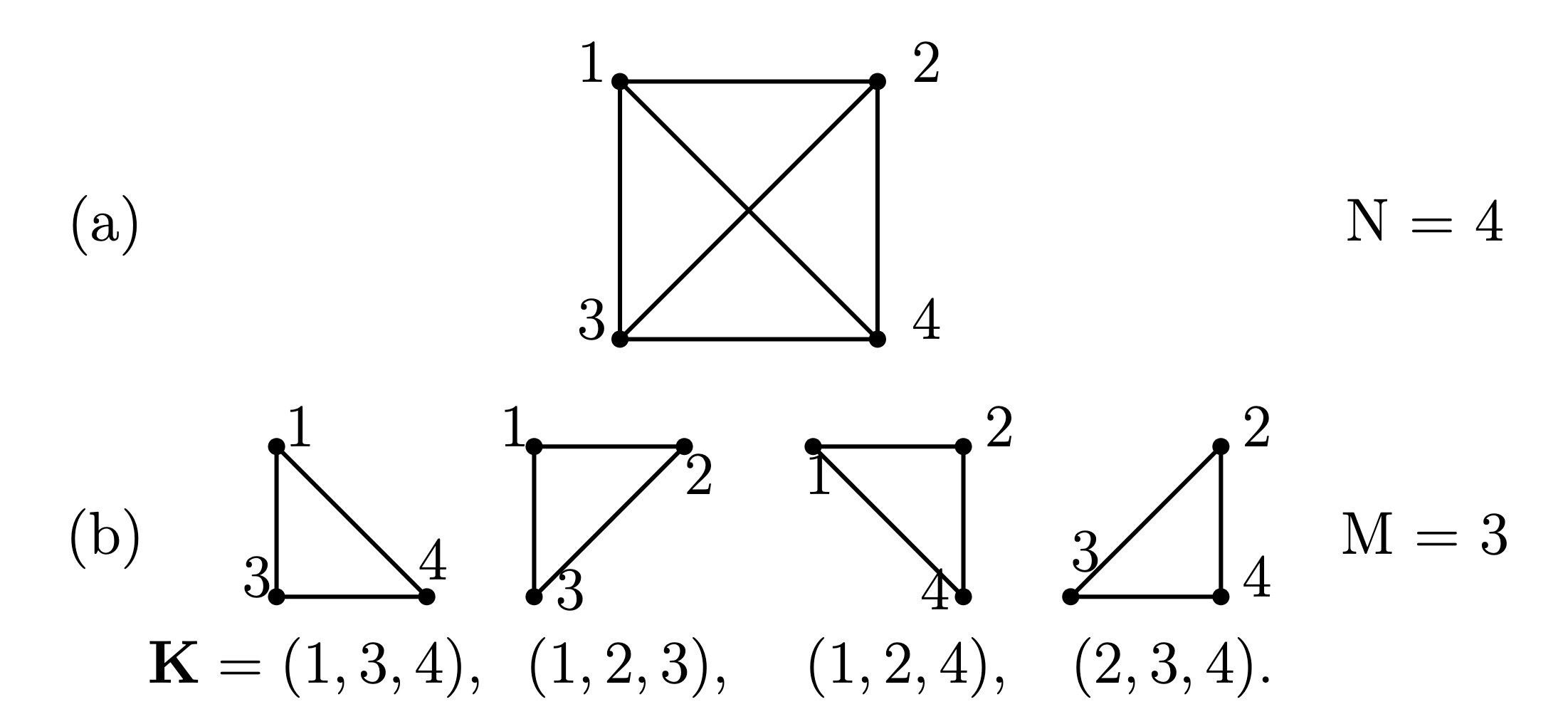}
    \caption{The model structure as a connected square (a) and the cliques as triangular graphlets (b). Index remains consistent between model and cliques. Cliques themselves are order invariant, but are used to construct the order dependent assignments, e.g. the clique with locations \{1, 3, 4\} has assignments K=(1, 3, 4), K=(3, 4, 1) etc.}
    \label{fig:tracks}
\end{figure}

For each clique, model locations are mapped to the measurement set by introducing a set of correspondences, or assignments, between each model location, $\mathbf{x}_n$, and measurement localisation, $\mathbf{l}_j$ in each clique. Note that the model locations, are shared across the cliques. The vector of correspondences between elements in the $\mathbf{L}$ measurement and $\mathbf{X}$ location sets is denoted
\[
\mathbf{K} = (k_1, k_2, \dots, k_M),
\]
where each $k_i \in \{1, 2, \dots, N\}$ for all $i = 1, \dots, M$, and $k_i \neq k_j,$ for all $ i \neq j$. Thus, the index of each assignment corresponds to the localisation and value corresponds to the location. This concept is depicted in Fig. \ref{fig:tracks}

As the order of assignments in $\mathbf{K}$ is informative, the list of possible assignment combinations for each dataset can be expressed as the permutations of size $M$ from the set $\{1,\dots,N\}$. There are therefore $(N)_M = \frac{N!}{(N - M)!}$ permutations of assignment sets.

\begin{figure}[ht!]
    \centering
    \includegraphics[width=0.75\linewidth]{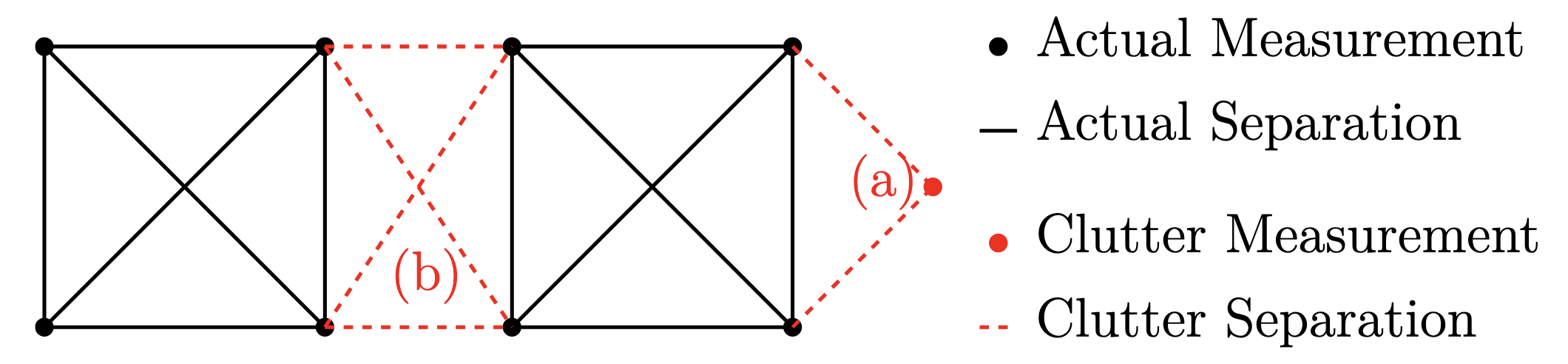}
    \caption{The two types of clutter in the model: (a) spurious detection, and (b) bridging.}
    \label{fig:typesclutter}
\end{figure}

A common problem in super-resolution imaging is the appearance of spurious detections in the data whose spatial arrangement is unknown, resulting in separations that are not otherwise accounted for by the model. The Voidwalker-Gibbs pipeline does substantially reduce this occurrence, but does not nullify it is a possibility. Where one or both of $\mathbf{x}_i$ or $\mathbf{x}_j$ are not incorporated in the model structure $\mathbf{X}$, their associated separation is termed a clutter separation. Additionally, close spatial proximity between distinct, disconnected structures may produce separations that bridge these independent structures. To handle this possibility, these bridge separations are also deemed clutter separations. Hence, there exist two distinct sources of clutter, depicted in Figure \ref{fig:typesclutter}. 

To account for clutter, an additional assignment value, $k_i = 0$ is introduced. Each non-clutter assignment $k_i \in \{1, 2, \dots, N\}$ can feature only once per assignment combination, however the clutter assignment $k_i = 0$ can appear more than once. This new, complete, set of possible assignments, $K_\text{all}$, is constructed \textit{a priori} and has size $|K_{\text{all}}|$\footnote{As the total number of possible assignment sets \(|K_{\mathrm{all}}|\) grows combinatorially in \(N\) and \(M\), explicitly computing the full probability distribution, $P(\mathbf{K}_n^*)$, becomes infeasible.  A sampler is adapted to batch process probabilities and retain the top-K \cite{koren2008factorization} highest probability assignments, binning all else into a non-top-K bin, and forming a probability distribution from the top-K, with this bin as a single entry. Sample then from this distribution, and sample uniformly in the non-top-K bin, if this bin is sampled. A Multiple-Try Metropolis correction step \cite{liu2000multiple} is incorporated to account for truncation bias as a result of top-$K$ rest-bin sampling.}, given by
\begin{equation}
    |K_{\text{all}}| = \sum_{q=0}^{M}\frac{N!}{(N+q-M)!}\frac{M!}{q!(M-q)!}, \text{ for } M\leq N.
    \label{Eq:ClutterPermutations}
\end{equation}

Under the assumption that the structure represented by the model is fully connected and the measurements are randomly drawn and abundant, each possible assignment set, $\mathbf{K}$, in $K_\text{all}$ would be expected to have an equal probability of appearing in the dataset. However, to ensure that the number of clutter assignments reflects prior beliefs regarding the expected proportion of clutter in a sample, the probability distribution over each possible assignment, $\mathbf{K}$, is constructed as follows.

A Beta prior is placed on the proportion of clutter locations, denoted $P_{\text{clutter}}$, so that

\[
P_{\text{clutter}} \sim \text{Beta}(a, b)
\]

where the shape parameters are initially set to $a = 1$ and $b = 9$ so that model will minimise $P_{\text{clutter}}$ and reflect that optimally there is no clutter and on average, 10\% clutter is expected in typical datasets. 

As the distribution of clutter measurements in the dataset is random, let the number of clutter measurements, $Q$, within a single assignment set of size $M$, be binomially distributed so that

\[
Q \sim \text{Binomial}(M, P_{\text{clutter}}).
\]

Therefore, $Q=0$ describes vectors with no clutter, $Q = 1$ describes vectors containing a single clutter assignment, $Q = 2$ describes vectors with two clutter assignments, and so on, for $Q = 0, 1, \dots, M$. Assignment combinations require at least 2 non-clutter assignments to provide valid separation information. Generally, a $d$-dimensional clique with fewer than $(M-(d+1))$ non-clutter assignments is equivalent to an all clutter assignment set as it contains insufficient  informative separations to constrain the structure of interest. To handle this, in the prior distribution, assignment probabilities of vectors with more than $(M-(d+1))$ clutter are assigned to the all clutter assignment set.

The probability distribution across all assignment vectors is defined by allocating the probability of having $Q=q$ clutter assignments evenly across all possible assignments containing $q$ clutter assignments. Therefore, the probability of selecting a specific assignment vector, $\mathbf{K}_i$, denoted $P(\mathbf{K}_i|P_\text{clutter})$, which has $q_i$ clutter assignments is given by
\[P(\mathbf{K}_i|P_\text{clutter}) = \frac{P(Q = q_i)}{\displaystyle\frac{N!}{(N+q_i-M)!}\frac{M!}{q_i!(M-q_i)!}} \text{ for } i = 1,\dots,|K_\text{all}|\text{ and } q_i \in \{1,\dots,M\}.\]
Substituting in the expression for $P(Q=q)$ from the binomial distribution and simplifying:

\[
P(\mathbf{K}_i|P_\text{clutter})  = \frac{\displaystyle{P_{\text{clutter}}}^{q_i}(1-P_{\text{clutter}})^{(M-q_i)}{(N+q_i-M)!}}{N!} \text{ for } i = 1,\dots,|K_{\text{all}}| \text{ and } q_i \in \{1,\dots,M\}.
\]

Consider now the distribution of separations from the Euclidean distance matrix of model locations, $R$. Considering point pairs independently, for two points chosen at random in a disk with radius $\rho =\frac{1}{2}R_{\text{max}}$, the distribution of a distance, $r$, between points is not uniform, but instead has the probability distribution function \cite{santalo2004integral}:

\begin{equation}\label{Eq:SeparationPrior}
    P(r) = \frac{4r}{\pi \rho^2} \cos^{-1}\left(\displaystyle\frac{r}{2\rho}\right) - \displaystyle\frac{2r^2}{\pi \rho^3}\sqrt{1 - \displaystyle\frac{r^2}{4\rho^2}}.
\end{equation}

$P(r)$ serves as the prior on each separation, $R_{ij}$, when locations $\mathbf{x}_i$ and $\mathbf{x}_j$ are considered to be within the model location set $\mathbf{X}$. Clutter separations follow a uniform distribution over $R_{\max}$, reflecting the maximal uncertainty in $r$ when localisation constraints do not apply. Therefore the prior distribution for each separation between locations $\mathbf{x}_i$ and $\mathbf{x}_j$ for all $i,j=1,\dots,N$, given their respective assignments denoted here for simplicity as $k_i$ for $\mathbf{x}_i$ and $k_j$ for $\mathbf{x}_j$, is given by
\[P(R_{ij}|k_i,k_j) = 
\begin{cases}
\displaystyle\frac{1}{R_{\max}} &\quad \text{ if } k_{i}=0 \text{ or } k_j =0  \text{ and } R_{ij}<R_{\max},\\
P(R_{ij}) &\quad \text{ if } k_i,k_j>0 \text{ and } R_{ij}<R_{\max},\\
0 &\quad \text{ if } R_{ij} > R_{\max},
\end{cases}\]
where $P(R_{ij})$ is as shown in Eq. (\ref{Eq:SeparationPrior}). The full set of all parameters to be estimated by the model is therefore

\[
\boldsymbol\theta = (R,\mathbf{K},P_\text{clutter}).
\]

The prior distribution, $P(\boldsymbol{\theta})$, is given by

\[
P(\boldsymbol\theta)=P(R|\mathbf{K})\times P(\mathbf{K}|P_\text{clutter})\times P(P_\text{clutter}),
\]

where pairwise independence is assumed for separations and assignments, so that
\begin{align*}
    P(R|\mathbf{K})&=\prod_{j=2}^{N}\prod_{i=1}^{j-1}P(R_{ij}|k_i,k_j),\\
P(\mathbf{K}|P_\text{clutter}) &= \prod_{i=1}^{|K_{\text{all}}|}P\left(\mathbf{K}|P_\text{clutter}\right).
\end{align*}

Consider the likelihood of the data; the Euclidean distance matrix of localisation separations, $S_n$, for $n=1,\dots,N_\text{cliques}$. Assume that for each matrix, each element $S_{ij}$, for $i,j=1,\dots,M$, follows a folded Gaussian distribution \cite{leone_folded_normal} with mean given by the assigned model separation, denoted here as $R_{ij}$ for simplicity, and variance $\sigma_{i}$ determined by the $r_{\Delta x}$ measure as described in Iyer \textit{et al}\cite{iyer2024DrugresistantEGFRmutations}, so that, for each $i,j\in\{1,\dots,M\}$, $i\neq j$ ,

\begin{equation}
 P(S_{ij}|R_{ij},\sigma_i^2) = \frac{1}{\sqrt{2\pi\sigma_i^2}} \left( \exp\left(-\frac{(S_{ij} - R_{ij})^2}{2\sigma_i^2} \right) + \exp\left(-\frac{(S_{ij} + R_{ij})^2}{2\sigma_i^2} \right) \right).
 \label{Eq:LikelihoodSinglePoint}
\end{equation}
%Our full model posterior can then be expressed as
%\[P(\boldsymbol{\theta}|S)=P(S|\boldsymbol{\theta}) \times P(\boldsymbol{\theta}).\]

Assuming that the localisation separations are pairwise independent, then
\begin{equation}
    P(S_n|\boldsymbol{\theta})=P(S_n|R)=\prod_{j=2}^{M}\prod_{i=1}^{j-1}P(S_{ij}|R_{ij},\sigma_i^2).
    \label{Eq:Likelihood}
\end{equation}

Thus, the full likelihood over all $N_\text{cliques}$ is given by

\[
P(\mathbf{S}|\boldsymbol{\theta})=\prod_{n=1}^{N_\text{cliques}}P(S_n|\boldsymbol{\theta}).
\]

The likelihood of the measurement separations, $S$, depends on the model separations, $R$, and so is then indirectly dependent on the assignments and amount of clutter in the model. This dependency is accounted for in the prior distribution by the $P(P_\text{clutter})$ term.

The full model posterior can then be expressed as
\[P(\boldsymbol{\theta}|\mathbf{S})\propto P(\mathbf{S}|\boldsymbol{\theta}) \times P(\boldsymbol{\theta}).\]

At each iteration, the Euclidean distance matrix of separations, $R$, is updated by proposing new locations for all $\mathbf{x}_i\in\mathbf{X}$. To achieve this, for each coordinate, $x_{i_k} \in \mathbf{x}_i$, a new value is proposed based on one of two Gaussian distributions or a uniform distribution. The Gaussian distributions are centred on the current coordinate value, $x_{i_k}^{(t)}$ at iteration $t$, and one of two possible variance values are considered and fixed \textit{a priori}, allowing for the moves of different magnitudes: a large move or a small move, with variances $\sigma^2_{\text{large}}$ or $\sigma^2_{\text{small}}$ respectively, or a uniform move across the domain.

The small move has variance $\sigma^2_{\text{small}} =  c\left(2.38/\sqrt{d}\right)) \sqrt{\sigma_{GT}^2 + (\sigma_{\text{measurement}}^2 / N_{\text{eff}}))}$, where $\sigma_{GT}$ represents the (sample dependent) uncertainty in positions in the structure to be reassembled, $\sigma_{\text{measurement}}$ is the average measurement uncertainty, $c\left(2.38/\sqrt{d}\right)$ is the optimal scaling factor for the proposal covariance in the random-walk Metropolis algorithm with Gaussian proposals in $d$ dimensions with optional scaling factor $c=1$ \cite{roberts1997weak}, and $N_{\text{eff}} = N_{\text{cliques}} \times (M/N)$ is the effective sample size.

The variance of the large move, $\sigma^2_{\text{large}} = R_{\max}/4$, is set such that in a single move, positions can be reasonably expected to traverse the portion of the domain with high probability of containing the structure of interest. For completeness, the new coordinate may also be drawn from a uniform distribution $U(1/R_{\max})$ to ensure the proposal has a non-zero probability of reaching any position in the domain in a single move. Therefore, for each coordinate $x_{i_k}$, a new value is drawn from one of:

\[
x_{i_k}^{*}\sim  \begin{cases}\text{N}\left(x_{i_k}^{(t)},\sigma_{\text{large}}^2\right) , \\
    \text{N}\left(x_{i_k}^{(t)},\sigma_{\text{small}}^2\right) , \\
    \text{Unif}(0,\rho),
\end{cases}
\]

where it is recalled that $\rho=\frac{1}{2}R_{\max}$.

\(P_{\text{clutter}}\) is initially drawn from \(\text{Beta}(1, 9)\), and new values for \(P_{\text{clutter}}\) are proposed from $\text{Beta}(a+Q,b+(M-Q))$, where $Q$ is number of clutter measurements. Thus the proposal distribution at iteration $(t+1)$ for $P_\text{clutter}$ is given by:

\[
P(P_\text{clutter}^*|P_\text{clutter}^{(t)}) \sim  \text{Beta}\left(a+Q^{(t)},b+(M-Q^{(t)})\right).
\]

Given $P_{\text{clutter}}$, the probabilities, $P(Q=q_i)$ and consequently $P(K_i|P_\text{clutter})$ can be drawn for all possible assignment sets, where $i=1,\dots,|K_{\text{all}}|$. To propose the new assignment set, denoted $\mathbf{K}_i$ for $i=1,\dots,|K_{\text{all}}|$, for each clique, a probability distribution, denoted $P(\mathbf{K}_i)$, is constructed as the product of relevant prior distributions and the likelihood of the subset of localisations corresponding to the assignment set. For each clique $n$, all possible assignment sets within clique $n$ have equal likelihood given by Eq. (\ref{Eq:Likelihood}), for $n=1,\dots,N_{\text{cliques}}$. This is then multiplied by the prior distribution of proposed parameter values for that assignment set, denoted here as $P(\boldsymbol{\theta}^*_{i})$, and given by

\[
P(\boldsymbol{\theta}^*_{i})=P(R|\mathbf{K}_i)\times P(\mathbf{K}_i|P_\text{clutter})\times P(P_\text{clutter}),
\]

To construct a probability distribution, we then normalize over all assignment sets within each clique $n$, and draw the proposed assignment set, $\mathbf{K}_n^*$, for each clique $n$ from this, so that

\[
P(\mathbf{K}_n^*)=\frac{P(S_n|\boldsymbol\theta^*)\times P(\boldsymbol\theta_n^*)}{\sum_{i=1}^{|K_{\text{all}}|}P(S_i|\boldsymbol\theta^*)\times P(\boldsymbol\theta_i^*)} \quad \text{and thus} \quad P(\mathbf{K}^*)=\prod_{n=1}^{N_\text{cliques}}P(\mathbf{K}_n^*).
\]

The un-normalised product of likelihood and prior for each assignment and each clique is retained to ensure detailed balance when calculating the acceptance ratio. The posterior distribution at iteration $(t+1)$ is then constructed for the proposed set of parameters over all cliques and accepted or rejected based on an acceptance ratio given by,

\[
\alpha = \frac{P(\boldsymbol{\theta^{*}}|\mathbf{S})\times P(\mathbf{K}^{(t)})\times P(P_\text{clutter}^{(t)}|P_\text{clutter}^*)}{P(\boldsymbol{\theta}^{(t)} |\mathbf{S})\times P(\mathbf{K}^*)\times P(P_\text{clutter}^*|P_\text{clutter}^{(t)})}.
\]

The ratio of the un-normalised proposal distribution for proposed and current assignment sets must also be multiplied to ensure detailed balance. Post-hoc, align all iterations to the \textit{Maximum a Posteriori} (MAP) to estimate uncertainties, and assess convergence of both model dimension and \(P_{\text{clutter}}\) by Gelman-Rubin.

\begin{figure}[ht]
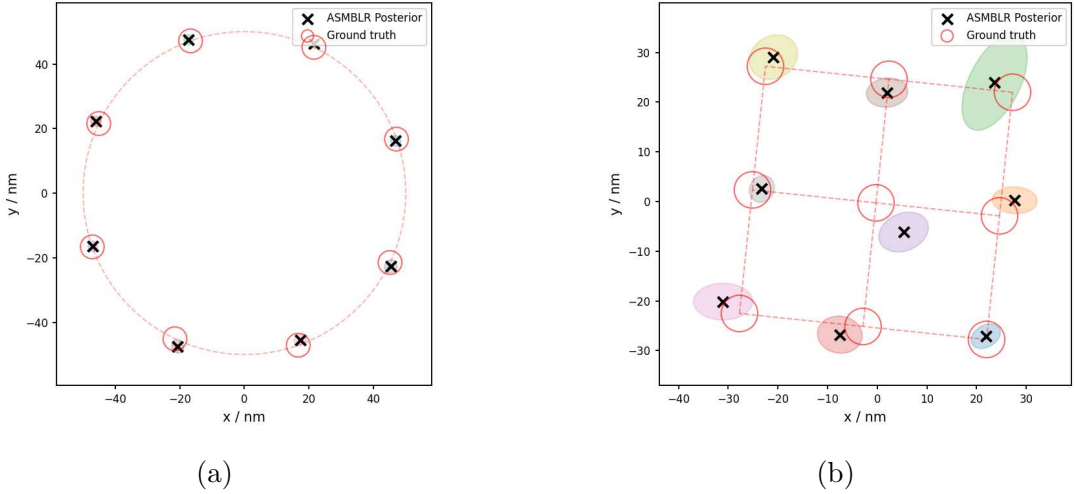

    \centering
    \subfloat[]{\includegraphics[width=0.5\linewidth]{1p0-mono-model.png}}
    \subfloat[]{\includegraphics[width=0.5\linewidth]{grid-di-model.png}}
    \caption{Template-free reconstructed molecule for inferred centres of (a) simulated Nup96 and (b) \(3\times3\) DNA-Origami.}
    \label{fig:methods-asmblr-chain-alignment}
\end{figure}

We validate ASMBLR reconstructions through three complementary statistical tests that collectively address whether inferred structures represent genuine geometric patterns rather than artifacts of template matching or overfitting to sparse data. Each test evaluates a distinct aspect of model-data agreement: per-vertex consistency, global coherence, and specificity of the discovered orientation.

\paragraph{Per-Vertex Hypothesis Tests with Bonferroni Correction} For each of the $k$ vertices in a reconstructed $k$-fold motif, we test whether the idealised model position is consistent with the corresponding posterior distribution. Let $\mathbf{m}_i$ denote the position of vertex $i$ in the idealised $k$-fold model (obtained via Hungarian algorithm\cite{kuhn1955hungarian} to minimise total Gaussian negative log-likelihood), and let $\boldsymbol{\mu}_i$ and $\boldsymbol{\Sigma}_i$ denote the empirical mean and covariance of vertex $i$ computed from post-burn-in MCMC samples.

We test the null hypothesis $H_0^{(i)}: \mathbf{m}_i \sim \mathcal{N}(\boldsymbol{\mu}_i, \boldsymbol{\Sigma}_i)$ for each vertex. The test statistic is the squared Mahalanobis distance:
\begin{equation}
D_i^2 = (\mathbf{m}_i - \boldsymbol{\mu}_i)^\top \boldsymbol{\Sigma}_i^{-1} (\mathbf{m}_i - \boldsymbol{\mu}_i),
\end{equation}
which measures the distance from $\mathbf{m}_i$ to $\boldsymbol{\mu}_i$ accounting for the shape and orientation of the posterior uncertainty ellipse. Under $H_0^{(i)}$, $D_i^2$ follows a $\chi^2$ distribution with 2 degrees of freedom (corresponding to the 2D spatial coordinates).

To control the family-wise error rate across $k$ simultaneous tests, we apply Bonferroni correction with adjusted significance level $\alpha_{\text{Bonf}} = \alpha / k$, where $\alpha = 0.05$. We compute the tail probability $p_i = P(\chi^2(2) \geq D_i^2)$ for each vertex and reject $H_0^{(i)}$ if $p_i < \alpha_{\text{Bonf}}$. A vertex is deemed consistent with the model if $p_i \geq \alpha_{\text{Bonf}}$.

\paragraph{Fisher's Combined Test for Global Consistency} While per-vertex tests assess individual positions, they do not evaluate whether all vertices are jointly consistent with the geometric constraint of $k$-fold rotational symmetry. We employ Fisher's method to combine the $k$ p-values $\{p_1, \ldots, p_k\}$ into a global test statistic:
\begin{equation}
\mathcal{F} = -2 \sum_{i=1}^k \log(p_i).
\end{equation}

Under the global null hypothesis $H_0^{\text{global}}$ (all individual null hypotheses $H_0^{(i)}$ hold simultaneously), and assuming independence of the p-values, $\mathcal{F}$ follows a $\chi^2$ distribution with $2k$ degrees of freedom. We compute the combined p-value as $p_{\text{Fisher}} = P(\chi^2(2k) \geq \mathcal{F})$.

\textbf{Interpretation:} Large values of $p_{\text{Fisher}}$ (e.g., $p_{\text{Fisher}} > 0.95$) indicate exceptionally tight agreement between the constrained model and posterior means, which arises when a genuine $k$-fold pattern exists in the data. This occurs because the mathematical constraint (only 3 free parameters: centre $x$, centre $y$, rotation $\theta$) naturally produces close alignment when the underlying structure exhibits true rotational symmetry. Conversely, $p_{\text{Fisher}} \approx 0.5$ indicates typical consistency, and $p_{\text{Fisher}} < 0.05$ suggests systematic deviation from the model.

\paragraph{Permutation Test Against Random Orientations} The per-vertex and Fisher tests establish that the model fits the data, but do not address whether the discovered orientation is uniquely determined by the data or whether any rotation would fit equally well. To distinguish genuine patterns from spurious alignments, we perform a permutation test comparing the observed alignment quality to a null distribution generated from random rotations.

Our test statistic is the sum of squared Mahalanobis distances:
\begin{equation}
Q = \sum_{i=1}^k D_i^2,
\end{equation}
where smaller values indicate better alignment. We compute $Q_{\text{obs}}$ for the optimal rotation $\theta^*$ discovered by ASMBLR. To construct the null distribution, we generate $N_{\text{perm}} = 10{,}000$ random rotations $\{\theta^{(1)}, \ldots, \theta^{(N_{\text{perm}})}\}$ uniformly sampled from $[0, 2\pi)$. For each $\theta^{(j)}$, we:
\begin{enumerate}
    \item Generate a $k$-fold model at rotation $\theta^{(j)}$: $\mathbf{m}_i^{(j)} = \mathbf{c} + R \begin{pmatrix} \cos(\theta^{(j)} + 2\pi i / k) \\ \sin(\theta^{(j)} + 2\pi i / k) \end{pmatrix}$, where $\mathbf{c}$ is the centre and $R$ is the radius estimated from the posterior means.
    \item Compute the optimal vertex-to-spot assignment via the Hungarian algorithm to minimize total Mahalanobis distance.
    \item Compute the null statistic $Q^{(j)} = \sum_{i=1}^k (D_i^{(j)})^2$ for this random rotation.
\end{enumerate}

The permutation p-value is the fraction of null statistics as extreme or more extreme than the observed:
\begin{equation}
p_{\text{perm}} = \frac{1}{N_{\text{perm}}} \sum_{j=1}^{N_{\text{perm}}} \mathds{1}(Q^{(j)} \leq Q_{\text{obs}}),
\end{equation}
where $\mathds{1}(\cdot)$ is the indicator function. Small values of $p_{\text{perm}}$ (e.g., $p_{\text{perm}} < 0.05$) indicate that the discovered orientation is significantly better than random, providing strong evidence for a genuine $k$-fold pattern.

\paragraph{Interpretation and Decision Criteria} We classify reconstruction quality based on the combined evidence from all three tests:
\begin{itemize}
    \item \textbf{Strongly Supported:} All vertices pass per-vertex tests, $p_{\text{Fisher}} > 0.5$, and $p_{\text{perm}} < 0.001$. The $k$-fold pattern is unambiguous.
    \item \textbf{Well-Supported:} $\geq (k-1)/k$ vertices pass, $p_{\text{Fisher}} > 0.05$, and $p_{\text{perm}} < 0.01$. The pattern is statistically clear despite some positional uncertainty.
    \item \textbf{Adequate:} $\geq (k-2)/k$ vertices pass, $p_{\text{Fisher}} > 0.01$, and $p_{\text{perm}} < 0.05$. The pattern is detectable but marginal.
    \item \textbf{Questionable:} Fewer than $(k-2)/k$ vertices pass or $p_{\text{perm}} \geq 0.05$. Insufficient evidence for $k$-fold symmetry.
\end{itemize}

The permutation test is the most critical diagnostic, as it directly addresses whether the reconstructed orientation represents genuine geometric regularity versus chance alignment to noise—the core concern raised by template-matching critiques. A large improvement ratio $Q_{\text{null}} / Q_{\text{obs}} \gg 1$ (where $Q_{\text{null}}$ is the mean of the null distribution) provides intuitive evidence that the discovered pattern is not arbitrary.

\section{Observations}
\subsection{GROUPA and BaGoL: Complementary Approaches to Emitter Estimation}
\label{subsec:groupa_bagol}

GROUPA and BaGoL \cite{fazel_bayesian_2019, fazel_high-precision_2022} represent fundamentally different approaches to the measurement-to-emitter clustering problem in SMLM, designed for distinct analytical scales and use cases. Here we clarify their relationship and explain why a direct quantitative comparison was not undertaken.

BaGoL (Bayesian Grouping of Localisations) operates at the single-structure level within manually defined regions of interest (ROIs). It employs a full Bayesian hierarchical model with RJMCMC to jointly infer:
\begin{itemize}
    \item The number of emitters $K$ within an ROI,
    \item Emitter positions $\{\boldsymbol{\mu}_k\}_{k=1}^{K}$,
    \item Assignment of each localisation to an emitter,
    \item Emitter-specific photophysical parameters.
\end{itemize}
The generative model assumes localisations arise from a mixture of Gaussians centred at true emitter positions, with measurement uncertainties propagated through the likelihood. BaGoL achieves sub-nanometre precision by pooling information across repeated observations of each emitter.

GROUPA (Grouping Observations Under Pairwise Associations) operates at the field-of-view level without requiring ROI pre-selection. Rather than fitting a global mixture model, GROUPA reduces the clustering problem to a series of pairwise hypothesis tests:
\begin{equation}
    \text{BF}_{ij} = \frac{P(\text{data} \mid H_0: \text{common emitter})}{P(\text{data} \mid H_1: \text{distinct emitters})} = I_{ij} \cdot V_{\text{eff}},
\end{equation}
where $I_{ij}$ is the overlap integral between localisation uncertainty distributions and $V_{\text{eff}}$ is the effective volume penalty (see Sec.~\ref{subsec:groupa}). These pairwise evidences are assembled into a weighted graph, and community detection via the Infomap algorithm \cite{Smiljanic2023MapEquation} partitions localisations into emitter groups.

The fundamental distinction lies in analytical scale:

\begin{table}[ht]
\centering
\begin{tabular}{lcc}
\toprule
\textbf{Characteristic} & \textbf{BaGoL} & \textbf{GROUPA} \\
\midrule
Input scope & Single ROI & Entire field-of-view \\
ROI selection & Manual, required & Not required \\
Typical input size & $10^1$--$10^2$ localisations & $10^3$--$10^5$ localisations \\
Output precision & Sub-nanometre & Nanometre-scale \\
Computational cost & High (RJMCMC per ROI) & Moderate (pairwise + graph) \\
Prior sensitivity & Moderate--High & Low \\
\bottomrule
\end{tabular}
\caption{Comparison of BaGoL and GROUPA operational characteristics.}
\label{tab:bagol_groupa}
\end{table}

BaGoL's RJMCMC sampler explores a model space that grows combinatorially with the number of localisations, making it computationally prohibitive for large fields without prior segmentation into ROIs. This segmentation, however, requires either manual intervention or a preliminary clustering step, introducing the very subjectivity that motivates alternative approaches.

GROUPA circumvents this by operating on pairwise relationships with computational complexity $\mathcal{O}(N \cdot \bar{k})$, where $N$ is the number of localisations and $\bar{k}$ is the average neighbourhood size under a spatial cutoff. The graph-based community detection then scales as $\mathcal{O}(E \log E)$ for $E$ edges, enabling field-wide analysis without ROI pre-selection.

A direct quantitative comparison between GROUPA and BaGoL would require:
\begin{enumerate}
    \item Applying BaGoL to the same synthetic datasets used for GROUPA benchmarking,
    \item Defining ROIs for BaGoL in a manner that does not presuppose knowledge of structure locations,
    \item Matching computational budgets or accepting substantial runtime disparities.
\end{enumerate}

The second requirement presents a methodological circularity: any automated ROI definition procedure would itself constitute a clustering algorithm, confounding the comparison. Manual ROI definition, while standard practice for BaGoL, is incompatible with the automated, discovery-focused workflow that GROUPA enables.

Moreover, the methods optimise for different objectives. BaGoL maximises localisation precision within known structures; GROUPA maximises partitioning accuracy across unknown structures. Comparing ARI scores between a method designed for precision and one designed for discovery would conflate distinct performance dimensions.

Rather than competing alternatives, GROUPA and BaGoL address complementary needs in SMLM analysis workflows:

\begin{enumerate}
    \item \textbf{Discovery-focused analysis}: GROUPA enables field-wide partitioning without prior structural knowledge, identifying candidate emitter groups for downstream analysis. This is appropriate when:
    \begin{itemize}
        \item Structure locations are unknown \textit{a priori},
        \item Large fields must be processed automatically,
        \item The goal is structural discovery rather than precision refinement.
    \end{itemize}
    
    \item \textbf{Precision-focused analysis}: BaGoL achieves maximal localisation precision within defined ROIs. This is appropriate when:
    \begin{itemize}
        \item Structures of interest have been identified (manually or via GROUPA),
        \item Sub-nanometre precision is required for downstream measurements,
        \item Computational cost is secondary to precision.
    \end{itemize}
    
    \item \textbf{Hybrid workflows}: A natural pipeline combines both approaches:
    \begin{enumerate}
        \item Apply GROUPA for field-wide emitter partitioning,
        \item Identify structures of interest via Voidwalker-Gibbs,
        \item Apply BaGoL to selected ROIs for precision refinement.
    \end{enumerate}
\end{enumerate}

This complementary relationship motivates our comparison of GROUPA against DBSCAN and HDBSCAN (Fig.~2 of the main text), which operate at the same analytical scale and address the same partitioning objective. BaGoL remains the method of choice for applications requiring maximal precision within pre-defined structures.

For completeness, we provide complexity estimates for both methods:

\textbf{BaGoL} (per ROI with $n$ localisations, $K$ emitters, $T$ MCMC iterations):
\begin{equation}
    \mathcal{O}\left(T \cdot n \cdot K\right) \text{ likelihood evaluations per ROI},
\end{equation}
with $T$ typically $10^4$--$10^5$ and the trans-dimensional moves (birth/death) requiring additional bookkeeping. For a field with $R$ ROIs, total complexity scales as $\mathcal{O}(R \cdot T \cdot \bar{n} \cdot \bar{K})$.

\textbf{GROUPA} (for $N$ localisations with average neighbourhood size $\bar{k}$):
\begin{equation}
    \mathcal{O}\left(N \cdot \bar{k}\right) \text{ pairwise Bayes factors} + \mathcal{O}\left(E \log E\right) \text{ community detection},
\end{equation}
where $E \leq N \cdot \bar{k}$ is the number of edges retained after thresholding. With spatial indexing (e.g., k-d trees), neighbourhood queries are $\mathcal{O}(\log N)$, yielding total complexity $\mathcal{O}(N \cdot \bar{k} \cdot \log N)$.

For typical SMLM datasets ($N \sim 10^4$, $\bar{k} \sim 10$, $R \sim 10^2$, $\bar{n} \sim 10^2$, $T \sim 10^4$), GROUPA requires $\sim 10^5$--$10^6$ operations while BaGoL requires $\sim 10^8$--$10^9$ operations, a difference of 2-3 orders of magnitude. This disparity underscores that the methods are designed for different scales rather than representing competing solutions to the same problem.

\subsection{Uniform Clique Sampling in SMLM}
The challenge of sampling structurally representative cliques from SMLM data is a practical bottleneck that significantly impacts downstream reconstruction quality. Here we provide additional mathematical detail and practical guidance on the clique sampling problem illustrated in Fig. 3(e) of the main text.

Consider a typical SMLM scenario where an 8-fold symmetric structure (such as Nup96) is imaged with incomplete labelling and background clutter. When sampling a clique of size $k$ from a radial neighbourhood, the probability of obtaining a "pure" clique - one where all members originate from the same underlying structure - depends on the composition of that neighbourhood.

\paragraph{Structurally Representative Clique.} A $k$-clique $C = \{e_1, \ldots, e_k\}$ sampled from an emitter set $X$ is \textit{structurally representative} if all members share the same true structural parent, i.e., $\text{parent}(e_i) = \text{parent}(e_j)$ for all $i, j \in \{1, \ldots, k\}$.

For uniform radial sampling within a cut-off distance $d_{\text{cut}}$, the neighbourhood around a seed emitter typically contains:

\begin{itemize}
    \item $s$ emitters from the same true structure as the seed
    \item $c_{\text{clutter}}$ spurious emitters (background noise, autofluorescence)
    \item $c_{\text{bridge}}$ emitters from neighbouring structures whose spatial extent overlaps the search radius
\end{itemize}

The total contamination is $c = c_{\text{clutter}} + c_{\text{bridge}}$, and the neighbourhood size is $n = s + c$.

When sampling uniformly without replacement from a neighbourhood of size $n$ containing $s$ true structure members, the probability of drawing a pure $k$-clique is given by the hypergeometric ratio:

$$P^{\text{rad}}_{\text{within}}(k; s, c) = \frac{\binom{s}{k}}{\binom{s+c}{k}} = \frac{s!/(s-k)!}{(s+c)!/(s+c-k)!} = \prod_{i=0}^{k-1} \frac{s-i}{s+c-i}$$

This expression has several important properties:

\begin{enumerate}
    \item \textbf{Monotonic decay in $k$:} As clique size increases, $P^{\text{rad}}_{\text{within}}$ decreases rapidly. Each additional member introduces another opportunity for contamination.
    \item \textbf{Sensitivity to contamination ratio:} The probability depends on $c/(s+c)$, the contamination fraction. Even modest contamination severely impacts large cliques.
    \item \textbf{Asymptotic behaviour:} For fixed contamination $c > 0$, $\lim_{k \to s} P^{\text{rad}}_{\text{within}} \to 0$, meaning pure cliques become increasingly rare as $k$ approaches the true structure size.
\end{enumerate}

Consider a fully labelled 8-fold ring ($s = 8$) with a neighbourhood contaminated by 3 spurious detections ($c = 3$), giving $n = 11$ total emitters. The probability of sampling a pure clique of various sizes is shown in Table~\ref{tab:clique_sampling_comparison}.

\begin{table}[ht]
  \centering
  \begin{tabular}{r r r r r r r r}
    \toprule
    $k$ &
    $T_k=\binom{s+c}{k}$ &
    $W_k=\binom{s}{k}$ &
    $B_k=T_k-W_k$ &
    $P^{\mathrm{rad}}_{\mathrm{within}}$ &
    $P^{\mathrm{rad}}_{\mathrm{between}}$ &
    $P^{\mathrm{VW}}_{\mathrm{within}}$ &
    Improvement \\
    \midrule
    2 & 55 & 28 & 27 & 0.509 & 0.491 & 0.810 & 1.6$\times$ \\
    3 & 165 & 56 & 109 & 0.339 & 0.661 & 0.729 & 2.1$\times$ \\
    4 & 330 & 70 & 260 & 0.212 & 0.788 & 0.656 & 3.1$\times$ \\
    5 & 462 & 56 & 406 & 0.121 & 0.879 & 0.590 & 4.9$\times$ \\
    6 & 462 & 28 & 434 & 0.061 & 0.939 & 0.531 & 8.7$\times$ \\
    7 & 330 & 8 & 322 & 0.024 & 0.976 & 0.478 & 19.9$\times$ \\
    8 & 165 & 1 & 164 & 0.006 & 0.994 & 0.430 & 71.7$\times$ \\
    \bottomrule
  \end{tabular}
  \caption{Probability of sampling a structurally pure $k$-clique under uniform radial sampling ($s=8$ true emitters, $c=3$ contaminants) versus Voidwalker-guided sampling (90\% assignment accuracy). $T_k$: total $k$-cliques in neighbourhood; $W_k$: within-structure cliques; $B_k$: between-structure/contaminated cliques; $P^{\mathrm{rad}}_{\mathrm{within}}$, $P^{\mathrm{rad}}_{\mathrm{between}}$: uniform sampling probabilities; $P^{\mathrm{VW}}_{\mathrm{within}}$: Voidwalker-guided probability ($0.9^k$); Improvement: ratio $P^{\mathrm{VW}}_{\mathrm{within}}/P^{\mathrm{rad}}_{\mathrm{within}}$.}
  \label{tab:clique_sampling_comparison}
\end{table}

This dramatic decay in $P^{\mathrm{rad}}_{\mathrm{within}}$ explains why uniform radial sampling becomes impractical for ASMBLR when contamination is present: the vast majority of sampled cliques will contain emitters from multiple sources, corrupting the internal geometry used for reconstruction. At $k=8$, fewer than 1\% of uniformly sampled cliques are structurally pure.

The Voidwalker-Gibbs pipeline provides per-emitter assignment probabilities to inferred structural centres, enabling \textit{assignment-guided} clique sampling. Rather than sampling uniformly from spatial neighbourhoods, cliques are drawn from emitters sharing the same centre assignment. Under the independence approximation for assignment correctness, if each emitter is correctly assigned with probability $a \in [0, 1]$, then:

$$P^{\text{VW}}_{\text{within}}(k; a) = a^k$$

This expression represents an upper bound on structural coherence achievable through assignment-guided sampling. The key observation from Fig.~3(e) is that the improvement factor grows rapidly with clique size, precisely where ASMBLR benefits most from consistent internal geometry. For $k = 8$ (full ring), Voidwalker-guided sampling with 90\% accuracy yields $P_{\text{within}} \approx 0.43$, a 72-fold improvement over uniform sampling in the contaminated neighbourhood.

Based on these analyses, we offer the following guidance for clique sampling in SMLM reconstruction:

\begin{enumerate}
    \item Prefer assignment-guided sampling when available; the improvement factor grows exponentially with clique size.
    \item Target moderate clique sizes ($k = 3$--$5$) that provide sufficient geometric constraints for reconstruction while maintaining reasonable purity probabilities above 50\%.
    \item Sample more cliques than strictly necessary, then use internal consistency metrics (e.g., variance in reconstructed separations) to identify and discard likely contaminated cliques.
    \item Account for imperfect assignment accuracy. Even at 90\% accuracy, approximately 57\% of sampled 8-cliques will contain at least one misassigned emitter. ASMBLR's robustness to moderate contamination partially mitigates this issue.
    \item The number of pure cliques available for reconstruction is approximately $N_{\text{cliques}} \times P_{\text{within}}(k; a)$; ensure this exceeds the minimum required for stable inference.
\end{enumerate}

\subsection{Information-Theoretic Limits of Under-sampled Data}
The super-structure discovery algorithm (Sec. \ref{subsec:assignments}) exhibits a sharp performance degradation at low labelling efficiencies that cannot be overcome by algorithmic refinement. This section characterises the information-theoretic basis for this limitation and provides practical guidance on when super-structure inference is feasible.

The super-structure discovery algorithm relies on comparing the marks associated with each structural centre, where a mark $m_i \in \Delta^{N-1}$ is a probability distribution over emitter assignments. The discriminative power of mark-based similarity depends on the information content of these distributions.

\paragraph{Shannon entropy of assignment distributions.} 
For a centre $i$ with $n_i$ assigned emitters from a total pool of $N$ emitters, the posterior responsibility distribution has Shannon entropy:

$$H(m_i) = -\sum_{n=1}^{N} m_i(n) \log m_i(n)$$

where $m_i(n)$ is the probability that emitter $n$ is assigned to centre $i$.

In the ideal case where assignments are deterministic (each emitter belongs to exactly one centre), $H(m_i) = 0$ for concentrated distributions and marks are maximally informative. As assignment uncertainty increases - due to overlapping structures, measurement noise, or sparse labelling - the entropy increases toward its maximum value $H_{\max} = \log N$.

Consider an 8-fold symmetric structure with labelling efficiency $\ell \in [0, 1]$. The expected number of observed emitters per structure is $\bar{n} = 8\ell$. Under the Voidwalker-Gibbs model, each observed emitter contributes probability mass to its parent centre's mark.

\paragraph{Low detection.} With $\bar{n} \approx 2$–$3$ emitters per structure:
\begin{itemize}
    \item Each centre's mark is spread across only 2–3 emitters
    \item The posterior responsibility for each emitter is diluted by assignment uncertainty
    \item Mark entropy approaches $H_{\max}$ as the distribution flattens
\end{itemize}

\paragraph{High detection.} With $\bar{n} \approx 7$–$8$ emitters per structure:
\begin{itemize}
    \item Marks are concentrated on a well-defined set of emitters
    \item Posterior responsibilities are sharply peaked
    \item Mark entropy is low, providing strong discriminative signal
\end{itemize}

The similarity between two centres is quantified via the Bhattacharyya coefficient:

$$k_B(i, j) = \sum_{n=1}^{N} \sqrt{m_i(n) m_j(n)}$$

For two distributions to be reliably distinguished as "similar" (same super-structure) vs. "dissimilar" (independent structures), their Bhattacharyya distances must separate into distinct populations.

As mark entropy increases, all pairwise Bhattacharyya distances converge toward a common value. In the limit where all marks are uniform (maximum entropy), $k_B(i,j) \to 1$ for all pairs, and no discrimination is possible.

We can formalise the minimum information required for super-structure detection. Let $\mu_{\text{same}}$ and $\mu_{\text{diff}}$ denote the mean Bhattacharyya distances for true super-structure pairs and independent centre pairs, respectively. Super-structure detection requires:

$$\frac{\mu_{\text{same}} - \mu_{\text{diff}}}{\sigma_{\text{pooled}}} > t_{\alpha}$$

where $\sigma_{\text{pooled}}$ is the pooled standard deviation and $t_{\alpha}$ is the critical threshold for significance level $\alpha$.

From Fig. 4 in the main text, this separation collapses below $\ell \approx 0.6$ for 8-fold structures. At $\ell = 0.3$:

\begin{itemize}
    \item Mean emitters per structure: $\bar{n} = 2.4$
    \item Effective bits of information per mark: $\lesssim 2$ bits
    \item Separation $(\mu_{\text{same}} - \mu_{\text{diff}})/\sigma_{\text{pooled}} \approx 0.3$–$0.5$ (indistinguishable from noise)
\end{itemize}

The performance boundary observed in Fig. 4 is not an algorithmic artefact but reflects a fundamental information-theoretic constraint:

\begin{enumerate}
    \item \textbf{Fixed information budget:} Each labelled emitter provides a finite amount of information about structural membership. At low labelling, this budget is insufficient to support hierarchical inference.
    \item \textbf{Compounding uncertainty:} Super-structure requires two levels of inference: (i) emitter-centre assignment, and (ii) centre to super-structure grouping. Uncertainty compounds across levels.
    \item \textbf{Pigeonhole principle:} With $k$ structures and $n \ll 8k$ total emitters, the mark matrices are inherently low-rank. The number of distinguishable mark patterns is bounded by $\binom{n}{\bar{n}}$, which may be smaller than the number of structure pairs requiring classification.
\end{enumerate}

\subsection{Structure-Dependent Fragility Under Incomplete Detection}

The performance boundaries documented throughout this work - particularly the divergent labelling requirements for centre detection versus super-structure inference - reflect deeper structure-dependent vulnerabilities that merit brief discussion.

Different molecular geometries degrade qualitatively differently under emitter loss. Consider two canonical cases: an 8-fold symmetric ring (such as Nup96) and a $3 \times 3$ grid lattice. At equivalent detection probability $p \approx 0.7$, the ring typically retains its characteristic topology - the circular arrangement with central void persists provided no gap exceeds a critical angular extent. The grid, however, may become unrecognisable depending on which emitters are lost: missing corner emitters preserve regularity, whereas losing the central emitter destroys the distance distribution that defines ``grid-ness.''

This asymmetry can be formalised through the concept of criticality distribution. Structures with high rotational symmetry exhibit distributed criticality: all emitters contribute equally, failure requires cumulative damage, and recognition probability varies smoothly with $p$. Structures with unique topological roles (central vertices, bridging elements) exhibit concentrated criticality: a small subset of emitters carries disproportionate structural information, and recognition probability displays step-like dependence on $p$.

For the 8-fold ring, topological recognition (preservation of the $H_1$ homological feature corresponding to the central hole) fails only when three or more consecutive emitters are undetected, creating a gap exceeding the ring diameter. The probability of such catastrophic gaps remains low ($< 0.1$) even at $p = 0.55$. For the $3 \times 3$ grid, the central emitter has betweenness centrality four times that of edge emitters; its loss alone (probability $1-p$) substantially degrades structural inference.

These observations suggest that minimum detection thresholds are not universal but structure-specific, and further suggest that symmetric ring-like architectures tolerate detection probabilities approximately 30-40\% lower than grid-like structures for equivalent recognition reliability. This has practical implications for experimental design: the labelling efficiency required to resolve a molecular architecture depends not only on emitter density but on the geometry of the target structure itself.

A comprehensive mathematical treatment of structural fragility - drawing on persistent homology, graph-theoretic centrality measures, and spatial point process theory - is beyond the scope of the present work but represents a natural extension of the framework developed here.

\section{Additional Figures: end-to-end}
The following figures display the end-to-end framework discussed in Sec. \ref{sec:methods} across a number of randomly selected synthetic Nup datasets.

\begin{figure}[ht]
    \centering
    \subfloat[]{\includegraphics[width=0.3\linewidth]{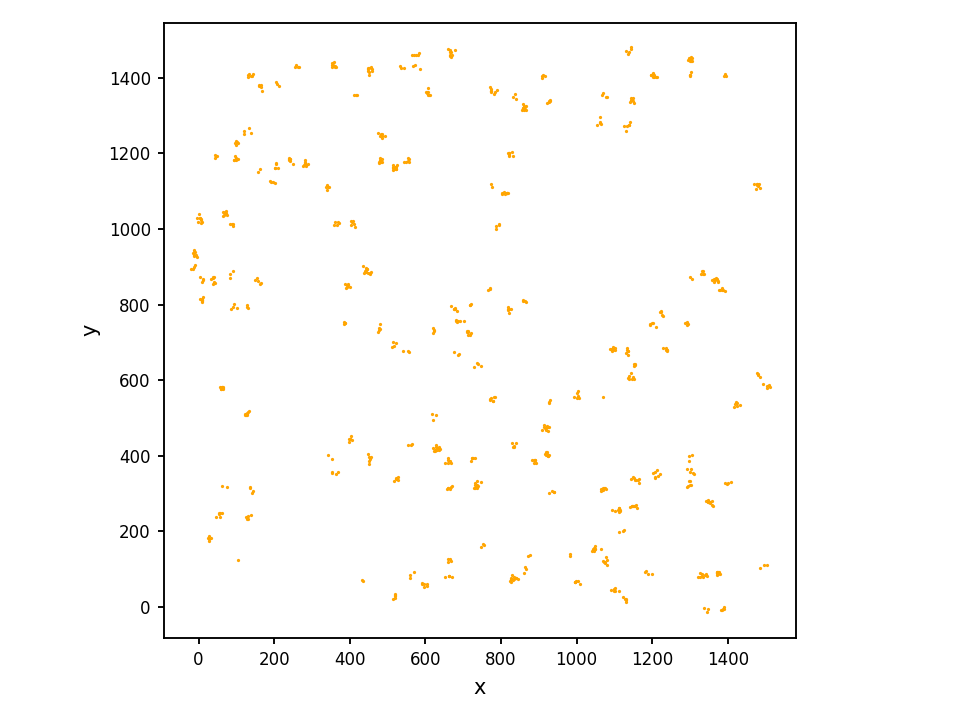}}
    \subfloat[]{\includegraphics[width=0.3\linewidth]{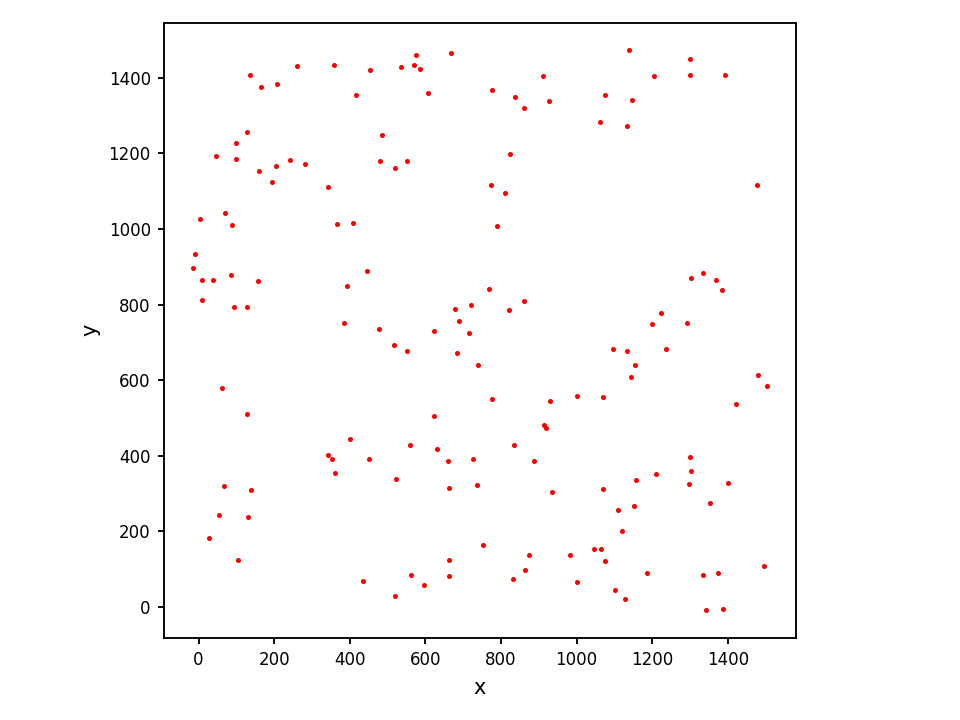}}
    \subfloat[]{\includegraphics[width=0.3\linewidth]{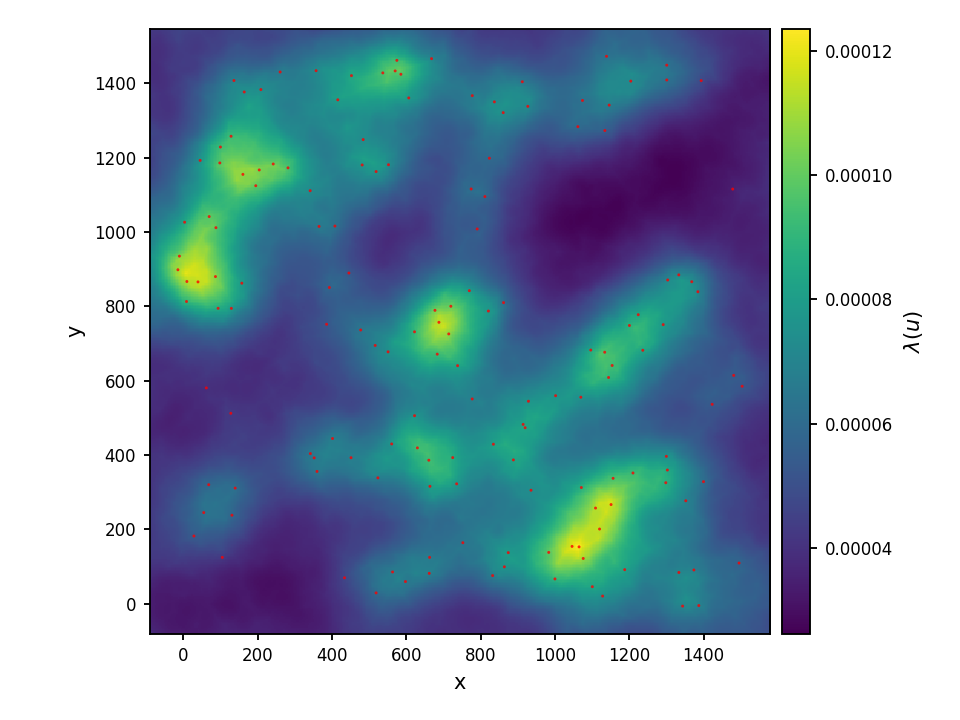}} \\
    \subfloat[]{\includegraphics[width=0.3\linewidth]{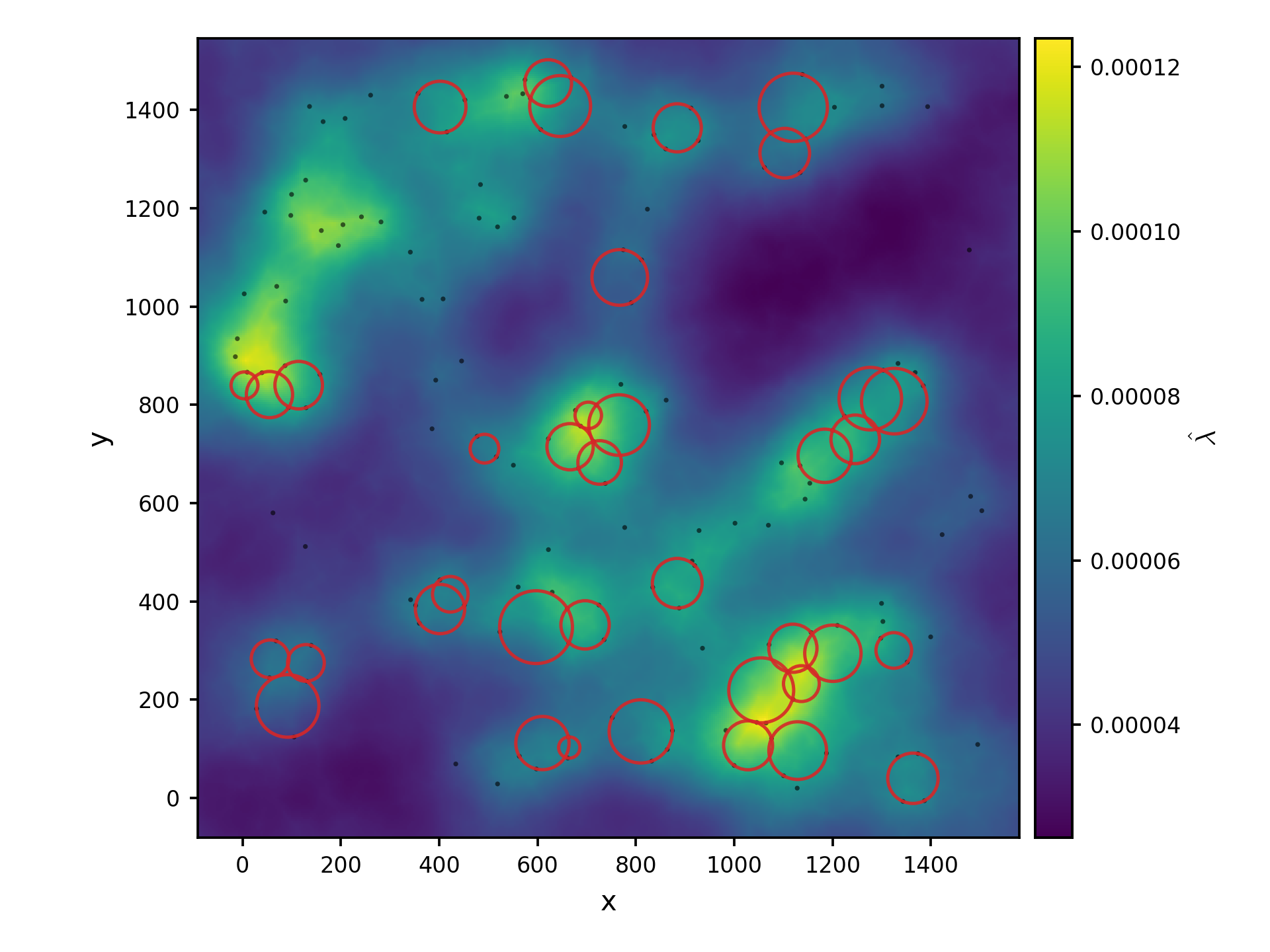}}
    \subfloat[]{\includegraphics[width=0.3\linewidth]{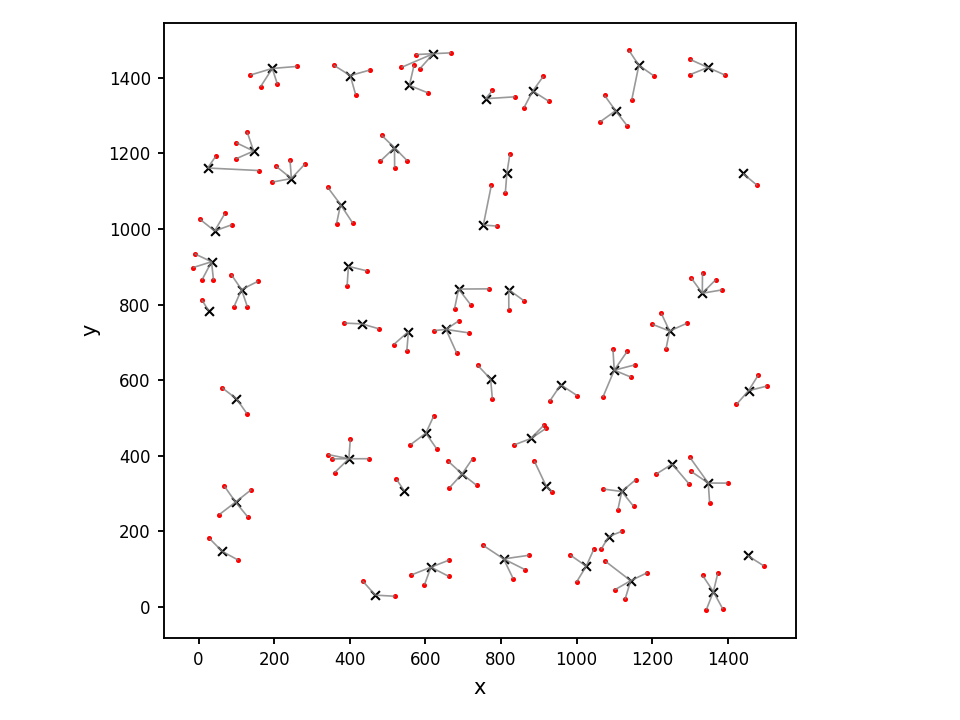}}
    \subfloat[]{\includegraphics[width=0.3\linewidth]{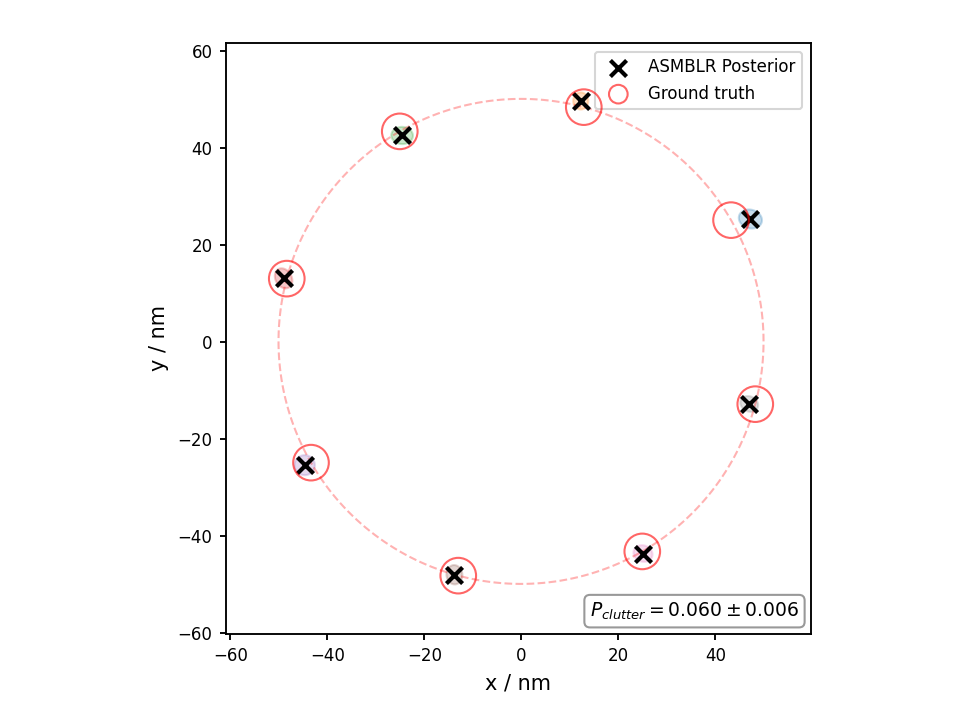}}
    \caption{0.3 labelling with 0.0 clutter. (a) Localisations. (b) GROUPA estimated emitters. (c) LGCP intensity map. (d) Voidwalker-discovered significantly empty space. (e) RJMCMC emitter-centre assignments. (f) ASMBLR reconstruction.}
    \label{fig:end-0p3-0p0}
\end{figure}

\begin{figure}[ht]
    \centering
    \subfloat[]{\includegraphics[width=0.3\linewidth]{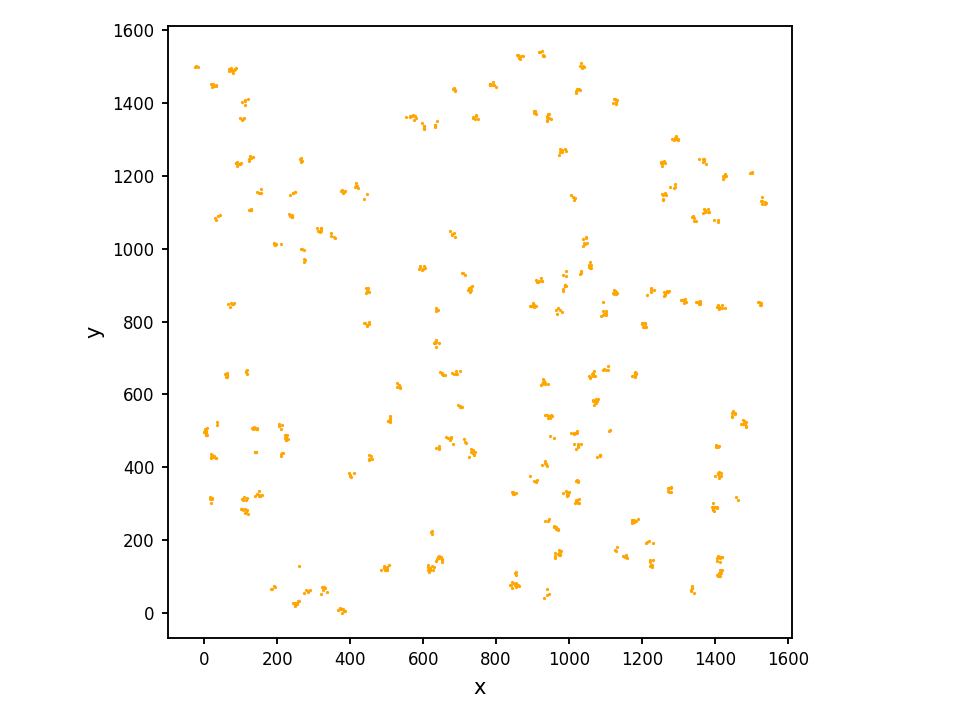}}
    \subfloat[]{\includegraphics[width=0.3\linewidth]{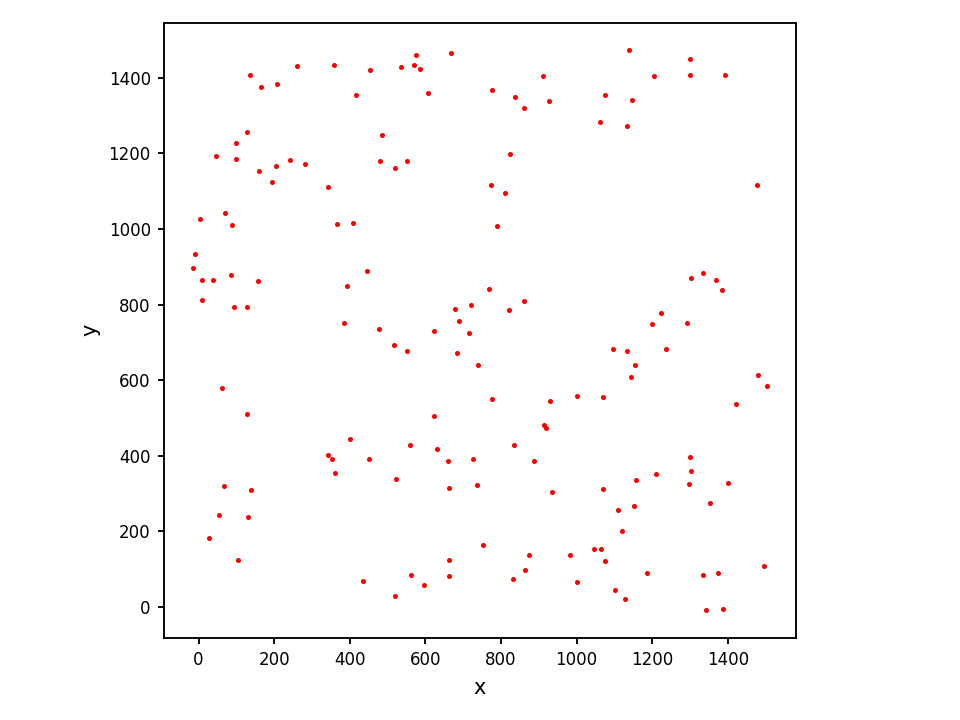}}
    \subfloat[]{\includegraphics[width=0.3\linewidth]{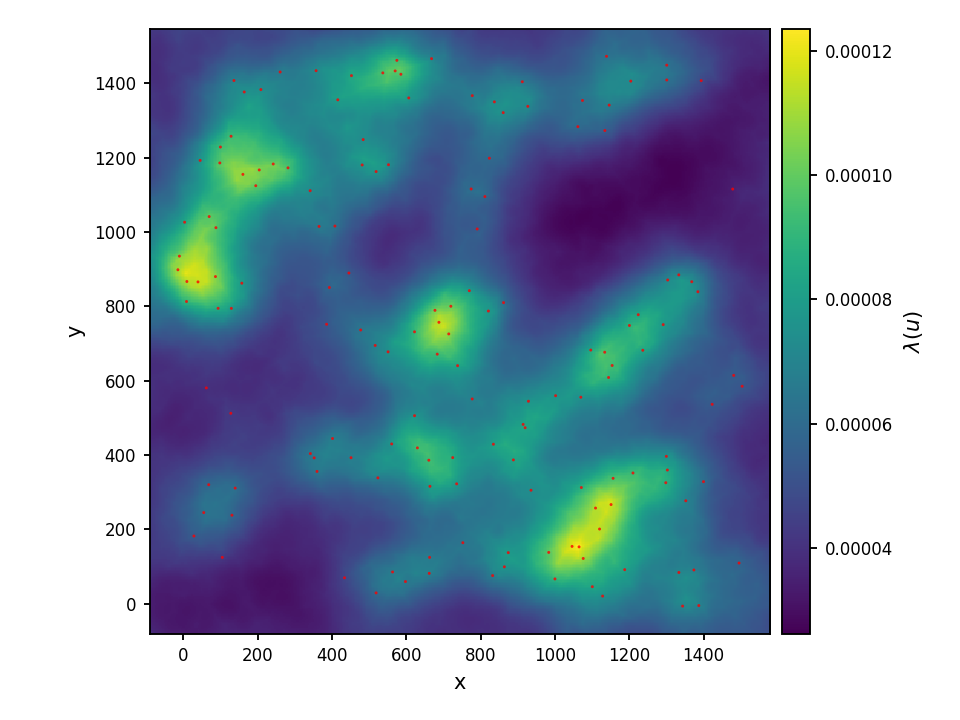}} \\
    \subfloat[]{\includegraphics[width=0.3\linewidth]{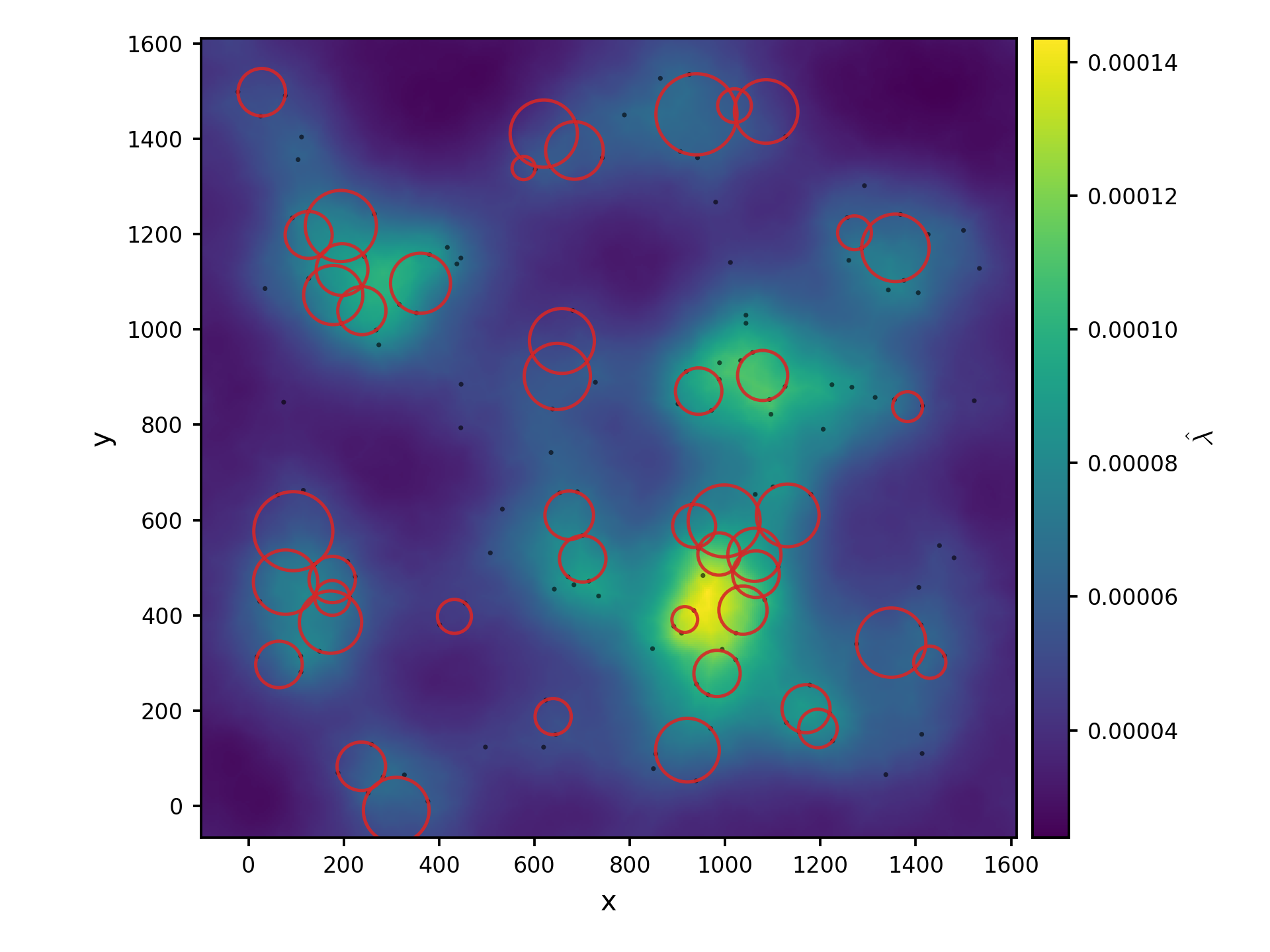}}
    \subfloat[]{\includegraphics[width=0.3\linewidth]{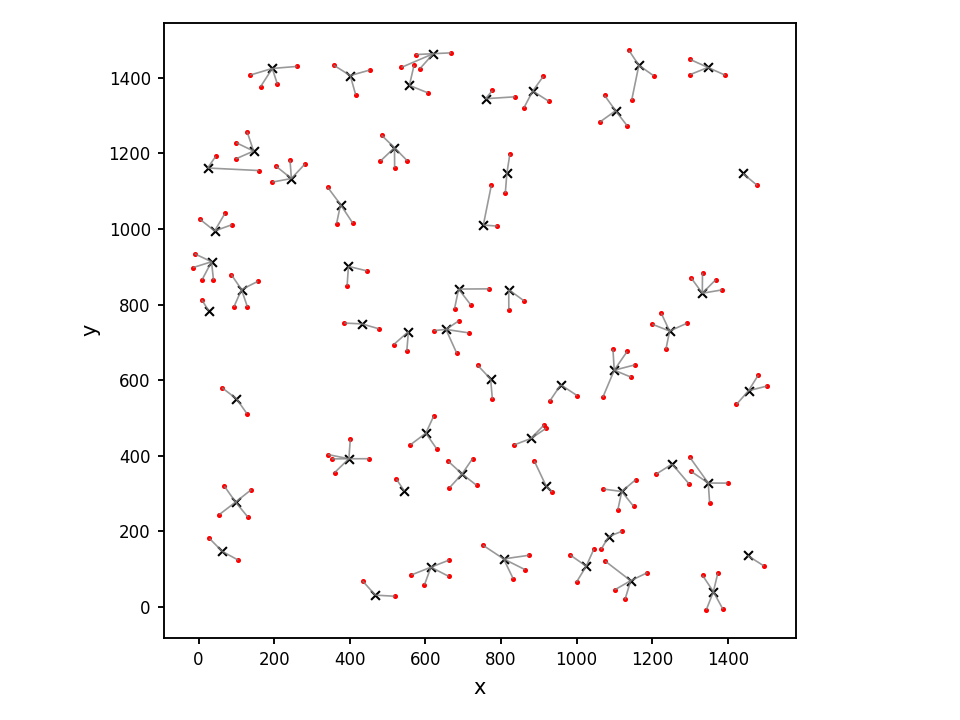}}
    \subfloat[]{\includegraphics[width=0.3\linewidth]{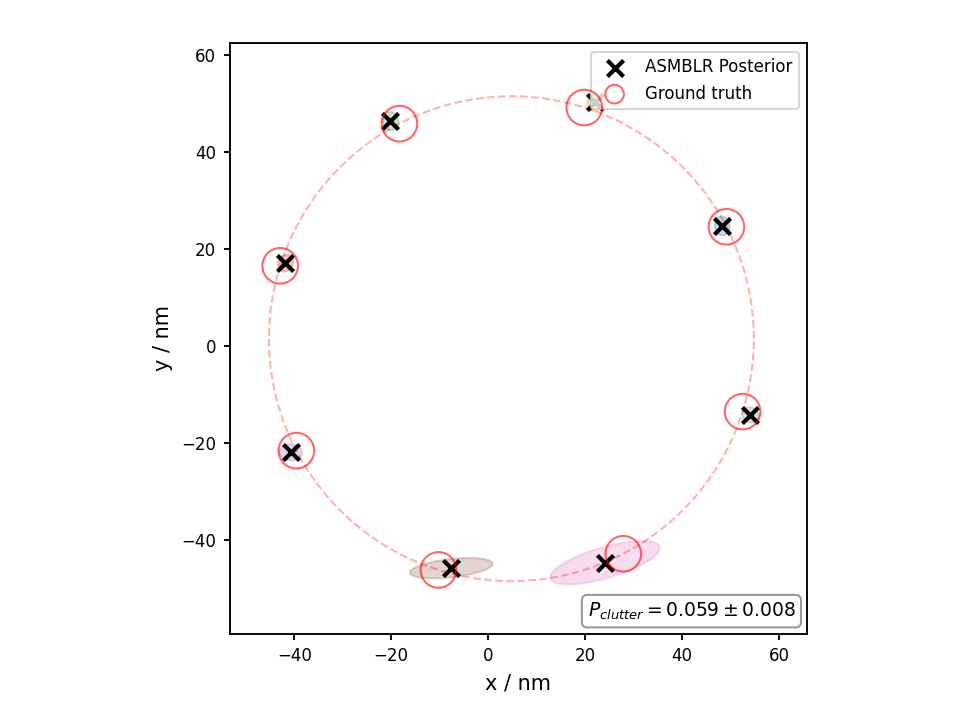}}
    \caption{0.3 labelling with 0.1 clutter. (a) Localisations. (b) GROUPA estimated emitters. (c) LGCP intensity map. (d) Voidwalker-discovered significantly empty space. (e) RJMCMC emitter-centre assignments. (f) ASMBLR reconstruction.}
    \label{fig:end-0p3-0p1}
\end{figure}

\begin{figure}[ht]
    \centering
    \subfloat[]{\includegraphics[width=0.3\linewidth]{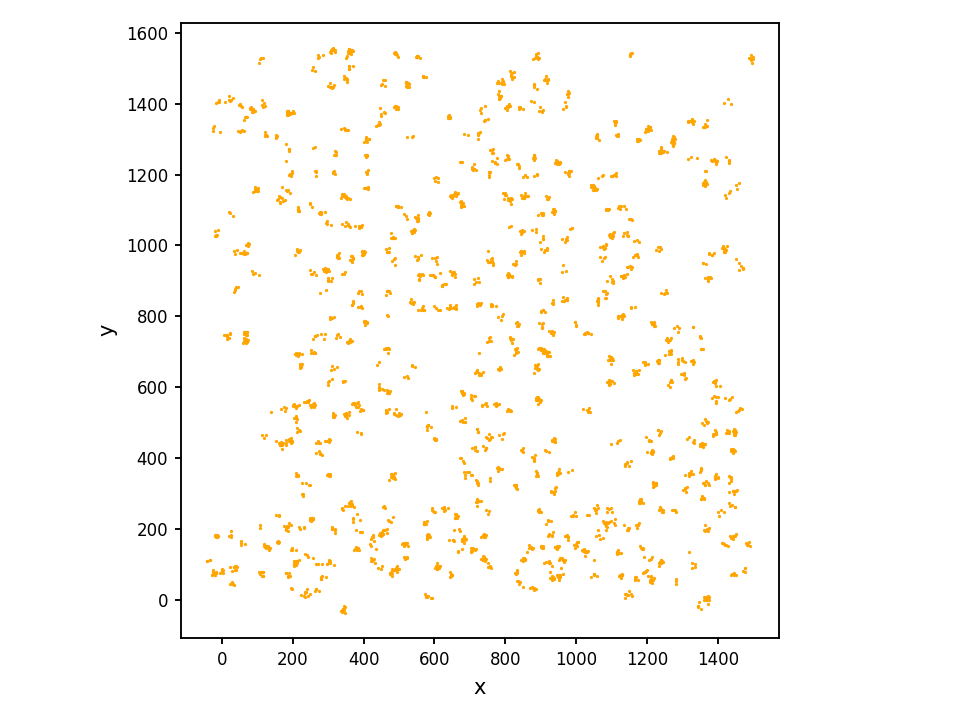}}
    \subfloat[]{\includegraphics[width=0.3\linewidth]{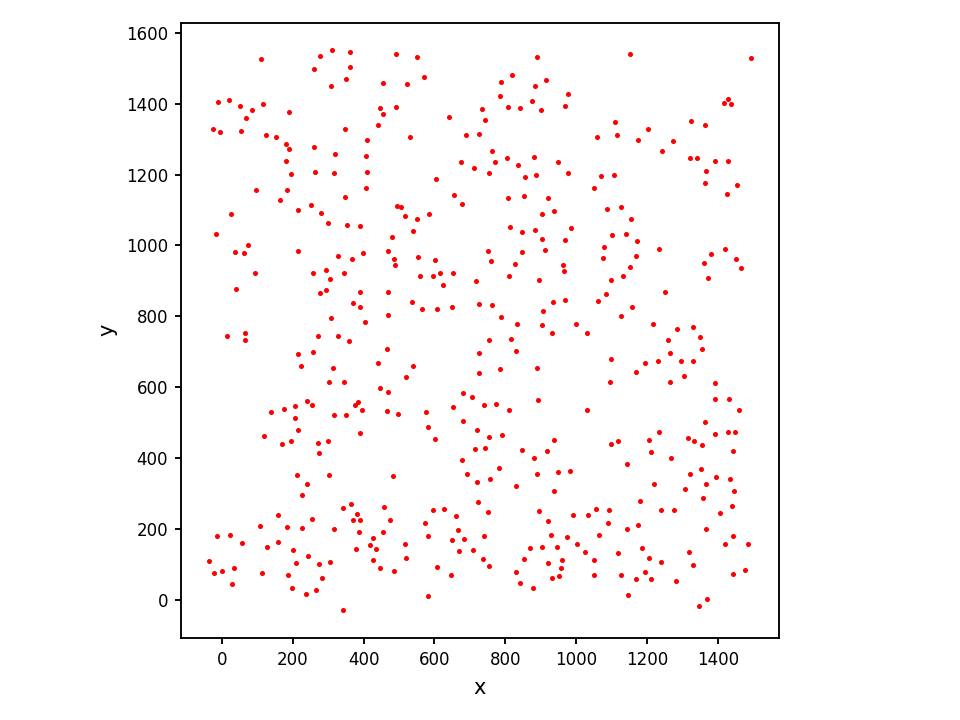}}
    \subfloat[]{\includegraphics[width=0.3\linewidth]{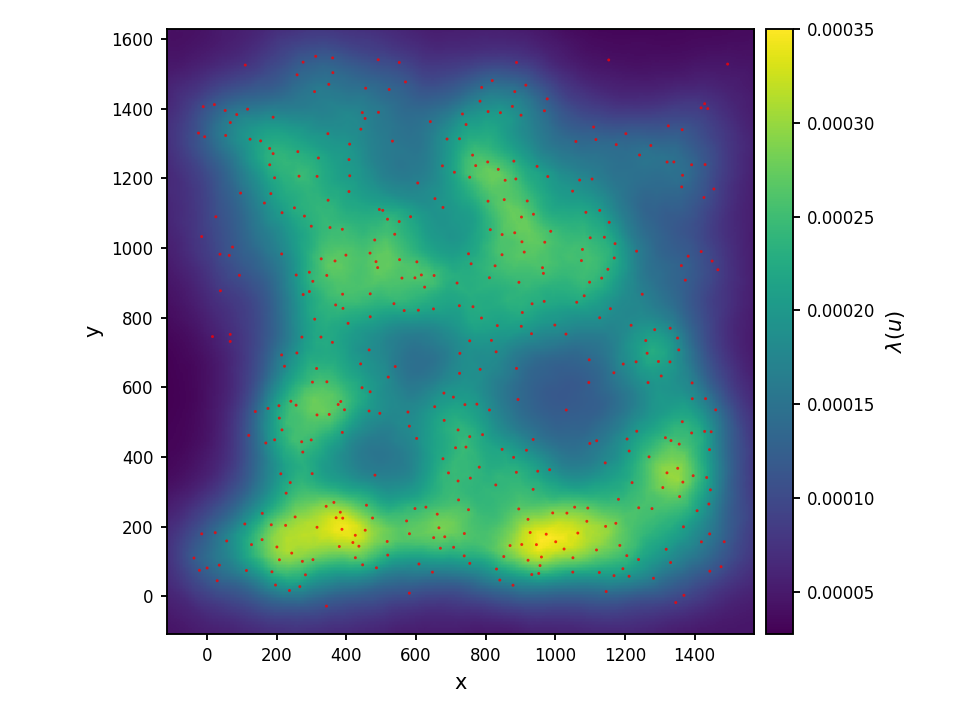}} \\
    \subfloat[]{\includegraphics[width=0.3\linewidth]{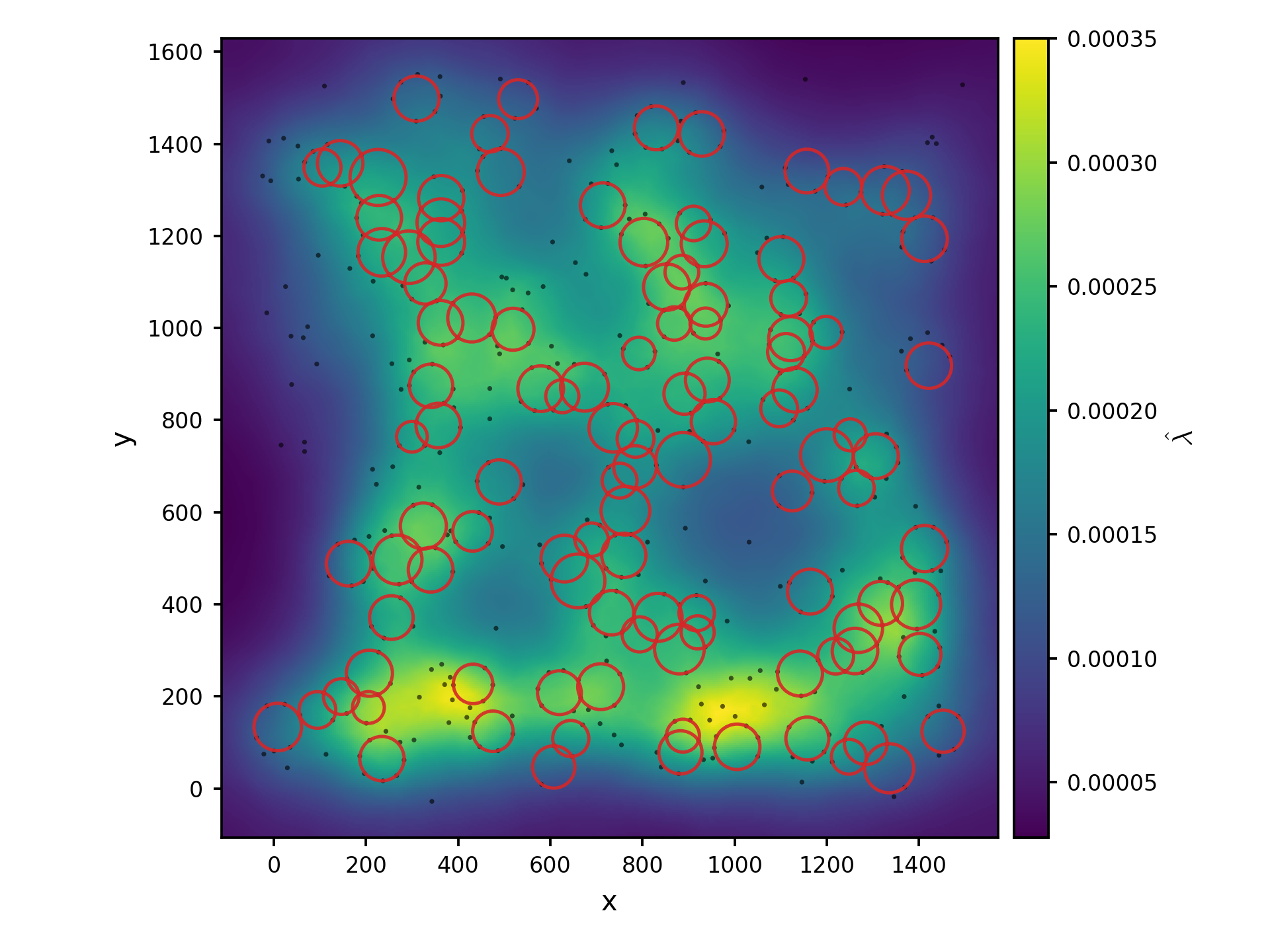}}
    \subfloat[]{\includegraphics[width=0.3\linewidth]{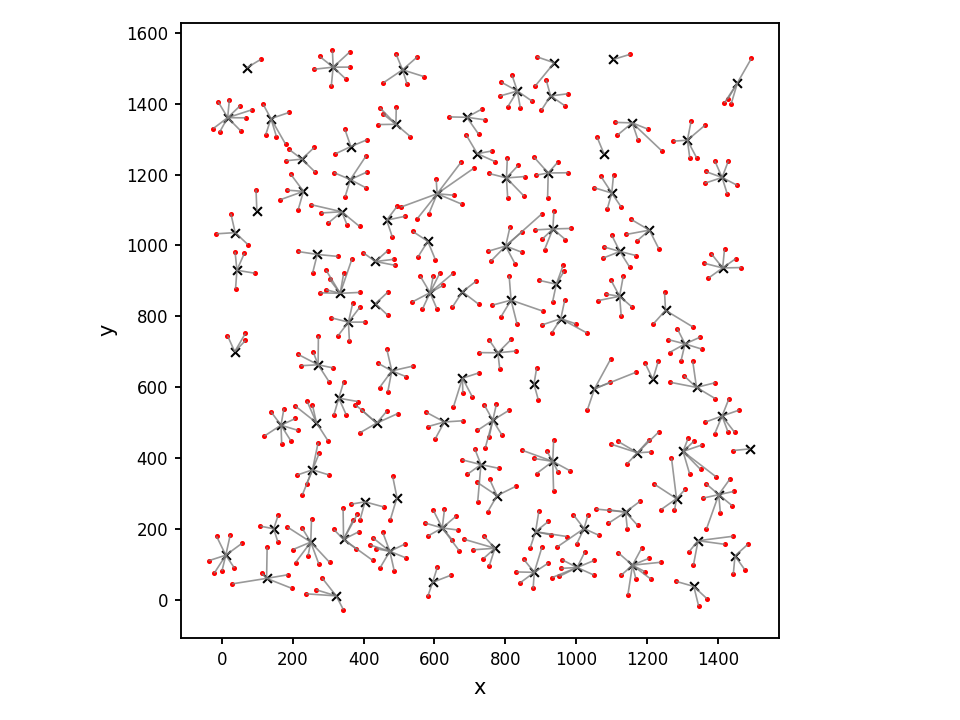}}
    \subfloat[]{\includegraphics[width=0.3\linewidth]{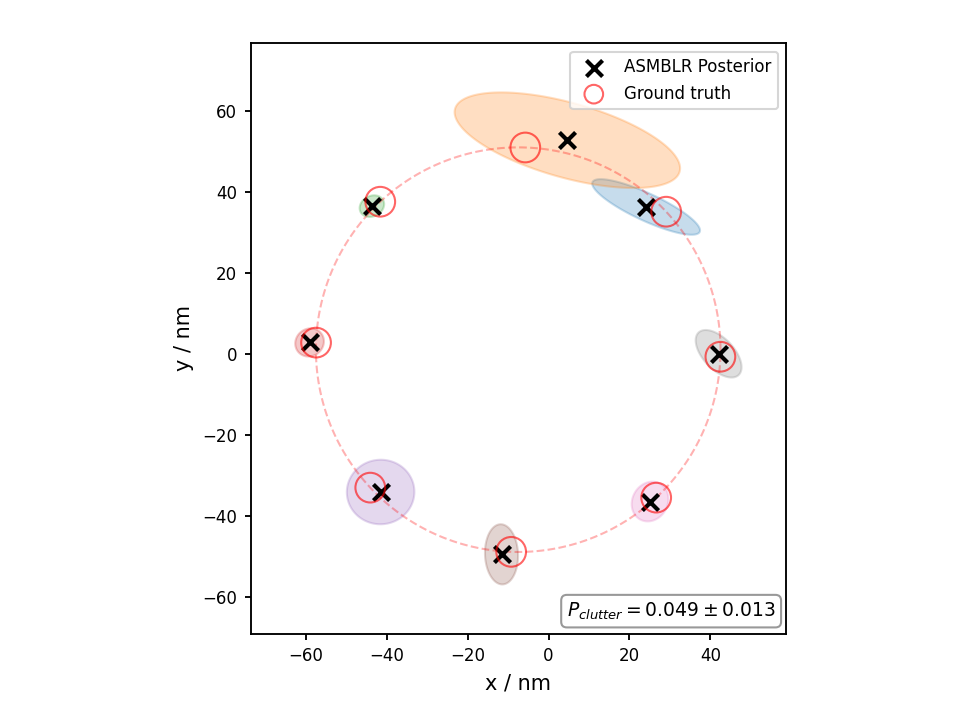}}
    \caption{0.6 labelling with 0.2 clutter. (a) Localisations. (b) GROUPA estimated emitters. (c) LGCP intensity map. (d) Voidwalker-discovered significantly empty space. (e) RJMCMC emitter-centre assignments. (f) ASMBLR reconstruction.}
    \label{fig:end-0p6-0p2}
\end{figure}

\begin{figure}[ht]
    \centering
    \subfloat[]{\includegraphics[width=0.3\linewidth]{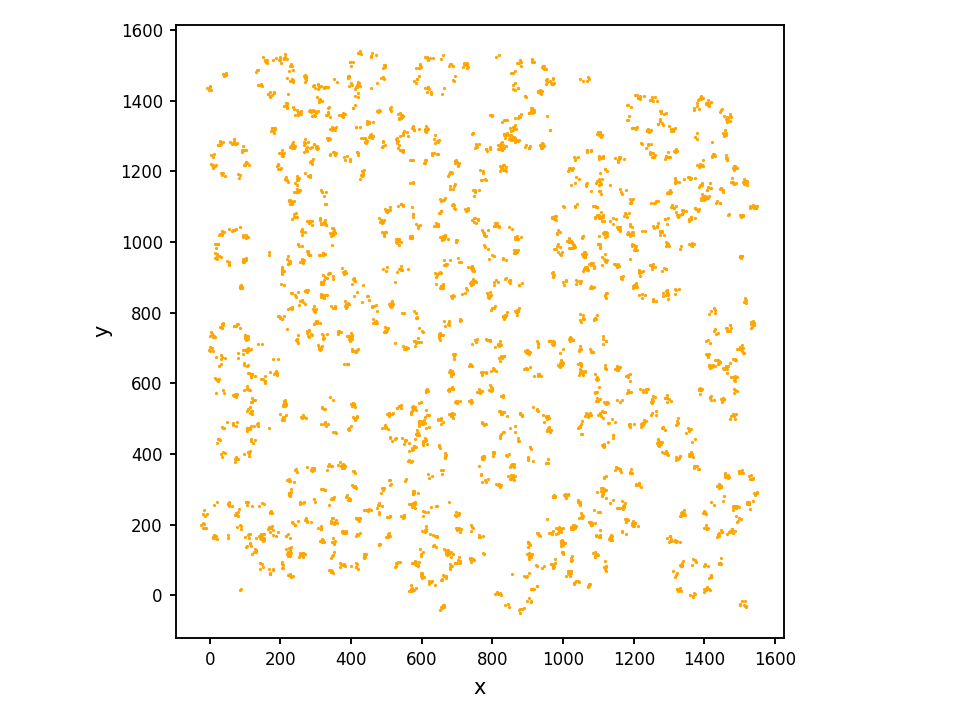}}
    \subfloat[]{\includegraphics[width=0.3\linewidth]{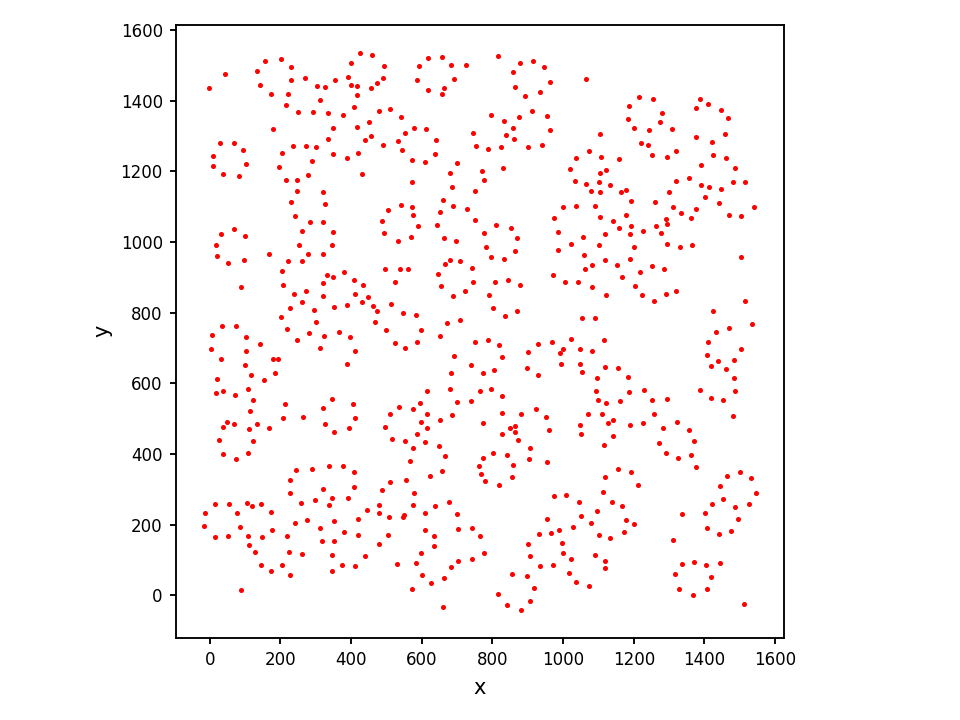}}
    \subfloat[]{\includegraphics[width=0.3\linewidth]{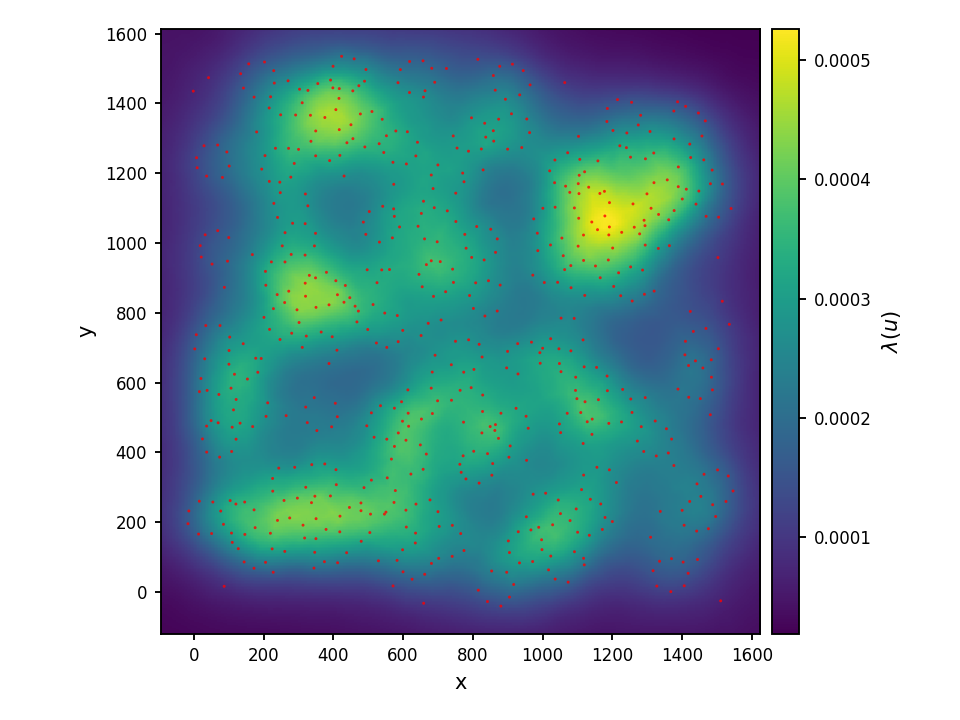}} \\
    \subfloat[]{\includegraphics[width=0.3\linewidth]{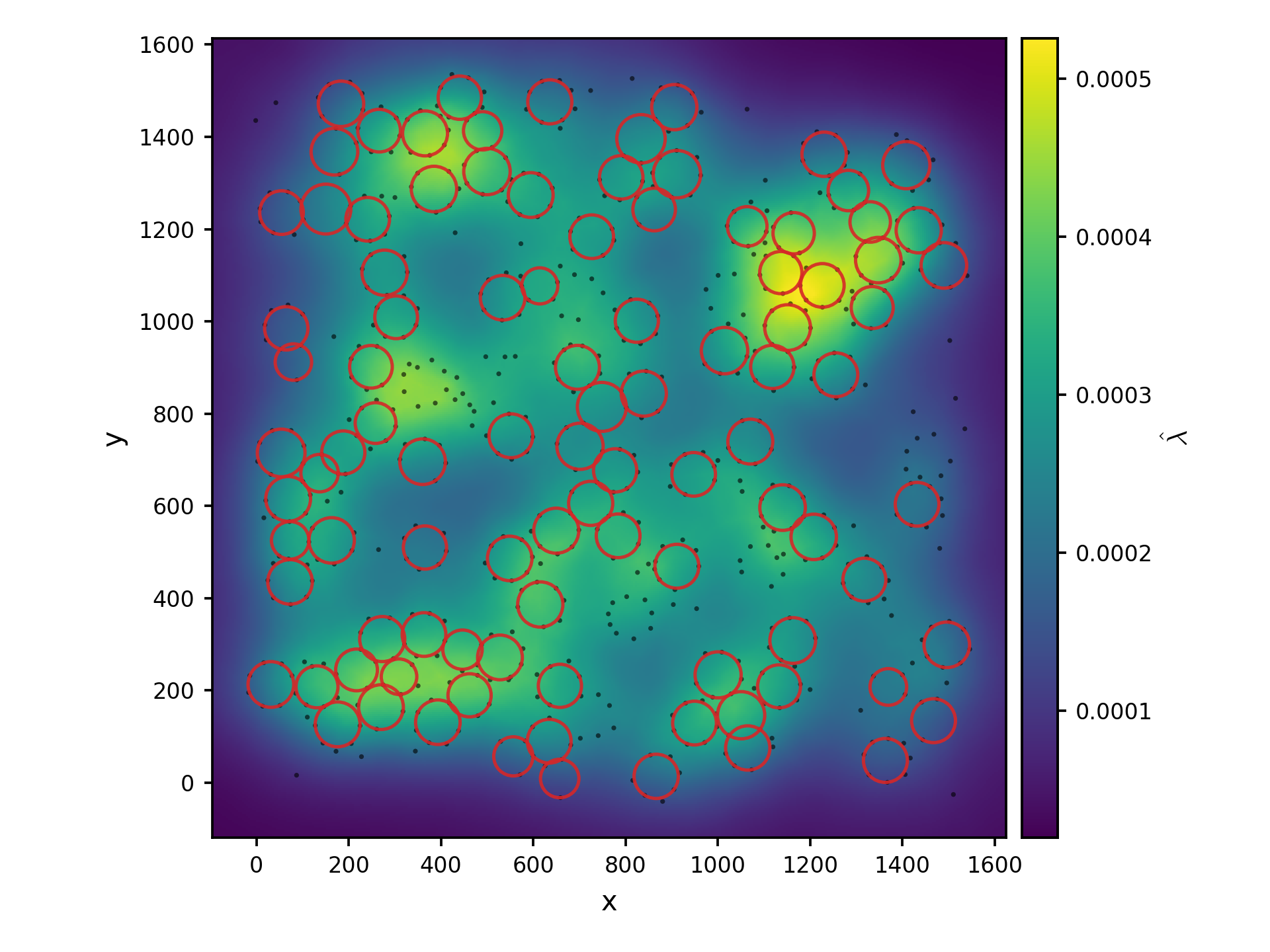}}
    \subfloat[]{\includegraphics[width=0.3\linewidth]{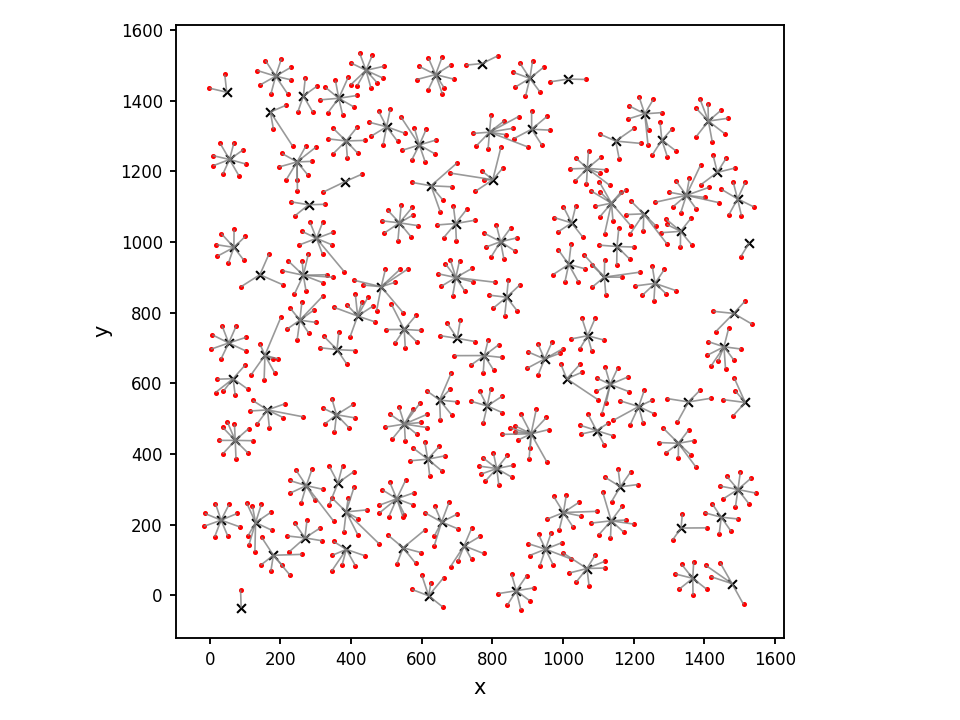}}
    \subfloat[]{\includegraphics[width=0.3\linewidth]{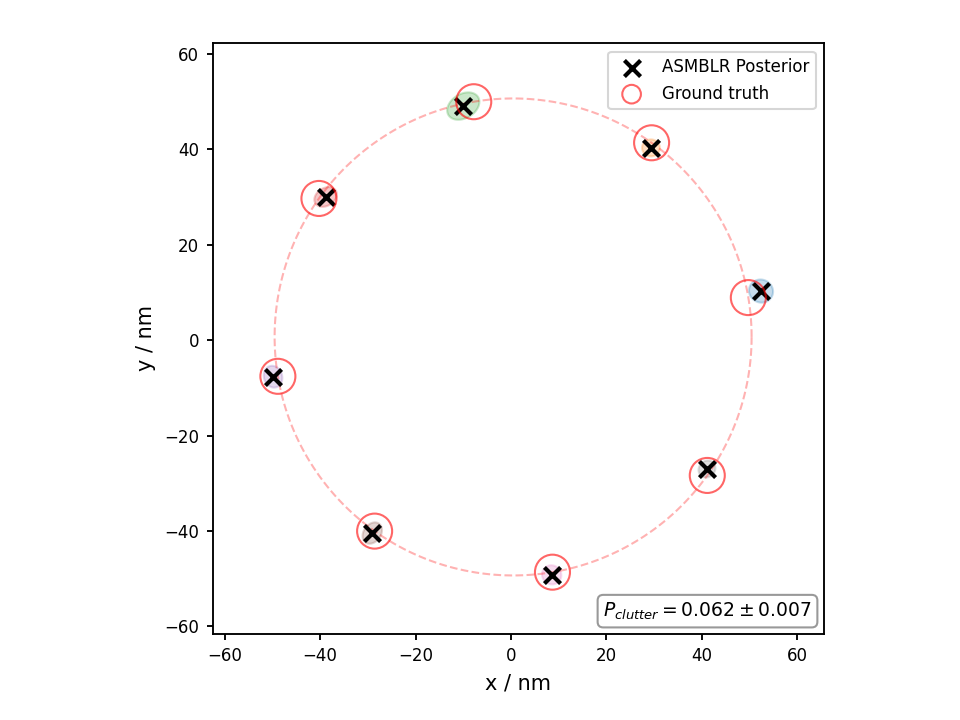}}
    \caption{0.9 labelling with 0.1 clutter. (a) Localisations. (b) GROUPA estimated emitters. (c) LGCP intensity map. (d) Voidwalker-discovered significantly empty space. (e) RJMCMC emitter-centre assignments. (f) ASMBLR reconstruction.}
    \label{fig:end-0p9-0p1}
\end{figure}

\begin{figure}[ht]
    \centering
    \subfloat[]{\includegraphics[width=0.3\linewidth]{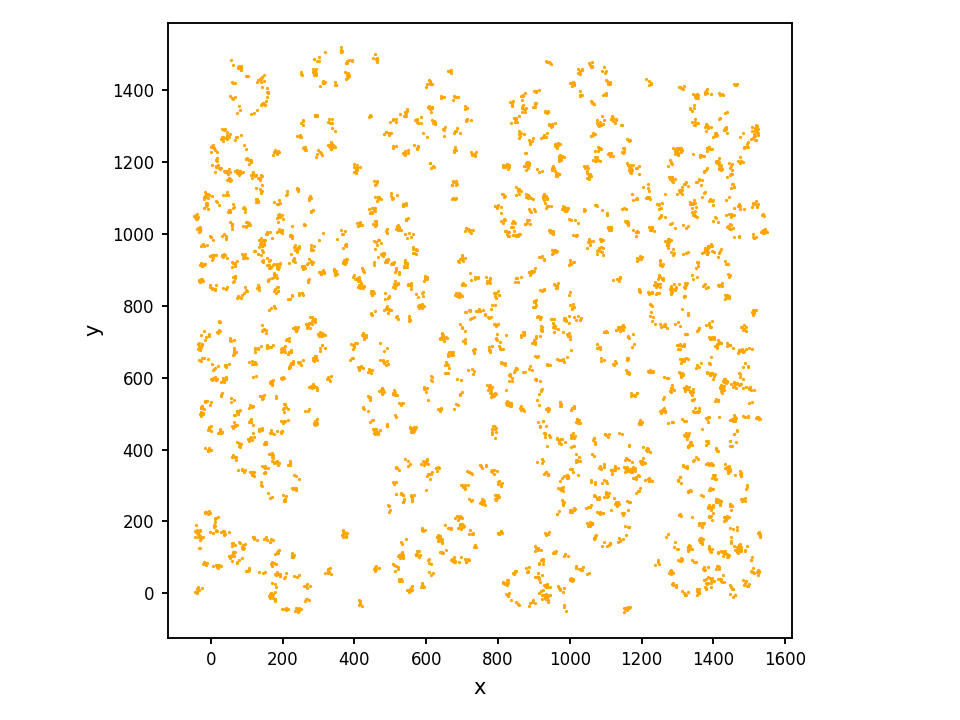}}
    \subfloat[]{\includegraphics[width=0.3\linewidth]{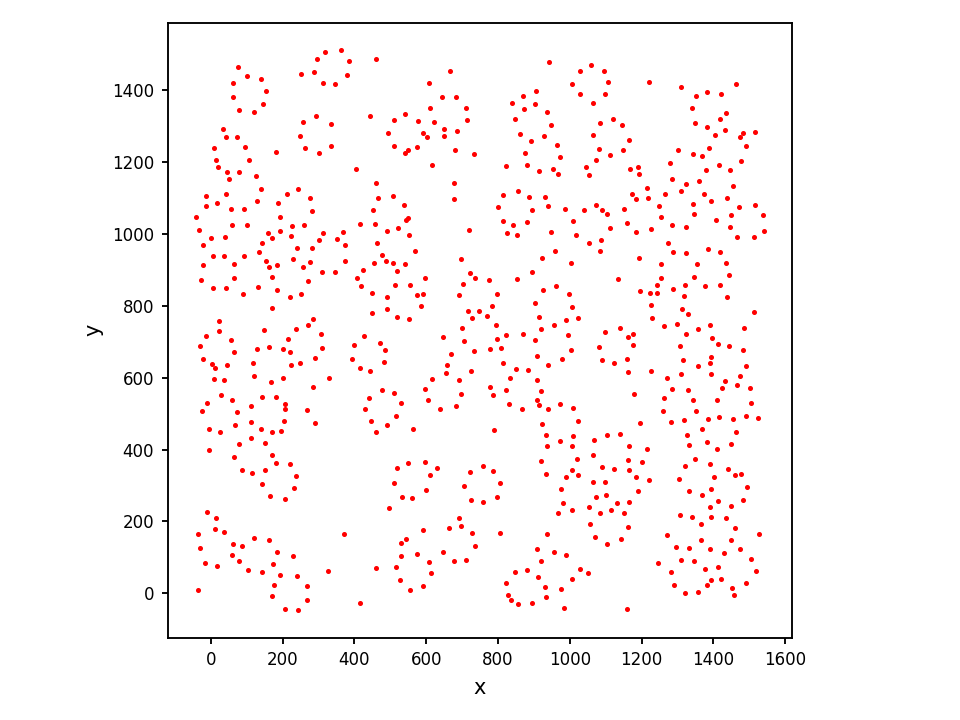}}
    \subfloat[]{\includegraphics[width=0.3\linewidth]{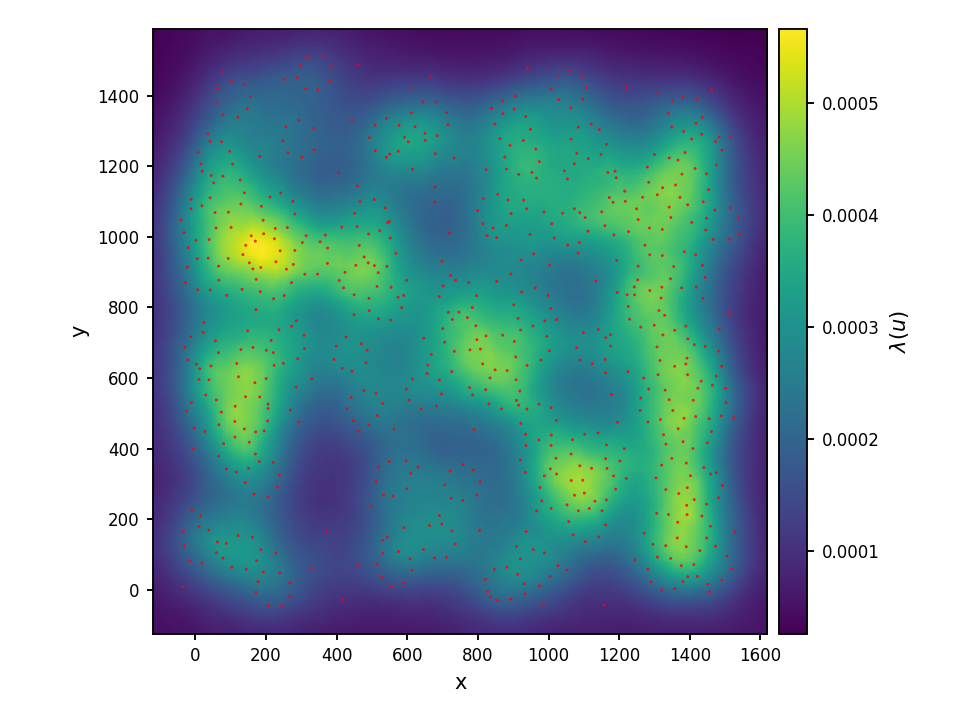}} \\
    \subfloat[]{\includegraphics[width=0.3\linewidth]{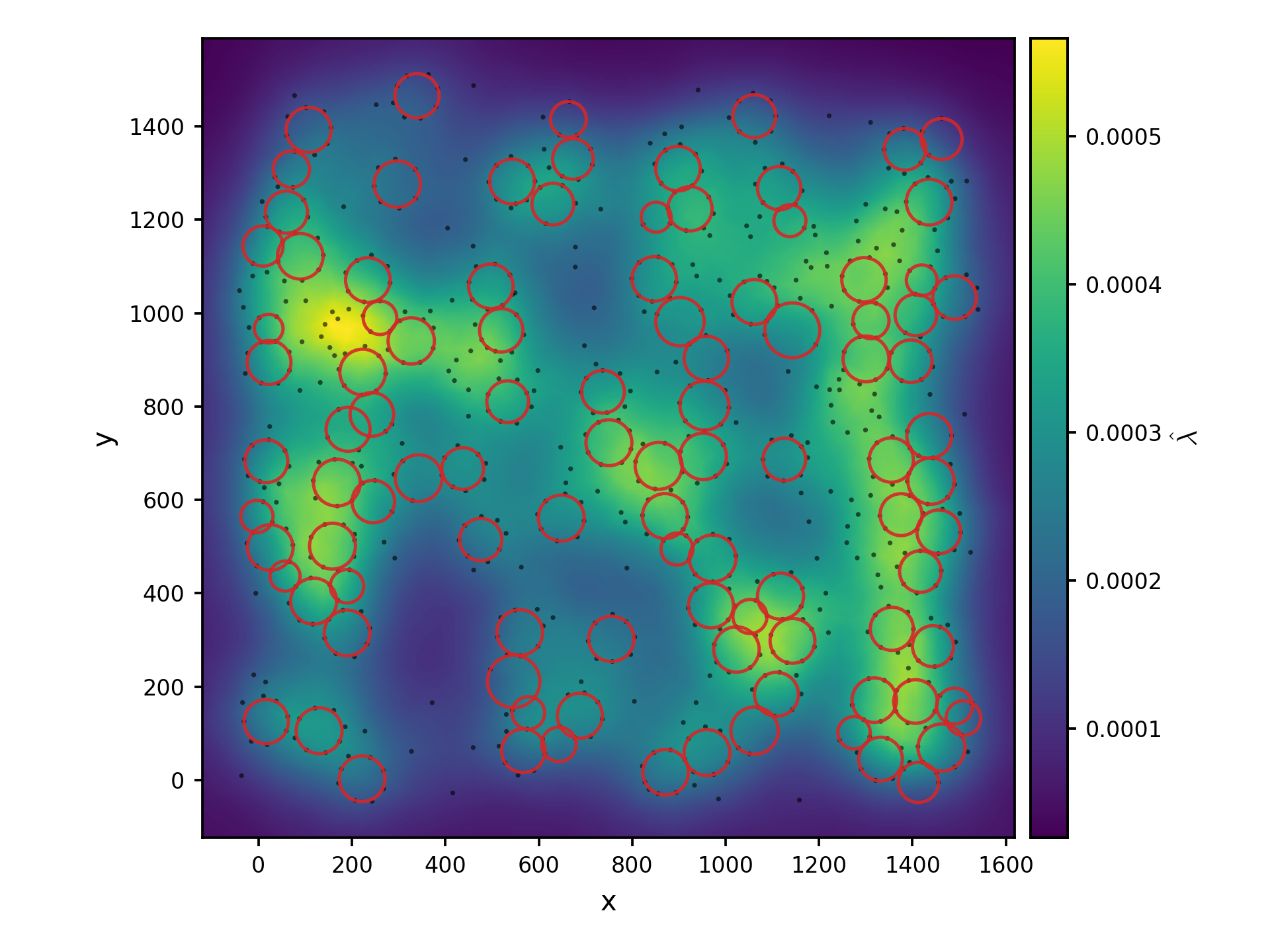}}
    \subfloat[]{\includegraphics[width=0.3\linewidth]{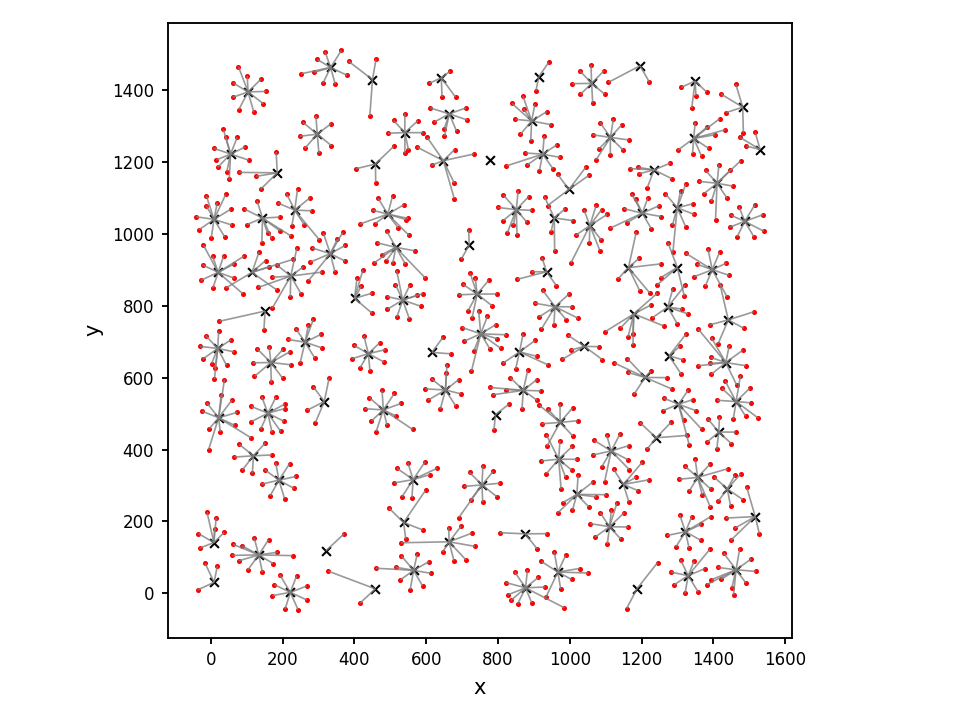}}
    \subfloat[]{\includegraphics[width=0.3\linewidth]{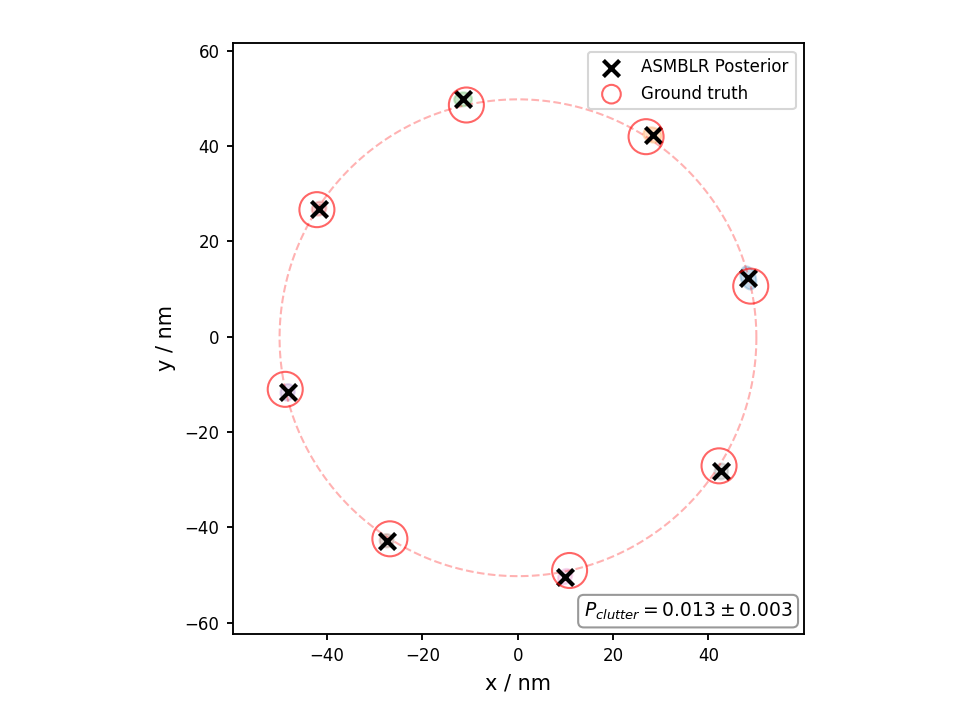}}
    \caption{1.0 labelling with 0.3 clutter. (a) Localisations. (b) GROUPA estimated emitters. (c) LGCP intensity map. (d) Voidwalker-discovered significantly empty space. (e) RJMCMC emitter-centre assignments. (f) ASMBLR reconstruction.}
    \label{fig:end-1p0-0p3}
\end{figure}

\FloatBarrier
\bibliographystyle{unsrt}
\bibliography{supplementary_references}{}